\documentclass[twocolumn, times]{aastex631}
\usepackage{amsmath}
\usepackage{soul}

\defcitealias{Main2021}{Paper~I}

\begin{document}

\title{Resolving the Emission Regions of the Crab Pulsar's Giant Pulses II. Evidence for Relativistic Motion.}
\shorttitle{Resolving the Emission Regions of the Crab Pulsar's Giant Pulses II.}

\author[0000-0003-4530-4254]{Rebecca Lin}
\affil{Department of Astronomy and Astrophysics, University of Toronto, 50 St. George Street, Toronto, ON M5S 3H4, Canada}

\author[0000-0002-5830-8505]{Marten H. van Kerkwijk}
\affil{Department of Astronomy and Astrophysics, University of Toronto, 50 St. George Street, Toronto, ON M5S 3H4, Canada}

\author[0000-0002-7164-9507]{Robert Main}
\affil{Max-Planck-Institut f\"{u}r Radioastronomie, Auf dem H\"{u}gel 69, 53121, Bonn, Germany}

\author[0000-0002-6317-3190]{Nikhil Mahajan}
\affil{Department of Astronomy and Astrophysics, University of Toronto, 50 St. George Street, Toronto, ON M5S 3H4, Canada}

\author[0000-0003-2155-9578]{Ue-Li Pen}
\affil{Institute of Astronomy and Astrophysics, Academia Sinica, Astronomy-Mathematics Building, No. 1, Sec. 4, Roosevelt Road, Taipei 10617, Taiwan}
\affil{Canadian Institute for Theoretical Astrophysics, 60 St. George Street, Toronto, ON M5S 3H8, Canada}
\affil{Canadian Institute for Advanced Research, 180 Dundas St West, Toronto, ON M5G 1Z8, Canada}
\affil{Dunlap Institute for Astronomy and Astrophysics, University of Toronto, 50 St George Street, Toronto, ON M5S 3H4, Canada}
\affil{Perimeter Institute of Theoretical Physics, 31 Caroline Street North, Waterloo, ON N2L 2Y5, Canada}

\author[0000-0001-6664-8668]{Franz Kirsten}
\affil{ASTRON, Netherlands Institute for Radio Astronomy, Oude Hoogeveensedijk 4, 7991 PD Dwingeloo, The Netherlands}
\affil{Department of Space, Earth and Environment, Chalmers University of Technology, Onsala Space Observatory, 439 92, Onsala, Sweden}

\correspondingauthor{Rebecca Lin}
\email{lin@astro.utoronto.ca}

\begin{abstract}
  The Crab Pulsar is the prime example of an emitter of giant pulses.
  These short, very bright pulses are thought to originate near the light cylinder, at $\sim\!1600{\rm\;km}$ from the pulsar.
  The pulsar's location inside the Crab Nebula offers an unusual opportunity to resolve the emission regions, using the nebula, which scatters radio waves, as a lens.
  We attempt to do this using a sample of 61998 giant pulses found in coherently combined European VLBI network observations at $18{\rm\;cm}$.
  These were taken at times of relatively strong scattering and hence good effective resolution, and from correlations between pulse spectra, we show that the giant pulse emission regions are indeed resolved.
  We infer apparent diameters of $\sim\!2000$ and $\sim\!2400{\rm\;km}$ for the main and interpulse components, respectively, and show that with these sizes the correlation amplitudes and decorrelation timescales and bandwidths can be understood quantitatively, both in our observations and in previous ones.
  Using pulse-spectra statistics and correlations between polarizations, we also show that the nebula resolves the nanoshots that comprise individual giant pulses.
  The implied diameters of $\sim\!1100{\rm\;km}$ far exceed light travel-time estimates, suggesting the emitting plasma is moving relativistically, with $\gamma\simeq10^{4}$, as inferred previously from drifting bands during the scattering tail of a giant pulse.
  If so, the emission happens over a region extended along the line of sight by $\sim\!10^{7}{\rm\;km}$.
  We conclude that relativistic motion likely is important for producing giant pulses, and may be similarly for other sources of short, bright radio emission, such as fast radio bursts.
\end{abstract}


\section{Introduction}\label{sec:intro}

The Crab Pulsar (\object{PSR B0531+21}) is the remnant of supernova SN 1054 and the central star in the \object{Crab Nebula}.
It powers the pulsar wind nebula (PWN) that fills the interior of the Crab Nebula.
The PWN is expanding into the freely expanding supernova ejecta \citep{Chevalier1977}, sweeping up ejecta into a dense thin shell of material.
This shell is subject to Rayleigh-Taylor instabilities leading to the filamentary structure seen in optical images of the Crab Nebula (e.g., \citealt{ChevalierGull1975, Jun1998}) and the observed acceleration of the Crab filaments \citep{Trimble1968}.

The pulsar itself was discovered by \cite{Staelin1968} through the detection of individual radio pulses and has since been studied extensively (for a review, see \citealt{Eilek2016}).
Its mean radio profile shows seven components, the dominant components of which are the main pulse (MP) and the interpulse (IP) at frequencies $<\!4{\rm\;GHz}$, which are separated by $\sim\!145\arcdeg$ in rotation phase.

The Crab Pulsar is unusual among pulsars in that it shows ``giant pulses'', extremely narrow and bright pulses that occur randomly within the phase windows of its MP and IP components \citep{Hankins2003}.
At lower observing frequencies ($<\!4{\rm\;GHz}$), the properties of the MP and IP giant pulses are quite similar, with pulses typically lasting overall a few microseconds, but comprised of numerous nanoshots that have durations down to the time resolution with which they have been observed, some clumping together in microbursts \citep{Sallmen1999, Hankins2007}.
At radio frequencies above $\sim\!4\rm{\;GHz}$, there is also an interpulse \citep{Moffett1996}. But it is offset by $\sim\!10\arcdeg$ in phase from the IP at low frequencies ($\sim\!0.3\!-\!3.5\rm{\;GHz}$) and has properties sufficiently different that it is almost certainly a different pulse component (e.g., \citealt{Hankins2016}).

The emission mechanism of giant pulses is unknown.
However, the strong alignment of the MP and (low frequency) IP components from radio to $\gamma$ observations to within $\sim\!2{\rm\;ms}$ \citep{Moffett1996} offers empirical evidence that both the radio and high energy emissions emanate from the same spatial region. For the MP, this is strengthened by the correlation between radio giant pulses and optical \citep{Shearer2003, Shearer2012} and X-ray pulses \citep{Enoto2021}.
Since pair production strongly absorbs $\gamma$-rays near the polar cap, giant pulse emissions likely occurs in other magnetospheric gaps \citep{Romani1995, Muslimov2004, Harding2008} or in regions beyond the light cylinder, $r_{\text{LC}} = cP/2\pi\approx1600{\rm\;km}$, \citep{Lyubarskii1996, Philippov2019} where plasma instabilities and magnetic reconnection can occur.

To help understand the emission mechanism, it would be useful to have constraints on the locations and sizes of the emission regions.
For this purpose, it may be possible to use the fact that pulsars scintillate, displaying intensity variations in time and frequency because of multipath propagations of their radio emission: if one can retrace those paths, they act like an interstellar interferometer with which the emission can be studied at extreme angular resolution.
\cite{Cordes1983} first used scintillation in this way to probe pulsar magnetospheres, by inferring limits on the transverse separations of the emitting regions of PSR~B0525+21 and PSR B1133+16.
Since then, this method has been applied to several other pulsars \citep{Wolszczan1987, Smirnova1996, Gupta1999, Pen2014}.

For the Crab Pulsar, observations of its scintillation show a fairly stable angular broadening, and a highly variable temporal broadening \citep{Rankin1973}.
VLBI measurements favor two screen locations with angular broadening originating from the interstellar medium (ISM) and (most) temporal scattering originating at the filaments inside the Crab Nebula \citep{Vandenberg1976}.
For resolving the Crab's emission, it is the nebular screen that is most relevant, as it is closer to the pulsar and thus gives higher spatial resolving power.
Hence, we focus on the nebular screen in this paper.

For individual Crab giant pulses, if they originated from the same physical location, one would expect that their radiation would follow the same paths through the scattering regions and thus that they would have imprinted on them the same impulse response function (IRF).
Hence, one would expect power spectra of pulses close in time to correlate strongly.
Indeed, \cite{Cordes2004} found that pulses close in time correlated strongly, with an average correlation coefficient of $1/3$, as would be expected for pulses that had different intrinsic frequency structure -- associated with the random nanoshots that they are comprised off -- but that were imprinted with the same IRF.
Similar levels of correlation were also found by \cite{Karuppusamy2010} for spectra of MPs and IPs in the same pulse rotation, as well as for spectra between different microbursts of individual MPs.

In contrast, \citeauthor{Main2021} (\citeyear{Main2021}, hereafter \citetalias{Main2021}) found a surprisingly low correlation coefficient of $\sim\!2\%$ when correlating power spectra of nearby MP-MP pairs and MP-IP pairs.
They suggested this could result if individual giant pulses arose in different parts of an extended emission region, one larger than the resolution of the scattering screen.
If so, then the nebular screen must have had lower effective resolution during the observations of \cite{Cordes2004} and \cite{Karuppusamy2010}; this is indeed consistent with those observations having had smaller scattering time when scaled to our observing frequency (see derivation in Section~\ref{sec:interstellar_interferometry}).
With the large sample of MPs and IPs in \citetalias{Main2021}, it was also possible to see differences between the scintillation patterns of the two components, hinting at a projected physical separation between the emitting regions of $50\!-\!400{\rm\;km}$.

Another result found in \citetalias{Main2021}, the importance of which we realized only later, was that the mean power in giant pulse spectra was approximately equal to the standard deviation.
This is not consistent with multiple nanoshots going through the same screen, as in that case the standard deviation should be $\sqrt{3}$ times larger than the mean (making the usual assumption that the IRF has Gaussian statistics; \citealt{Rickett1990}).
Instead, it suggests that the emission locations of the individual nanoshots are also separated sufficiently for them to be resolved by the nebular screen, and that, therefore, the scintillation pattern imprinted on each nanoshot is different.
This has the paradoxical implication that the nanoshots, which occur within a few $\mu{\rm s}$ in the giant pulse and thus must be causally related to each other, appear to have separations on the sky in excess of the $\sim\!300{\rm\;km}$ resolution found in \citetalias{Main2021}, i.e., far more than can naively be understood from light-travel time arguments.

In this paper, we use another dataset to get a more complete picture of the Crab's emission regions.
We begin with a brief review on how scattering screens close to pulsars can be seen as magnifying lenses in Section~\ref{sec:interstellar_interferometry}, and infer the expected resolution for the Crab Pulsar.
In Section~\ref{sec:observations} we describe our multi-telescope data sets and in Section~\ref{sec:data_reduction} their reduction, including the coherent combination of the data from different telescopes.
We measure scattering times from our data in Section~\ref{sec:scattering_timescale}, and correlations between pulses and polarizations in Section~\ref{sec:analysis}.
The latter show that in our data both giant pulses and their constituent nanoshots are imprinted with different IRFs.
In Section~\ref{sec:emission_regions}, we discuss the ramifications for the sizes of the emission regions and for the screen.
We address the paradox of the resolved nanoshots and conclude it is most easily resolved if the plasma emitting the pulses moves highly relativistically, as has also been suggested by \cite{Bij2021} based on drifting frequency structure in the scattering tail of giant pulses.
We finish with an outlook in Section~\ref{sec:future_work}.

\section{Interstellar Interferometry}\label{sec:interstellar_interferometry}

A scattering screen can be seen as a lens, which yields a certain resolution at the pulsar, over which the interference pattern changes by order unity.
The resolution depends on where the scattering occurs, improving as the screen is placed closer to the pulsar.
If the probability of scattering is normally distributed around the line of sight -- which yields an exponential scattering tail as is often observed -- the variance of the distribution is related to the exponential decay time $\tau$ by,
\begin{equation}
  \sigma_L = d_{\text{s}}\left(\frac{c\tau}{d_{\text{eff}}}\right)^{1/2},
\label{eqn:screen_size}
\end{equation}
where $d_{\text{s}}$ is the distance to the scattering screen and $d_{\text{eff}}=d_{\text{p}}d_{\text{s}}/(d_{\text{p}}-d_{\text{s}})$ the effective distance, with $d_{\text{p}}$ the distance to the pulsar.
For these normally distributed scatterers, the corresponding angular resolution is $\lambda/2\pi\sigma_L$, where $\lambda$ is the observing wavelength.
Hence, the physical resolution at the pulsar is given by \citep{cordes1998},
\begin{equation}
  \sigma_x = (d_{\text{p}}-d_{\text{s}})\frac{\lambda}{2\pi\sigma_L}
  = \frac{\lambda}{2\pi}\left(\frac{d_{\text{p}} - d_{\text{s}}}{c\tau}\frac{d_{\text{p}}}{d_{\text{s}}}\right)^{1/2}.
\label{eqn:resolution}
\end{equation}

In the Crab Nebula, as discussed in \citetalias{Main2021}, the only conceivable location for scattering is in the optically emitting filaments, as only their densities are high enough (of order $n_{\text{e}}\approx10^3{\rm\;cm^{-3}}$ \citealt{Osterbrock1957}).
\cite{Lawrence1995} and \cite{Martin2021} show that these filaments reside within $0.5\!-\!2{\rm\;pc}$ of the pulsar and \cite{Trimble1973} give a range of distances to the pulsar from $1.4\!-\!2.7{\rm\;kpc}$.
For an estimate of the resolution $\sigma_x$ of the lens at the pulsar, we use a nominal pulsar distance of $d_{\text{p}}=2{\rm\;kpc}$ and distance of the screen from the pulsar $d_{\text{p}}-d_{\text{s}}=1{\rm\;pc}$ along with a geometric time delay $\tau = 1{\rm\;\mu s}$ at $\lambda=18{\rm\;cm}$ \citepalias{Main2021}, to infer $\sigma_x\approx290{\rm\;km}$.

The above only holds if the scattering time is dominated by delays in the nebula.
It is known, however, that there is a contribution from the interstellar medium, which may be important at times that the scattering in the nebula is relatively weak \citep{Rankin1973,Vandenberg1976}.
We can estimate its contribution by assuming that the interstellar scattering time does not vary much, so that an upper limit to its contribution is set by the lowest observed scattering times.
\cite{Losovsky2019} find that at $111{\rm\;MHz}$ the scattering time varies between 10 and $115{\rm\;ms}$, while Serafin-Nadeau et al. (in preparation) find that at $600{\rm\;MHz}$ the lowest scattering times are $\sim\!10{\rm\;\mu s}$.
Scaling with $\nu^{-4}$, both results imply an upper limit of $\sim\!0.2{\rm\;\mu s}$ at $\lambda=18{\rm\;cm}$.
If the observed scattering time is close to this range, the resolution due to the nebular screen will be poorer than inferred from Equation~\ref{eqn:resolution}.

We can verify the above result by noting that the interstellar screen dominates the angular broadening of the Crab Pulsar, and at $\lambda=18{\rm\;cm}$, \citet{Rudnitskii2016} measure full width at half maximum $0.5\lesssim w\lesssim1.3{\rm\;mas}$.
The range likely reflects that the screen is not isotropic, but for an estimate we nevertheless assume an isotropic screen with a normal distribution, such that the lens size is $\sigma_L=d_{\text{s}}(w/\sqrt{8\log2})$.
For a screen halfway to the pulsar, Equation~\ref{eqn:screen_size} then would imply a range in scattering time of $0.2\lesssim\tau\lesssim1.5{\rm\;\mu s}$.
In reality, since the Crab is relatively far above the Galactic plane, the interstellar screen is likely closer to us than halfway, which would make the result more consistent with the upper limit above.
Assuming that the actual value of the interstellar scattering time is close to the upper limit, $\tau=0.2{\rm\;\mu s}$, then to get the maximum observed angular broadening, $w=1.3{\rm\;mas}$, requires a screen distance of $0.24{\rm\;kpc}$.
Without any nebular contribution, this geometry would imply a resolution  $\sigma_x\simeq80,000{\rm\;km}$.

\section{Observations}\label{sec:observations}

\begin{deluxetable*}{clclcccccc}[ht!]
\tabletypesize{\small}
\tablecaption{Observation and Giant Pulse Log\label{table:log}}
\tablenum{1}
\tablehead{\colhead{Observation}&
  &
  \colhead{$t_{\text{exp}}$ \tablenotemark{a}}&
  &
  \colhead{DM \tablenotemark{c}}&
  \multicolumn{5}{c}{\dotfill Giant Pulses \tablenotemark{d}\dotfill}\\[-.7em]
  \colhead{code}&
  \colhead{Date}&
  \colhead{(h)}&
  \colhead{Telescopes used \tablenotemark{b}}&
  \colhead{(${\rm pc\;cm^{-3}}$)}&
  \colhead{$N$}&
  \colhead{$N_{\text{MP}}$}&
  \colhead{$N_{\text{IP}}$}&
  \colhead{$r_{\text{MP}}$ (${\rm s^{-1}}$)}&
  \colhead{$r_{\text{IP}}$ (${\rm s^{-1}}$)}}
\startdata
EK036 A & 2015 Oct 18-19   & 3.27 & Ef, Bd, Hh, Jb, O8, Sv, Wb, Zc     & 56.7772 & 21735 & 18941 & 2794 & 1.6082 & 0.2372 \\
EK036 B & 2016 Oct 31-Nov1 & 1.65 & Ef, Bd, Hh, O8, Sv, Wb, Zc         & 56.7668 & 18891 & 15176 & 3715 & 2.5618 & 0.6271 \\
EK036 C & 2017 Feb 25      & 1.15 & Ef, Bd, Hh, Jb, O8, Sv, Wb, Zc     & 56.7725 &  8399 &  7203 & 1196 & 1.7433 & 0.2895 \\
EK036 D & 2017 May 28      & 1.25 & EF, Bd, Hh, Jb-II, O8, Sv, Wb, Zc  & 56.7851 & 12973 & 10725 & 2248 & 2.3881 & 0.5006
\enddata
\tablenotetext{a}{ Total on-source time, i.e., excluding telescope setup and calibration.}
\tablenotetext{b}{ Telescope abbreviations are: Ef: Effelsberg; Bd: the $32{\rm\;m}$ at Badary; Hh: the $26{\rm\;m}$ in Hartebeesthoek; Jb: the Lovell telescope; Jb-II: Mark II Telescope at the Jodrell Bank Observatory; O8: the $25{\rm\;m}$ at Onsala; Sv: the $32{\rm\;m}$ at Svetloe; Wb: a single dish from the Westerbork Synthesis Radio Telescope; and Zc: the $32{\rm\;m}$ at Zelenchukskaya.
Other telescopes participated in some of these observation runs, but we did not use their data because of a variety of problems.}
\tablenotetext{c}{ Inferred from giant pulses (see Section~\ref{subsec:pipeline}).}
\tablenotetext{d}{ The number of giant pulses and their rates (per second) listed here are found using a detection threshold of $8\sigma$ on coherently summed data in a $16{\rm\;\mu s}$ window (see Section~\ref{subsec:giant_pulses}). This corresponds to a limiting flux of about $15\!-\!18{\rm\;Jy}$, depending on the number of telescopes combined.}
\end{deluxetable*}

We analyse a total of $7.31{\rm\;hr}$ of European VLBI Network (EVN) dual-polarization data, taken at four epochs between 2015 Oct and 2017 May (see Table~\ref{table:log}).
For our analysis, we use data only from the up to 8 telescopes that had relatively clean signal in both polarizations and covered the full frequency range of $1594.49\!-\!1722.49{\rm\;MHz}$.
At each telescope, real-sampled data in both circular polarizations were recorded in either 2-bit MARK 5B or VDIF format, covering the frequency range in either eight contiguous $16{\rm\;MHz}$ wide bands or four contiguous $32{\rm\;MHz}$ wide bands.
During each observation run, the telescopes regularly switched to calibrator sources resulting in short breaks in our data.

The use of multiple telescopes at baselines of up to $10,000{\rm\;km}$ as an interferometer allows us to achieve a resolution of $\sim\!4{\rm\;mas}$.
For our purposes here, this is useful as it resolves the radio-bright nebula of angular size $\sim\!6\arcmin\times4\arcmin$, the dominant source of noise.
The angular resolution is not high enough, however, to resolve the nebular or interstellar scattering screens (which at our frequency have sizes of $\sim\!0.005$ and $\sim\!0.5$--$1.3{\rm\;mas}$, respectively; \citealt{Vandenberg1976,Rudnitskii2016}).

\section{Data Reduction}\label{sec:data_reduction}

In order to combine individual telescope data coherently and obtain sensitive measurements of giant pulses, we first need to align the voltage data in both time and phase.
The largest delays come from differences in path length travelled by the signal to each telescope.
We used the software program {\sc CALC10}\footnote{\url{https://space-geodesy.nasa.gov/techniques/tools/calc_solve/calc_solve.html}} \citep{Ryan1980} through a wrapper from the Super FX Correlator ({\sc SFXC}; \citealt{Keimpema2015}) to calculate these geometric delays.
The geocentric frame was chosen as the reference frame and the geometric delays were generated at one second intervals which we interpolate with an \cite{Akima1970} spline in our beamformer pipeline.
The second largest delays come from differences between each telescope's local clock.
We obtained the clock offset and rate information from the post-observation VEX file\footnote{\url{https://vlbi.org/vlbi-standards/vex/}} which were obtained from standard VLBI clock searching techniques. With these corrections, giant pulses common between telescopes are aligned to within $\sim\!10{\rm\;ns}$ (see Figure~\ref{fig:fringedelay}).
Lastly, cables and electronic components, and the atmosphere also introduce time delays and phase rotations.
Since the time-averaged emission of the Crab Pulsar is not very bright relative to the nebula, we determined these time delays and phase rotations using just the giant pulses themselves.

We present our beamformer pipeline in Section~\ref{subsec:pipeline}. In Section~\ref{subsec:giant_pulses} and Section~\ref{subsec:fringe_solution} we describe how we identified giant pulses from our data, and how we then use these giant pulses to find fringe solutions respectively. We present results of our coherently combined data in Section~\ref{subsec:combined_data}.

\subsection{Beamformer Pipeline}\label{subsec:pipeline}
Our pipeline closely follows that used for the Large European Array for Pulsars (LEAP), as described by \cite{Bassa2016, Smits2017} and SFXC as described by \cite{Keimpema2015}.

We first bring the signal from each telescope to the geocentric frame using the pre-determined geometric and clock delays.
Given a geocentric time $t_{\text{geo}}$, the corresponding time at a telescope is
\begin{equation}
 t_{\text{tel}} = t_{\text{geo}} + \tau_{\text{geo + clock, tel}}, \label{eqn:tel_time}
\end{equation}
where $\tau_{\text{geo + clock, tel}}$ is the geometric and clock delay for the telescope of interest.
We use {\sc Baseband} \citep{VanKerkwijk2020} and {\sc Pulsarbat} \citep{Mahajan2022} to read in the baseband data to the nearest integer time sample, flip lower-sidebands so that frequencies are all in increasing order, and convert the real-sampled data to complex\footnote{The conversion is done by computing the analytic representation of the signal via a Hilbert transform, removing the negative frequency components, and then shifting the signal down in frequency by half the bandwidth. $-B/2$, where $B$ is the bandwidth of the signal (either $16$ or $32{\rm\;MHz}$ for our data).}.
This real-to-complex conversion shifts the frequency from the edge of each sub-band to the center.
Since the integer geometric and clock delay compensation was applied when reading in the data (before the real-to-complex conversion), we need to shift the integer delay compensation frequency to the center of the sub-band. This was done by applying the shift
\begin{equation}
    \phi = -\pi N^{\text{int}}_{\text{geo + clock, tel}}, \label{eqn:phase_shift}
\end{equation}
where $N^{\text{int}}_{\text{geo + clock, tel}}$ is the geometric and clock delay in integer number of samples.
The residual fractional delay manifests itself as a phase error across the frequency band and is removed by rotating the signal by
\begin{equation}
    \phi(\nu) = -2\pi\frac{ N^{\text{frac}}_{\text{geo + clock, tel}}}{\text{SR}}\nu, \label{eqn:fractional_delay}
\end{equation}
where $N^{\text{frac}}_{\text{geo + clock, tel}}$ is the fractional delay, $\nu$ is the baseband frequency and $\text{SR}$ is the sampling rate.
Due to varying radial velocities between the different telescopes, the signals are Doppler shifted relative to each other; we correct for this frequency shift by a phase rotation in the time domain,
\begin{equation}
    \phi(t) = 2\pi t\dot{\tau}_{\text{geo + clock, tel}}\nu_{\text{sky}}, \label{eqn:fringe_stopping}
\end{equation}
where $\dot{\tau}_{\text{geo + clock, tel}}$ is the time derivative of the geometric and clock delays (which encodes the Doppler shift), and $\nu_{\text{sky}}$ is the central observing frequency of the sub-band.
Lastly, the above time shifts in the down-converted baseband signal does not correct for the phase rotation of the radio waves at each telescope at the central observing frequency. We corrected for this by rotating the phase of the signal by,
\begin{equation}
    \phi = 2\pi \tau_{\text{geo + clock, tel}}\nu_{\text{sky}},
\end{equation}
a process called fringe stopping.

In the corrections outlined above, only single values of the delay and its rate are used.
This is only acceptable at short time scales where the delay is approximately constant.
Thus, we apply our corrections in data chunks of no more than 128 samples ($8{\rm\;us}$), which ensures drifts in delay of less than 0.000192 sample ($0.012{\rm\;ns}$) even for the maximum, $\sim\!1.5{\rm\;\mu s/s}$ delay rate on terrestrial baselines.

After applying the geometric and clock delays, we removed radio frequency interference (RFI) from our data in chunks of $2^{24}$ samples ($1.048576{\rm\;s}$).
We start by removing obvious RFI and those mentioned in each telescope's log files.
To detect residual RFI, we first channelized our data into 8192 frequency channels per sub-band.
We then normalized these by the square root of the time-averaged power spectrum, thus correcting for the fact that the passband is not perfectly flat, with roll-offs at the edges of the band and some other structures.
RFI spikes and highly variable channels were flagged by comparing with a 128-channel
median-filter ($0.25{\rm\;MHz}$), removing all signals above a $5\sigma$ cut-off.
We also normalized time-variability resulting from instrumentation by dividing by the square root of the frequency averaged power spectrum smoothed over $8.192{\rm\:ms}$.

Next, we coherently de-dispersed the data.
For the dispersion measure (DM), we started with initial guesses from Crab monitoring data\footnote{\url{http://www.jb.man.ac.uk/~pulsar/crab.html}} \citep{Lyne1993}, and then adjusted the value to ensure the profiles of bright giant pulses were aligned in frequency, leaving us with final values listed in Table~\ref{table:log}.

At this point, we incoherently summed the data from each telescope and searched for giant pulses (see Section~\ref{subsec:giant_pulses}) as the telescope data are aligned to within a sample in time.

Once we have modelled the delays and phase rotations resulting from instruments and atmospheric variations using giant pulses (using Effelsberg as a reference; see Section~\ref{subsec:fringe_solution}), we applied the fringe delay and phase solutions
\begin{equation}
    \phi(\nu) = -2\pi\nu\tau_{\text{fringe delay}} - \phi_{\text{fringe phase}} \label{eqn:fringe_solutions}
\end{equation}
to the baseband data.

After all the delay and phase corrections are applied we weighed the data from each telescope by the amplitude of their gains to ensure maximum S/N in the coherently added baseband data (given that the nebula dominates the system temperature, the different telescopes have very similar gain; see below).
A final normalization of the coherently summed voltage data was performed such that the intensity of the noise level is unity in each sub-band and polarization.
Finally, we applied a parallactic angle correction to the coherent beam, as appropriate for the Crab at our reference telescope (Effelsberg). The rotation measure (RM) towards the Crab Pulsar is $\sim45{\rm\;rad/m^{2}}$ \citep{Sobey2019}. Thus, at our observing wavelength of $18{\rm\;cm}$ the expected phase rotation across our total bandwidth is $\sim0.1{\rm\;rad}$.
Since this is small and we do not detect RM or cable delay in our sub-bands, we do not correct for it.

\subsection{Giant Pulse Search}\label{subsec:giant_pulses}

We searched for giant pulses in two separate passes, the first using incoherently summed data to determine fringe solutions and the second using coherently added data. In both passes we summed the power over both polarizations and all eight sub-bands.
Peaks above $8\sigma$ in a $16{\rm\;\mu s}$ wide running average of the intensity time stream (with the window size roughly matched to the typical width of a giant pulse) were flagged as potential giant pulses. We define the start time of a giant pulse as the start of the first window where we detect it.

We created a folded pulse profile from our list of potential giant pulses using polyco files generated with {\sc Tempo2} \citep{Hobbs2012}.
The polyco file contains a polynomial model of the pulsar phase as a function of time at the geocenter for our central observing frequency and observation window.
From the folded pulse profile, we determined the MP and IP phase windows and any potential giant pulses that did not fall within these phase windows were discarded. As we had removed most of the RFI from our data, there were only about $\sim\!10$ false detections per observation; visual inspection showed these were clearly RFI.
Given the narrowness of the pulse windows, our sample should thus contain no spurious pulses.

In Table~\ref{table:log}, we list the resulting number of MP and IP detections for each observation in our coherent data, as well as their rate of occurrence.
One sees that the rates vary significantly between the observations, something which has been seen before \citep{Bera2019}, but is not understood.

\subsection{Fringe Solution}\label{subsec:fringe_solution}

\begin{figure}
\centering
\includegraphics[width=0.47\textwidth,trim=0 0 0 0,clip]{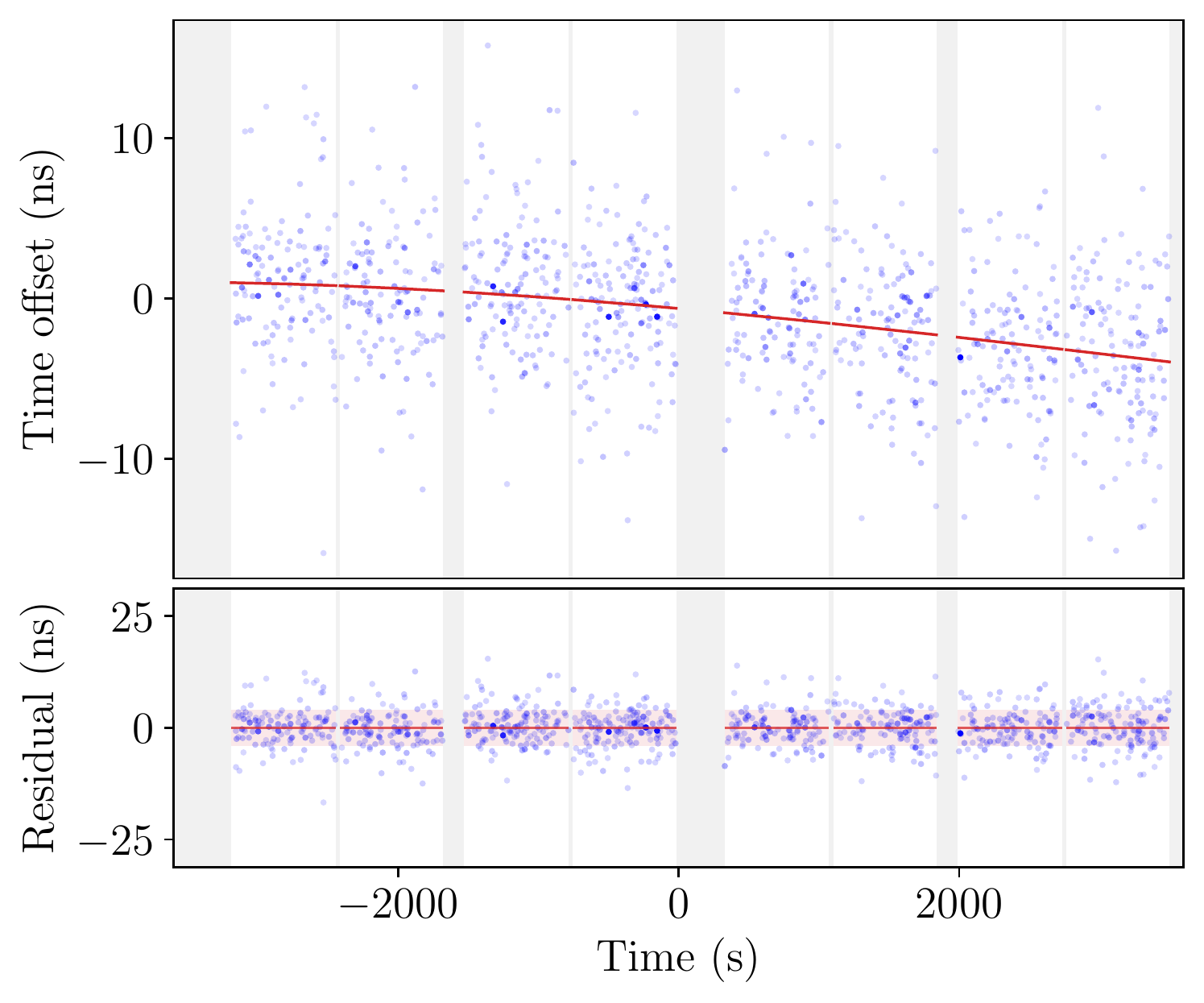}
\caption{
  Time delays between Ef and Bd from EK036~B for the frequency band $1594.49\!-\!1610.49{\rm\;MHz}$ in left circular polarization, after correcting for geometric and clock delays.
  The gray shaded regions indicate when the telescope was off-source.
  {\em Top:\/} Time offset due to remaining instrumental and ionospheric variations between the two telescopes, as tracked by giant pulses.
  The opacity of the individual points scales with the square root of the S/N of the giant pulse.
  A polynomial fit of the time offset is represented by the solid red line.
  {\em Bottom:\/} Residuals of our delay fit.
  The pink shaded range shows the root-mean-square scatter.
  The range of the y-axis corresponds to one time sample ($62.5{\rm\;ns}$).
  \label{fig:fringedelay}}
\end{figure}

\begin{figure}
\centering
\includegraphics[width=0.47\textwidth,trim=0 0 0 0,clip]{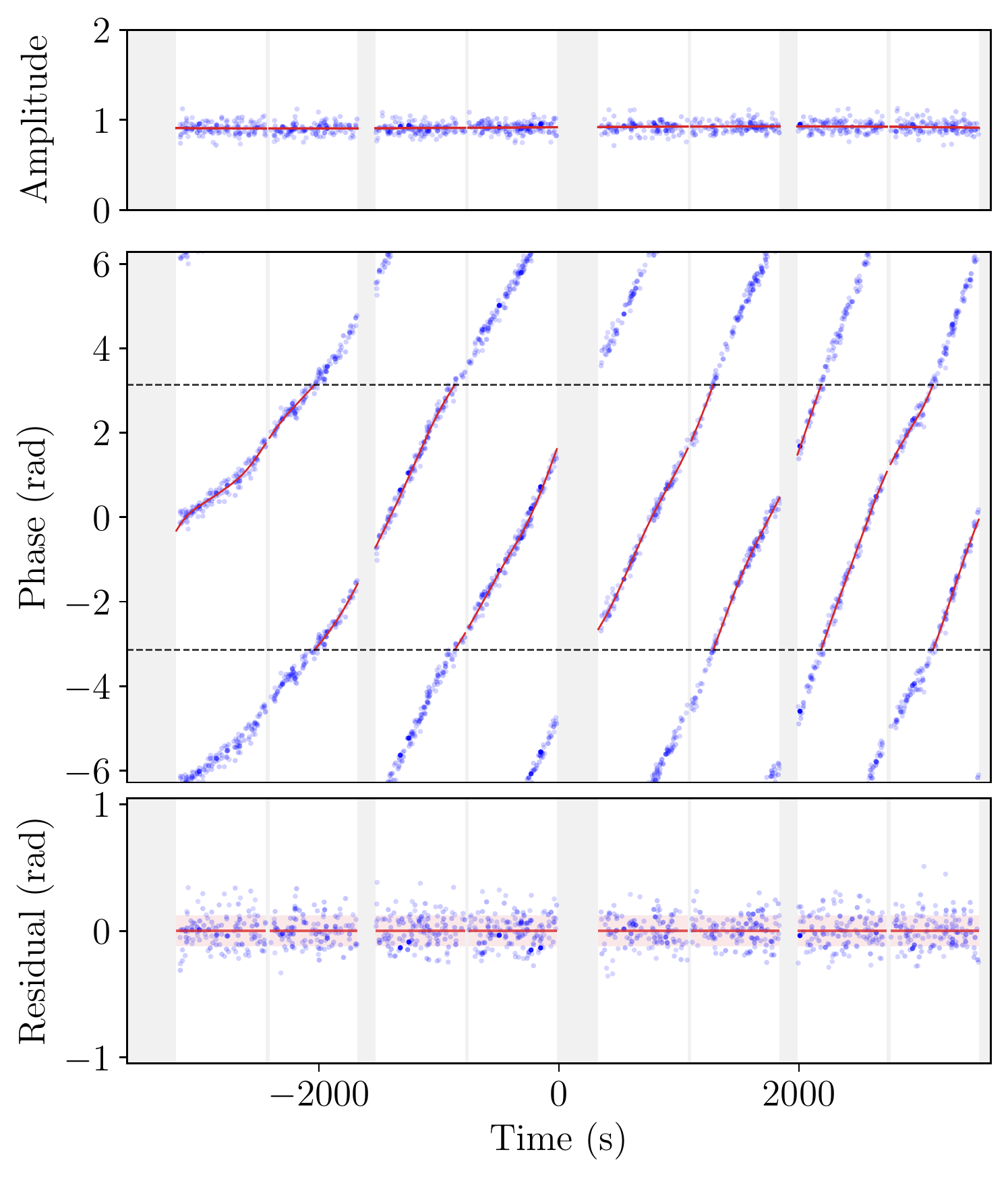}
\caption{
  Complex gain of Bd relative to Ef from EK036~B for the frequency band $1594.49\!-\!1610.49{\rm\;MHz}$ in left circular polarization.
  The gray shaded regions indicate when the telescope was off-source.
  {\em Top\/}: Relative amplitudes as inferred from giant pulses.
  The opacity of individual points indicate the S/N of the giant pulse.
  A third-degree polynomial fit of the amplitude is represented by the solid red line.
  Note that the amplitude is very close to unity, as expected given that the dominant source of noise for each telescope comes from the Crab Nebula.
  {\em Middle\/}: Relative fringe phases determined from performing an eigenvalue decomposition on matrices of visibilities of the individual giant pulses.
  The red line shows our fit.
  {\em Bottom\/}: Residuals of our phase fit, with the root-mean-square scatter indicated by the pink-shaded range.
  \label{fig:fringephase}}
\end{figure}

To determine the delay and phase models for coherent combination, we used giant pulses with a signal-to-noise $S/N>50$ (as measured on the incoherently summed data).
We chose Effelsberg to be the reference location and time standard because of its relatively clean signal.

In the first processing stage, after correcting for the geometric delays and clock offsets (see \ref{subsec:pipeline}), there will be further delays due to instrumental effects and ionospheric variations.
To measure these, we correlated giant pulses observed at each individual telescope with the reference telescope in voltage, and fit the resulting offsets using a polynomial across each observation, weighting the results for each pulse by its S/N.

Since each polarization and sub-band may be affected by instrumentation and the atmosphere differently, we modelled them separately.
We achieved fringe delay solutions for each baseline to Effelsberg with average root-mean-square deviation of $5{\rm\;ns}$, i.e., better than 8\% of a $62.5{\rm\;ns}$ time sample.
An example is shown in Figure~\ref{fig:fringedelay}.

In the second processing stage, after correcting for the fringe delays, we determined and modelled the fringe amplitudes and phases.
To solve for the time-varying complex telescope gains in each polarization and sub-band, we first cross-correlate the voltage series of giant pulses between pairs of telescopes, integrating over time, to form complex visibilities.
A matrix of complex visibilities was then created for each giant pulse,
\begin{equation}
    \mathbf{V} = \begin{bmatrix}
    0 & V_{1,2} & \cdots & V_{1,n}\\
    V_{2,1} & 0 & \cdots & V_{2,n}\\
    \vdots & \vdots & \ddots & \vdots \\
    V_{n,1} & V_{n,2} & \cdots & 0
    \end{bmatrix},
\label{eqn:visibility_matrix}
\end{equation}
where $V_{i,j}$ denotes the visibility of telescope pairs and the telescopes are numbered from 1 to $n$.
Note that auto-correlation data were not included (the diagonals are set to 0) and a minimum of 4 telescopes were used to create the visibility matrix in order to satisfy both closure phase and closure amplitude \citep{Thompson2017}, ensuring that any residual telescope based errors cancelled out.
We then performed an eigenvalue decomposition on these visibility matrices.
The eigenvectors corresponding to the dominant eigenmode give the relative complex gain between telescopes up to a complex constant.
The gain amplitudes were modelled by fitting a polynomial across the whole observation and the complex gains were modeled by a sum of sinusoids which we then convert to phases. We show an example of our gain calibration in Figure~\ref{fig:fringephase}.
The average root-mean-square deviation for our phase model is $0.11{\rm\;rad}$ (implying $S/N\simeq9$ per sub-band, polarization and telescope, consistent with what is expected given $S/N\simeq50$ for the incoherently summed signal over 8 telescopes, 8 sub-bands and 2 polarizations, taking into account that this is measured over $16{\rm\;\mu{}s}$, while most signal is within the scattering time of $\sim\!5{\rm\;\mu{}s}$).

\subsection{Combined Data}\label{subsec:combined_data}

\begin{figure}
\centering
\includegraphics[width=0.47\textwidth,trim=0 0 0 0,clip]{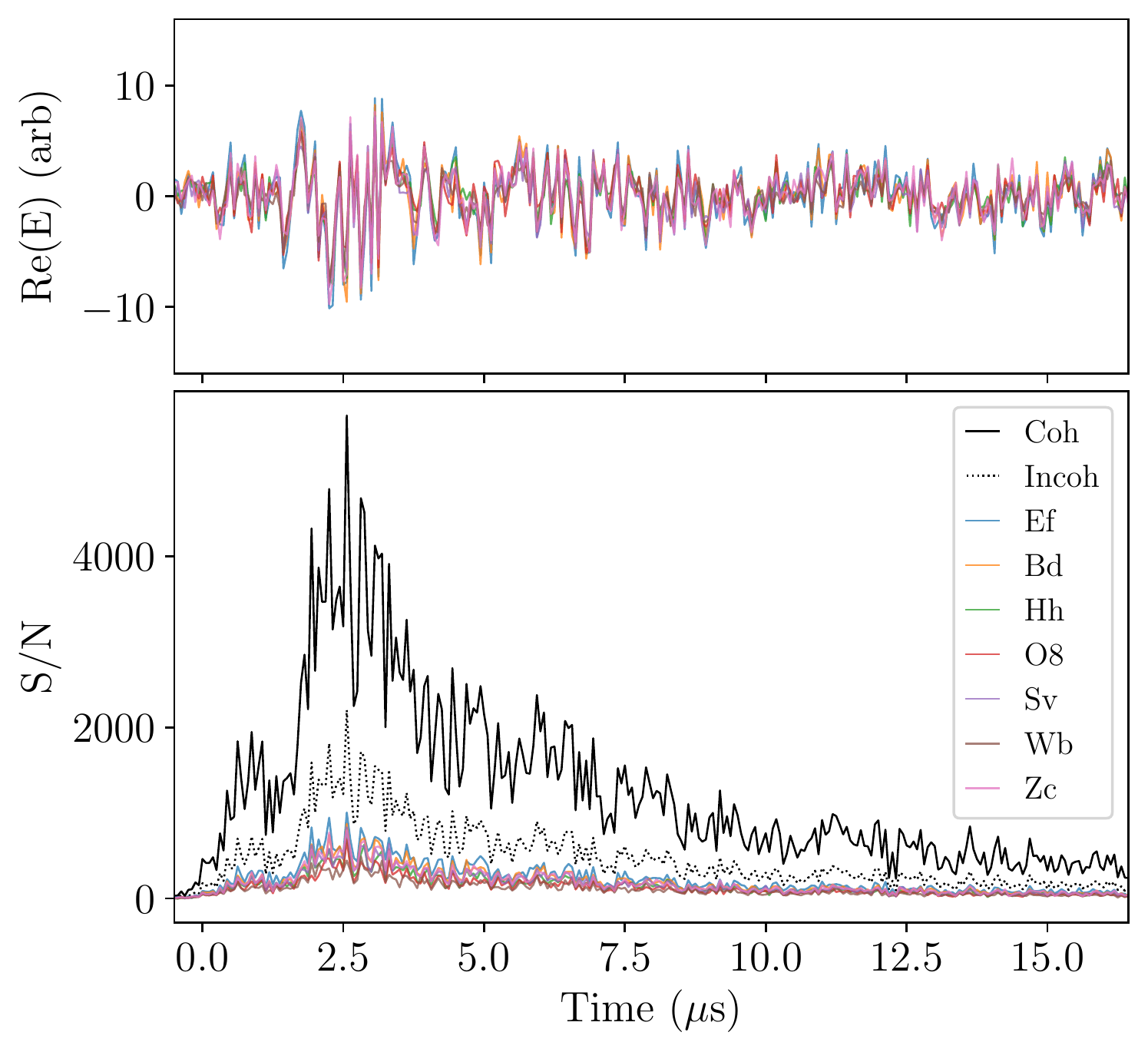}
\caption{
    {\em Top:\/} Delay and phase corrected complex baseband data in the frequency band $1594.49\!-\!1610.49{\rm\;MHz}$ and left circular polarizatin, of the brightest giant pulse in EK036~B at multiple telescopes.
    {\em Bottom:\/} Pulse profile of this pulse (which happens to be a MP) as detected at each telescope (coloured lines), and after summing incoherently (dotted black line) and coherently (solid black line).
    For these profiles, the intensities in all sub-bands and both polarizations were summed, and were divided by the background, so that the profile is in $S/N$ units.
    The time resolution is $62.5{\rm\;ns}$.
    \label{fig:coherance}}
\end{figure}

\begin{figure*}
\centering
\includegraphics[width=0.985\textwidth,trim=0 0 0 0,clip]{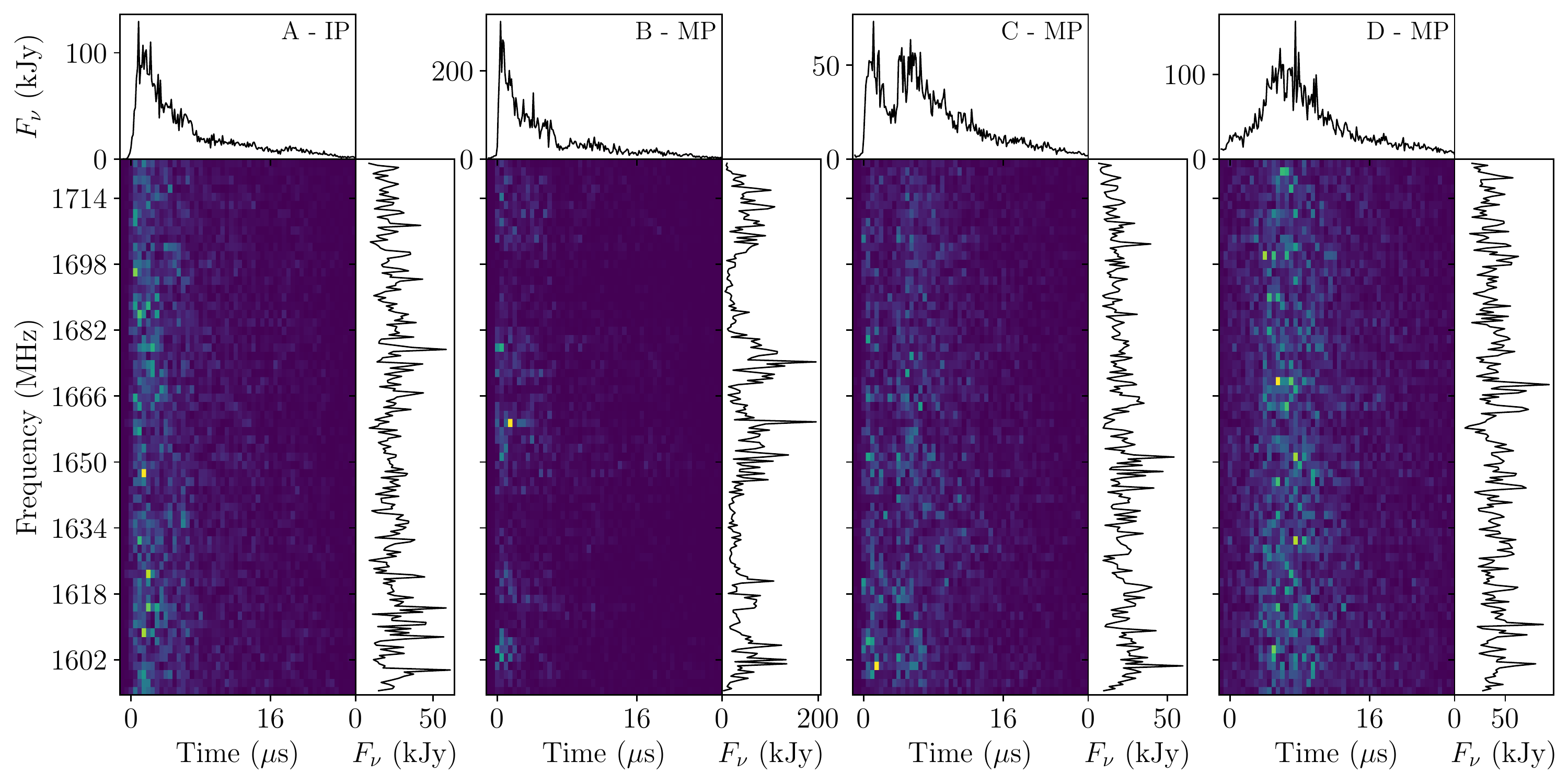}
\caption{
    {\em Images:\/} Waterfalls of giant pulses with $500{\rm\;ns}$ time resolution and $2{\rm\;MHz}$ frequency resolution from EK036~B.
    {\em Top panels:\/} Pulse profiles in $250{\rm\;ns}$ bins.
    {\em Right panels:\/} Pulse spectra containing emission from $0\!-\!16{\rm\;\mu s}$, in $500{\rm\;kHz}$ channels.
    \textbf{A}: an IP arriving with a strong initial burst;
    \textbf{B}: a MP pulse with banded spectra;
    \textbf{C}: a MP with two distinct peaks; and
    \textbf{D}: a MP that increases slowly in emission power, likely because it also has multiple components.
    \label{fig:waterfall_zoo}}
\end{figure*}

\begin{figure}
\centering
\includegraphics[width=0.47\textwidth,trim=0 0 0 0,clip]{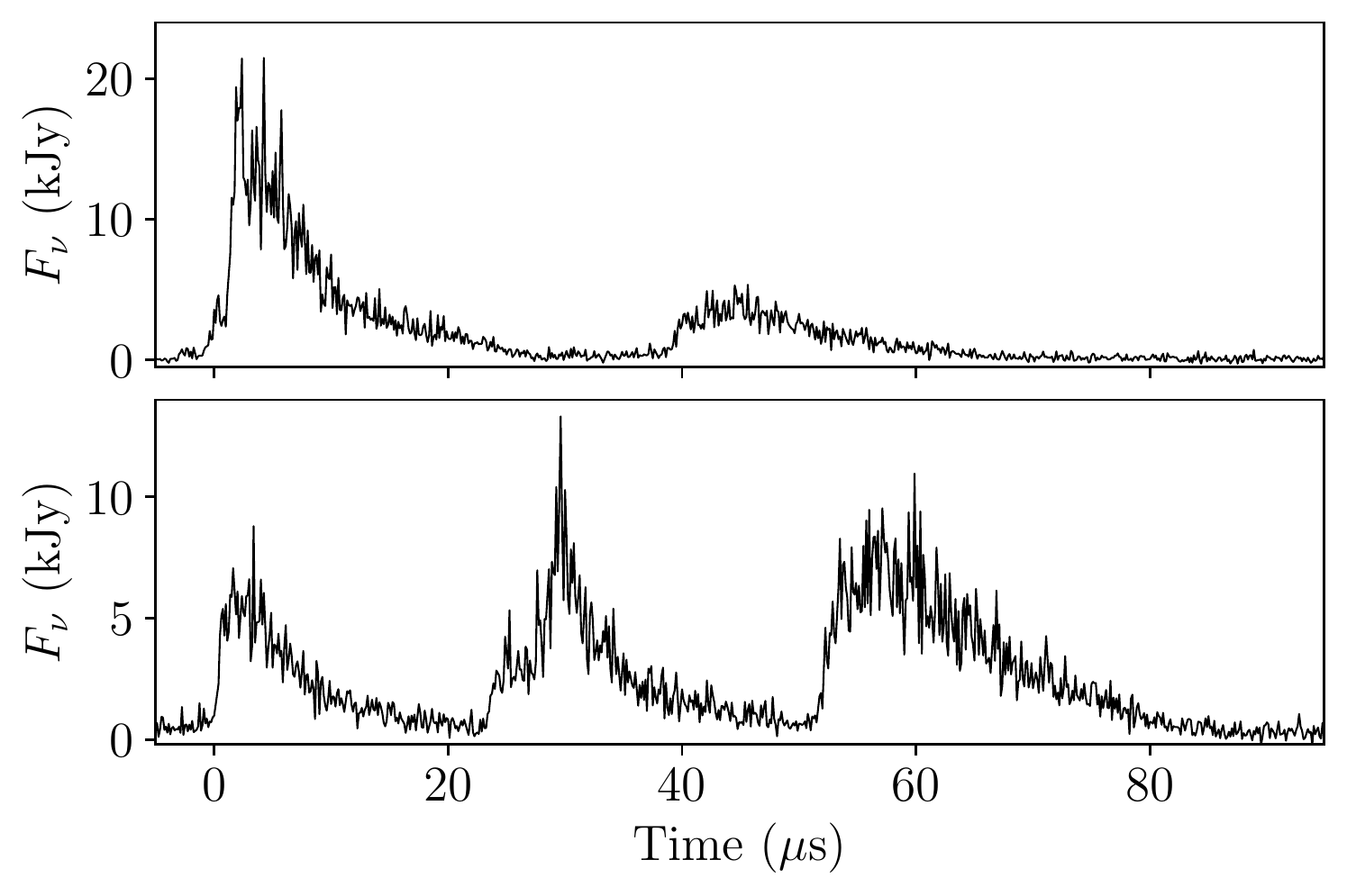}
\caption{
  Pulse profiles showing distinct microbursts using time bins of $125{\rm\;ns}$ (from EK036~B).
  Likely, these consist of multiple giant pulses occurring in the same rotation by chance.
  \label{fig:microbursts}}
\end{figure}

Our pipeline achieves the expected coherence, as can be seen from the example giant pulse shown in Figure~\ref{fig:coherance}: the voltage series from multiple telescopes align well after delay and phase corrections, and the coherently summed giant pulse profile has S/N higher by the number of telescopes than the profiles of the single dishes, much better than the incoherently summed giant pulse profile where the S/N increase only by the square root of the number of telescopes.
In the single-dish data, one sees that the intensities are very similar in units of the off-pulse noise, independent of telescope aperture.
This reflects that the total system noise is roughly the same, as it is dominated by the radio emission from the Crab Nebula itself, which has a flux density of $S_{\text{CN}}\approx833{\rm\;Jy}$ at our observing frequency \citep{Bietenholz1997}, while the nominal system equivalent flux (SEFD) for a telescope ranges between $S_{\text{tel}}\approx19$ and $450{\rm\;Jy}$\footnote{\url{http://old.evlbi.org/cgi-bin/EVNcalc}}, with an average $\langle S_{\text{tel}}\rangle\simeq300{\rm\;Jy}$. Thus, the SEFD for our tied-array beam, which largely resolves out the nebula, can be estimated as $(S_{\text{CN}} + \langle S_{\text{tel}}\rangle)/N\approx\!140\!-\!160{\rm\;Jy}$ (depending on the number of telescopes $N$ used).

In Figure~\ref{fig:waterfall_zoo}, we highlight an IP and some other interesting giant pulses that we found.
Two pulses with distinct multi-burst structure are shown in Figure~\ref{fig:microbursts}; we will return to the properties of pulses with multiple bursts in Section~\ref{subsec:correlation_microbursts}.

\section{Scattering Timescale}\label{sec:scattering_timescale}

\begin{deluxetable}{ccccc}
\tabletypesize{\small}
\setlength\tabcolsep{2.9pt}
\tablecaption{Scattering Timescales and Intrinsic Widths.\label{table:profiles}}
\tablenum{2}
\tablehead{\colhead{Observation}&
  \multicolumn{2}{c}{\dotfill MP \dotfill}&
  \multicolumn{2}{c}{\dotfill IP \dotfill}\\[-.7em]
  \colhead{Code}&
  \colhead{$\tau$ (${\rm\mu s}$)}&
  \colhead{$p$ (${\rm\mu s}$)}&
  \colhead{$\tau$ (${\rm\mu s}$)}&
  \colhead{$p$ (${\rm\mu s}$)}}
\startdata
EK036 A & $5.08\pm0.15$ & $0.76\pm0.05$ & $3.75\pm0.14$ & $0.43\pm0.07$ \\
EK036 B & $8.90\pm0.19$ & $0.63\pm0.08$ &  $5.8\pm0.3$  & $0.34\pm0.05$ \\
EK036 C & $6.17\pm0.16$ & $1.00\pm0.10$ &  $5.1\pm0.2$  & $0.62\pm0.15$ \\
EK036 D & $8.45\pm0.14$ & $2.31\pm0.06$ &  $6.8\pm0.2$  & $1.94\pm0.18$ \\
\enddata
\tablecomments{The scattering timescales, $\tau$, and intrinsic widths, $p$, of stacked MP and IP giant pulse profiles are presented here. The uncertainties from bootstrapping are likely underestimated, mostly because of the presence of echoes in the scattering tail (see Section~\ref{sec:scattering_timescale}).}
\end{deluxetable}

\begin{figure}
\centering
\includegraphics[width=0.47\textwidth,trim=0 0 0 0,clip]{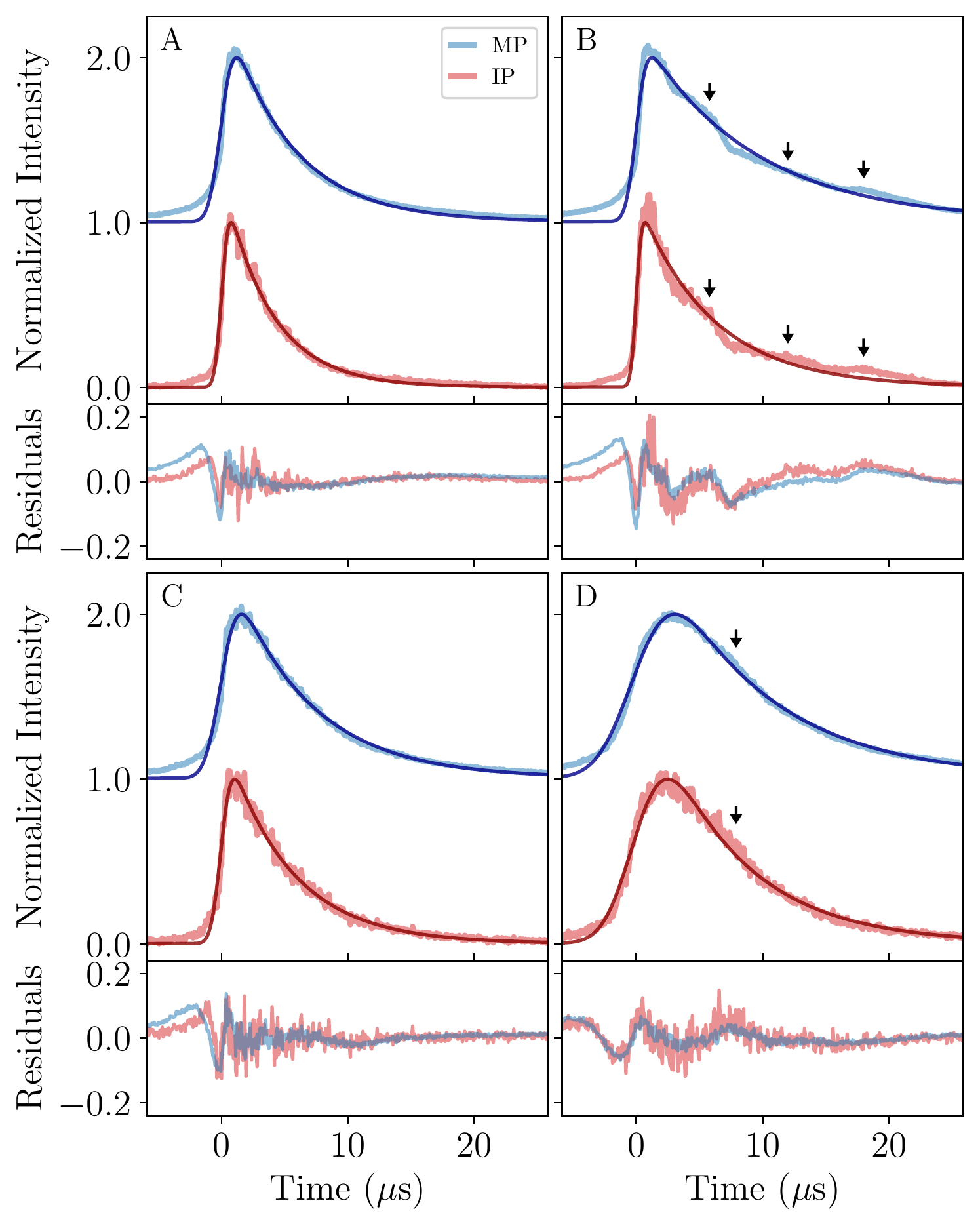}
\caption{
  Stacked MP (blue line) and IP (red line) components normalized to the peak of the exponentially modified Gaussian fit (dark blue and dark red lines, respectively) and centred so that $t_{0}$ is at $0{\rm\;\mu s}$.
  Fit values are given in Table~\ref{table:profiles}.
  We also show the residuals between the best-fit model and the data for each of our observations.
  \textbf{A}: Stacked pulses from EK036~A.
  \textbf{B}: Stacked pulses from EK036~B.
  It is clear that the same echo features, which are indicated by arrows, (around $\sim\!6$, $12$ and $18{\rm\;\mu s}$) appear in both MP and IP components.
  \textbf{C}: Stacked pulses from EK036~C.
  \textbf{D}: Stacked pulses from EK036~D showing an echo feature around $\sim\!8{\rm\;\mu s}$ as well as possibly one near a few ${\rm \mu s}$ that causes the profile to be substantially broader than is the case in the other epochs.
    \label{fig:tau_sigma_fit}}
\end{figure}

Emission from the Crab Pulsar passes through two scattering screens, one originating at the optical filaments in the Crab Nebula and one in the ISM.
The temporal broadening is usually dominated by the nebular screen \citep{Vandenberg1976}, and one sees its effect in Figures~\ref{fig:coherance}--\ref{fig:microbursts} in the roughly exponential tail shared by all pulses, with a timescale of $\sim\!5{\rm\;\mu s}$.
One also sees that the pulses have internal structure on a similar timescale.

To measure the timescale more precisely -- and thus help determine the resolution at the pulsar (Section~\ref{sec:interstellar_interferometry}) -- we construct average pulse profiles for the MP and IP components separately, by aligning and stacking giant pulses with S/N above 100 (removing profiles with obvious multiple bursts like those shown in Figure~\ref{fig:microbursts}).
For the alignment, we fit the pulse intensities with an exponentially modified Gaussian of the form,
\begin{eqnarray}
    I(t) &=& \frac{Ap}{\tau}\sqrt{\frac{\pi}{2}}\exp \left(\frac{p^{2}}{2\tau^{2}} - \frac{t-t_{0}}{\tau}\right)\nonumber \\
    &&\times\text{erfc}\left(\frac{p}{\sqrt{2}\tau} - \frac{t-t_{0}}{\sqrt{2}p}\right) + C,
\label{eqn:taufit}
\end{eqnarray}
where $A$, $t_0$, and $p$ are the amplitude, centroid, and standard deviation of the Gaussian, respectively, $\tau$ is the scattering timescale, and $C$ is the background intensity.
We then create our stacked pulse profiles by adding MPs and IPs aligned using the individual $t_{0}$, and fit the same exponentially modified Gaussian profile above to the stacks.

Our pulse stacks and fits are shown in Figure~\ref{fig:tau_sigma_fit} and relevant fit parameters are given in Table~\ref{table:profiles}.
One sees that the fits are not formally acceptable: they do not capture the relatively slow rise in intensity, suggesting that a Gaussian is not a good model for the average giant pulse profile, and the scattering tail shows bumps inconsistent with smooth exponential decay.
Hence, the formal errors of the fits have little meaning, and we instead use bootstrapping, determining the standard deviations in the fit parameters from 8192 sets of pulse profiles created using random selection with replacement from our input profiles.

Note that the bad fits are not surprising: the individual pulses are made up of nanoshots and their average distribution does not have to resemble a Gaussian, and the scattering tail can have structure arising from individual scattering filaments.
Indeed, the scattering tail for EK036~B shows clear evidence for echoes (see Figure~\ref{fig:tau_sigma_fit}B), with both the MP and IP profile showing similar structure in the scattering tail.
These echo features appear on all high S/N giant pulse profiles in EK036~B (at lower S/N, it is difficult to make out echoes by eye).
Echoes have previously been seen at lower frequencies, from 327 to $1400{\rm\;MHz}$ \citep{Backer2000, Lyne2001, Crossley2007, Driessen2019}, so the present study extends it to $1660{\rm\;MHz}$.
Echoes were not seen at $4.9{\rm\;GHz}$ during a period when echoes were present at $1.4{\rm\;GHz}$ \citep{Crossley2007}, but this may just reflect that echoes are frequency dependent, as expected if they arise in refraction in nebular structures.

In EK036~A and EK036~C, the modified exponential fits the scattering tail quite well with no apparent echoes. In EK036~D, however, there are deviations from an exponential scattering tail.
Furthermore, the inferred intrinsic  profile is much broader, with a larger $\sigma$.
Since it seems unlikely the actual intrinsic average giant pulse profile changed, this may reflects an echo as well, one that is at a delay of only a few microseconds.

For all our observations, we see that the sharp rise of the model does not fit the intrinsic pulse structure of giant pulses well, especially for the MP profiles.
Comparing MP and IP profiles, one sees that the former are systematically broader, with larger fitted $\sigma$ and $\tau$.
The larger $\sigma$ suggests MPs have intrinsically longer durations over which nanoshots are emitted than IPs.
The longer scattering times likely also reflect some intrinsic difference in emission, e.g., that the nanoshots in MPs have a more skewed intensity distribution, falling off more slowly at the tail end, and that this skewed distribution leads to fits with $\tau$ biased high.

From the above, we conclude that among our measurements of the scattering time, those from the IP are probably more reliable, although even for those the errors are likely underestimated.
For the purpose of a rough estimate, however, taking the scattering time in all epochs to be $\tau\simeq5{\rm\;\mu s}$ should be good to about 50\%.

\section{Correlations}\label{sec:analysis}

\begin{deluxetable*}{llcccccc}[ht]
\tabletypesize{\small}
\tablecaption{Correlation Characteristics of Main Pulse, Main Pulse Pairs and Main Pulse, Interpulse Pairs\label{table:correlations}}
\tablenum{3}
\tablehead{\colhead{Observation/Reference}&
  \colhead{Correlation pair}&
  \colhead{$A$ ($\%$)}&
  \colhead{$t_{\text{scint}}$ (${\rm s}$)}&
  \colhead{$\nu_{\text{decorr}}$ (${\rm MHz}$)}&
  \colhead{$\Delta t_{0}$ (${\rm s}$)}&
  \colhead{$\Delta\nu_{0}$ (${\rm MHz}$)}&
  \colhead{$\rho_{\text{f,t}}$}}
\startdata
EK036 A\dotfill & MP-MP                               & $0.701\pm0.016$ & $9.3\pm0.2$    & $0.492\pm0.011$ & \nodata       & \nodata        & $-0.32\pm0.03$ \\
                & MP-IP                               & $0.52\pm0.05$   & $12.1\pm1.0$   & $0.47\pm0.04$   & $+0.1\pm0.7$  & $-0.03\pm0.03$ & $-0.25\pm0.11$ \\[0.8ex]
EK036 B\dotfill & MP-MP                               & $0.815\pm0.013$ & $10.45\pm0.16$ & $0.463\pm0.007$ & \nodata       & \nodata        & $-0.14\pm0.02$ \\
                & MP-MP$_{1^{\text{st}}\text{ half}}$ & $1.01\pm0.02$   & $10.1\pm0.2$   & $0.509\pm0.010$ & \nodata       & \nodata        & $-0.01\pm0.03$ \\
                & MP-MP$_{2^{\text{nd}}\text{ half}}$ & $0.94\pm0.04$   & $10.1\pm0.5$   & $0.26\pm0.012$  & \nodata       & \nodata        & $-0.39\pm0.05$ \\
                & MP-IP                               & $0.65\pm0.03$   & $11.4\pm0.5$   & $0.49\pm0.02$   & $-0.4\pm0.4$  & $+0.05\pm0.02$ & $-0.17\pm0.06$ \\
                & MP-IP$_{1^{\text{st}}\text{ half}}$ & $0.74\pm0.04$   & $12.6\pm0.7$   & $0.59\pm0.03$   & $-1.1\pm0.5$  & $+0.10\pm0.03$ & $-0.06\pm0.08$ \\[0.8ex]
EK036 C\dotfill & MP-MP                               & $0.52\pm0.03$   & $8.7\pm0.6$    & $0.35\pm0.02$   & \nodata       & \nodata        & $+0.04\pm0.09$ \\[0.8ex]
EK036 D\dotfill & MP-MP                               & $0.48\pm0.02$   & $10.8\pm0.4$   & $0.342\pm0.014$ & \nodata       & \nodata        & $-0.27\pm0.05$ \\[0.8ex]
                & MP-IP                               & $0.46\pm0.06$   & $12.3\pm1.5$   & $0.25\pm0.03$   & $-0.3\pm1.1$  & $+0.03\pm0.03$ & $-0.16\pm0.17$ \\[0.8ex]
\cite{Main2021}\dotfill
                & MP-MP                               & $1.80\pm0.03$   & $9.24\pm0.13$  & $1.10\pm0.02$   & \nodata       & \nodata        & \nodata\\
                & MP-IP                               & $0.97\pm0.07$   & $10.7\pm0.8$   & $1.44\pm0.10$   & $+1.0\pm0.5$  & $-0.34\pm0.09$ & \nodata\\[0.8ex]
\cite{Cordes2004}\dotfill
                & MP-MP                               & $\sim\!30$      & $\sim\!27$     & $\sim\!0.6$     & \nodata       & \nodata        & \nodata
\enddata
\tablecomments{
  The scintillation timescale $t_{\text{scint}}$ and frequency decorrelation $\nu_{\text{decorr}}$ are defined as the values where the correlation function drops to $1/e$ and $1/2$, respectively.
  For EK036~C, there were too few MP-IP pairs to derive a meaningful correlation.
  For \cite{Cordes2004}, the values are approximate, as they were interpolated between their $1.48$ and $2.33{\rm\;GHz}$ observations (assuming $\nu_{\text{decorr}}\propto\nu^{4}$ and $t_{\text{scint}}\propto\nu$).
}
\end{deluxetable*}

For regular pulsars, auto-correlations of dynamic spectra can be used to infer the scintillation bandwidth and timescale, and to learn about possible spatial offsets between pulse emission regions.
For the Crab Pulsar's randomly occurring giant pulses, we follow \cite{Cordes2004} and \citetalias{Main2021} and construct the normalized time and frequency correlation by first correlating pairs of giant pulse power spectra and then binning the correlations by the time separation between the giant pulses.

The correlation coefficient $\rho(P_{1},P_{2})$ between two power spectra $P_{1}$ and $P_{2}$ sampled at $k$ frequencies can be estimated with,
\begin{eqnarray}
    r(P_{1}, P_{2}) &=& \frac{\frac{1}{k-1}\sum_{i=1}^{k} (P_{1,i}-m_{1})(P_{2,i}-m_{2})}{s_{1}s_{2}} \nonumber \\
    &&\times \frac{m_{1}m_{2}}{(m_{1} - 1)(m_{2}-1)},
\label{eqn:r_power}
\end{eqnarray}
where $m$ and $s$ are estimates of the average and standard deviation of the power spectra, respectively.
The second term accounts for noise biases (see \citetalias{Main2021} and Appendix~\ref{sec:appendix}), using that the mean and standard deviation of the noise power spectra are $1$ as we have normalized our data by the background noise.

To construct our correlations, we create power spectra of giant pulses, discarding the $15\%$ of each sub-band near each edge where little signal is detected because the passband rolls off.
We then correlate spectra in each sub-band separately over frequency, taking into account that for larger frequency offsets fewer points contribute, giving individual estimates of the correlation as a function of frequency offset $\Delta\nu$ and time offset $\Delta t$.
Next, we construct the normalized 2-D correlations by binning all pulse pairs by time separation, using a bin width of $1{\rm\;s}$. We average pairs across sub-bands and polarizations using optimal weights,
\begin{equation}
    w_{1,2}=\frac{(m_{1}-1)^2(m_{2}-1)^2}{m_1^2m_2^2}, \label{eqn:r_power_weight}
\end{equation} appropriate for our case where the correlations are low (see Appendix~\ref{sec:appendix}, Equation~\ref{eqn:rho_weights}).

Below, in Section~\ref{subsec:correlation_pulsepairs}, we discuss the correlations between MP-MP and MP-IP pairs of pulses and in Section~\ref{subsubsec:correlation_havles} we take a look at differences between the correlations for the first and second halves of pulses.
Next, in Section~\ref{subsec:correlation_microbursts}, we focus on the correlations between multiple bursts within a given pulse,
and in Section~\ref{subsec:correlation_polarizations}, we look at correlations between the left and right circular polarizations.
We make qualitative comparisons with previous work and with expectations based on the scintillation screen's resolution, but leave the interpretation in terms of physical properties of the emission regions to Section~\ref{sec:emission_regions}.

\subsection{Correlations of Pulse Pairs in Time and Frequency}\label{subsec:correlation_pulsepairs}

\figsetstart
\figsetnum{7}
\figsettitle{MP-MP and MP-IP correlations}

\begin{figure}
\centering
\includegraphics[width=0.47\textwidth,trim=0 0 0 0,clip]{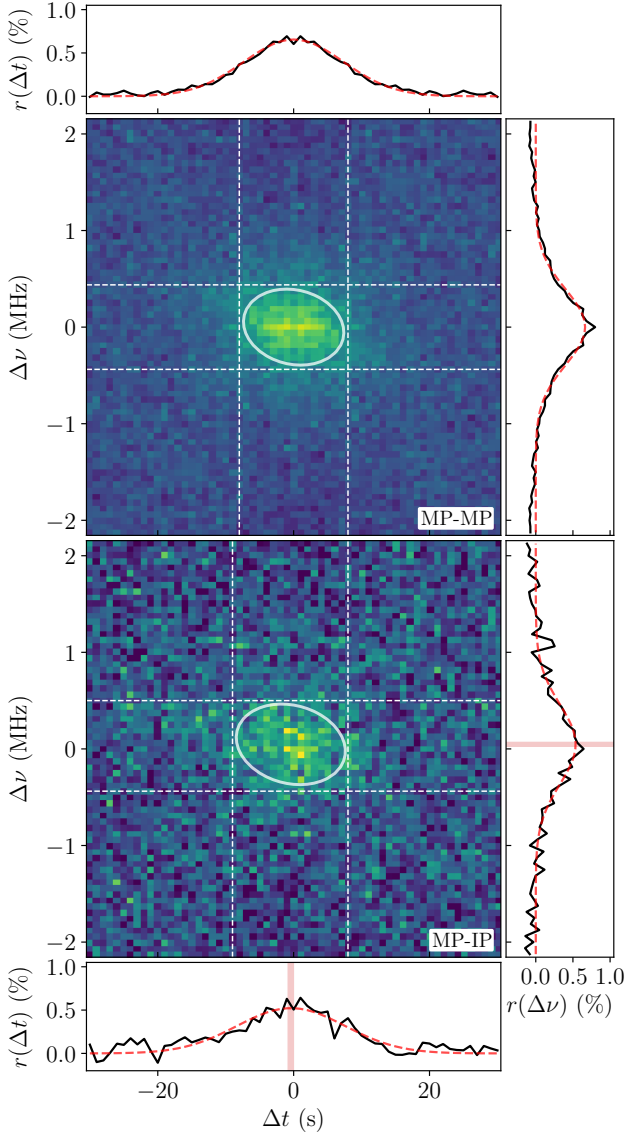}
\caption{
    Cross-correlations of giant pulse spectra between MP-MP ({\em top\/}) and MP-IP pairs ({\em bottom\/}) from EK036~B.
    The MP-MP correlation is symmetric by construction while the MP-IP correlation is not.
    For both, we correlated spectra of pulse pairs that have frequency resolution of $62.5{\rm\;kHz}$ and binned these in $1{\rm\;s}$ time bins.
    The correlations are fitted with a two-dimensional Gaussian (as described in Section~\ref{subsec:correlation_pulsepairs}), with the white contour representing $1\sigma$ away from the peak
    (fitted values are listed in Table~\ref{table:correlations}).
    The attached panels show the average correlation within the corresponding regions marked by dashed white lines and the red line shows the fit, also averaged over the dashed white lines.
    The red bars indicate the best-fit time and frequency offsets in the MP-IP correlation.
    Here, IP precedes MP, though the difference from zero is not significant.
    The complete figure set (4 images) is available in the online journal.
    \label{fig:corr_mpmp_mpip_full}}
\end{figure}

\figsetgrpstart
\figsetgrpnum{7.1}
\figsetgrptitle{EK036A}
\figsetplot{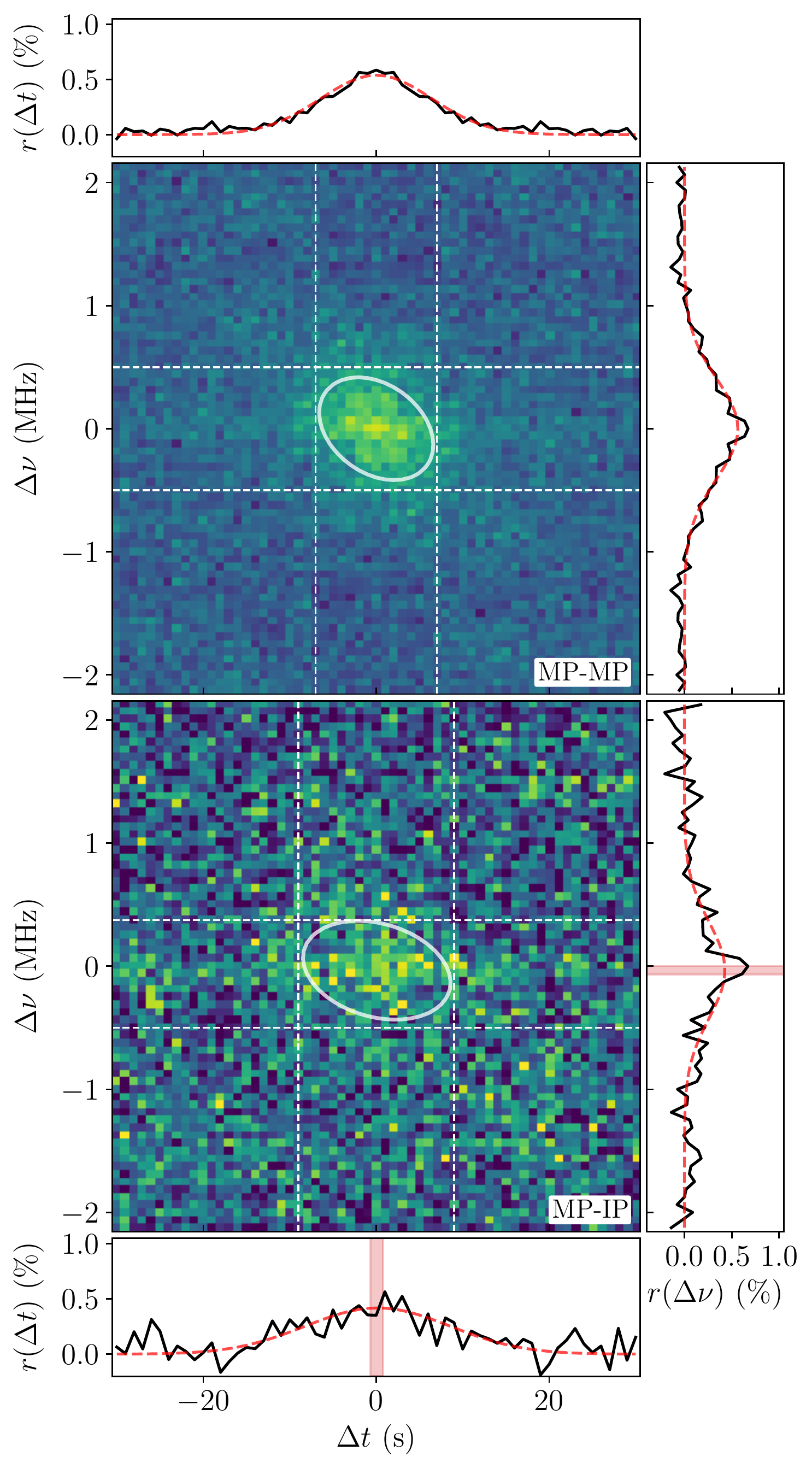}
\figsetgrpnote{Cross-correlations of giant pulse spectra between MP-
MP and MP-IP from EK036 A. The layout, time binning and Gaussian fits are as in Figure~\ref{fig:corr_mpmp_mpip_full}}
\figsetgrpend

\figsetgrpstart
\figsetgrpnum{7.2}
\figsetgrptitle{EK036B}
\figsetplot{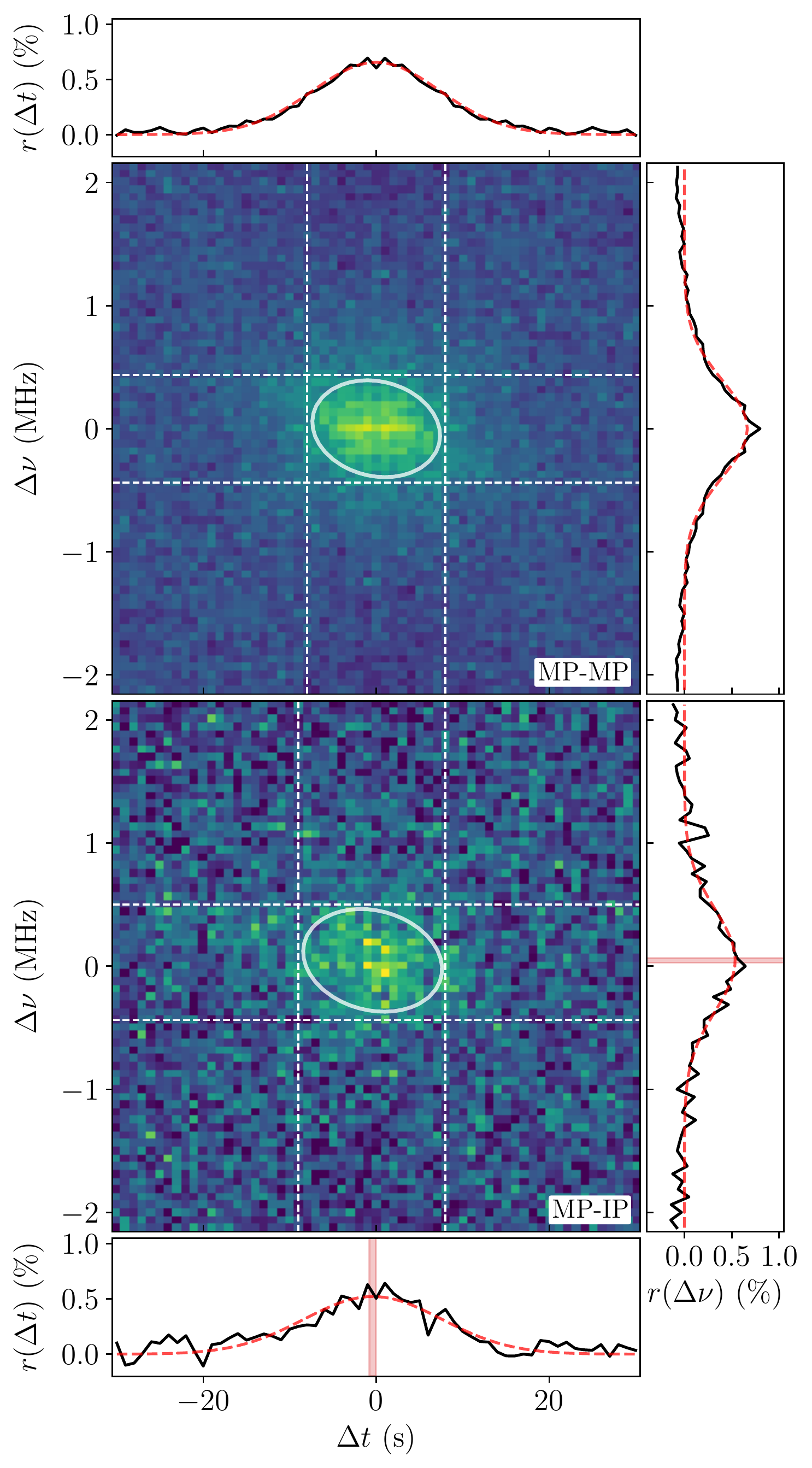}
\figsetgrpnote{Cross-correlations of giant pulse spectra between MP-
MP and MP-IP from EK036 B. The layout, time binning and Gaussian fits are as in Figure~\ref{fig:corr_mpmp_mpip_full}}
\figsetgrpend

\figsetgrpstart
\figsetgrpnum{7.3}
\figsetgrptitle{EK036C}
\figsetplot{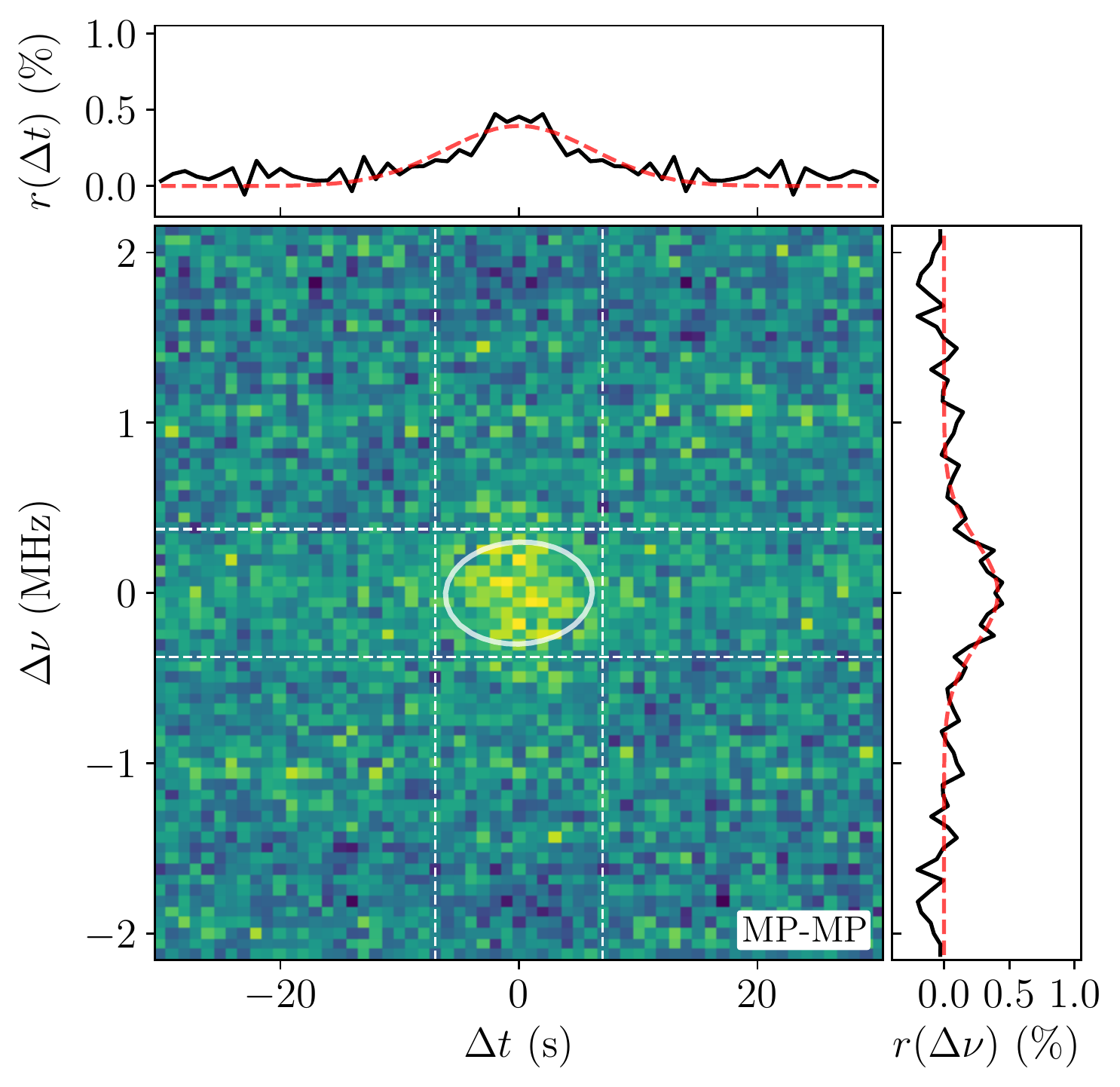}
\figsetgrpnote{Cross-correlations of giant pulse spectra between MP-
MP from EK036 C. The layout, time binning and Gaussian fits are as in Figure~\ref{fig:corr_mpmp_mpip_full}}
\figsetgrpend

\figsetgrpstart
\figsetgrpnum{7.4}
\figsetgrptitle{EK036D}
\figsetplot{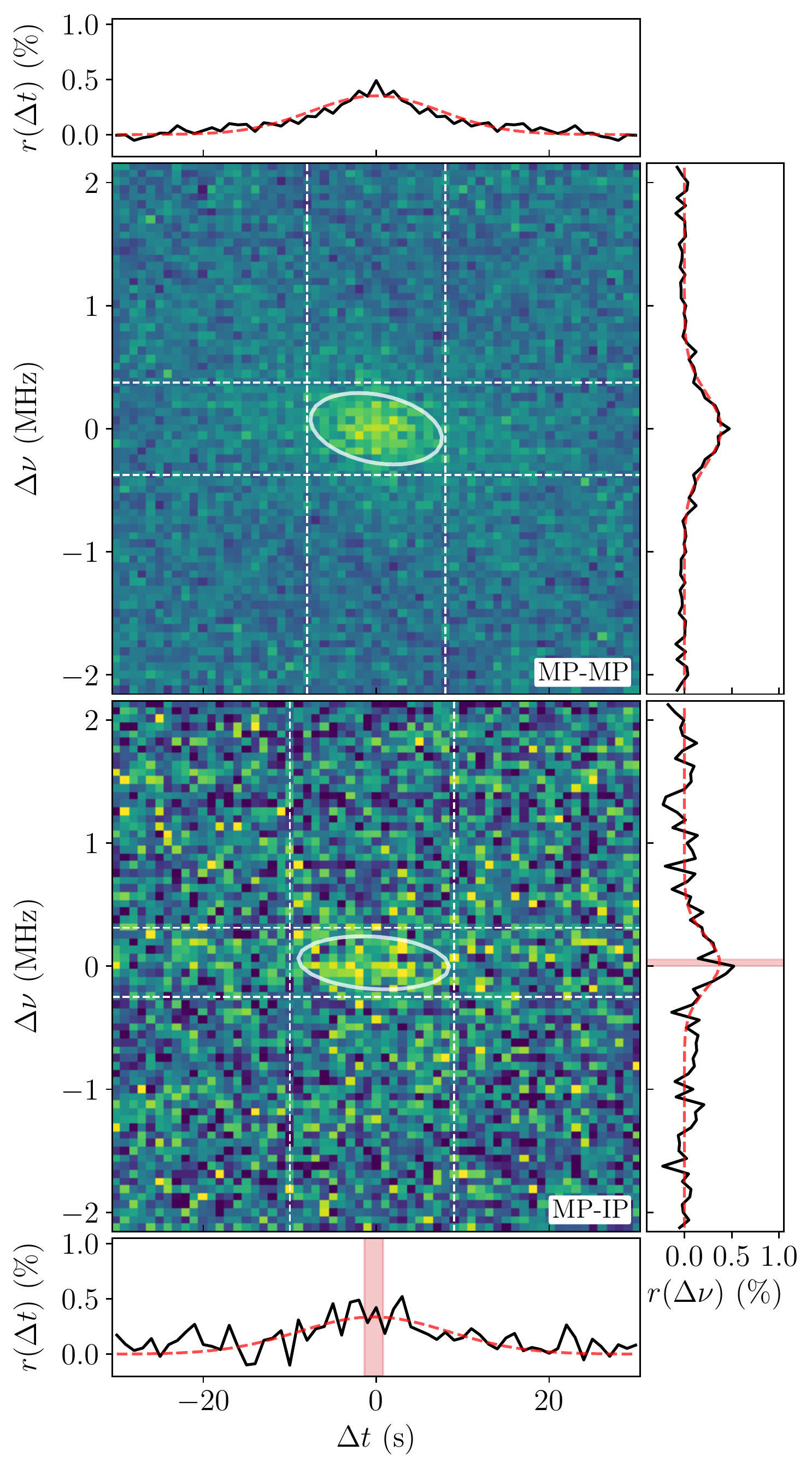}
\figsetgrpnote{Cross-correlations of giant pulse spectra between MP-
MP and MP-IP from EK036 D. The layout, time binning and Gaussian fits are as in Figure~\ref{fig:corr_mpmp_mpip_full}}
\figsetgrpend
\figsetend

\begin{figure}
\centering
\includegraphics[width=0.47\textwidth,trim=0 0 0 0,clip]{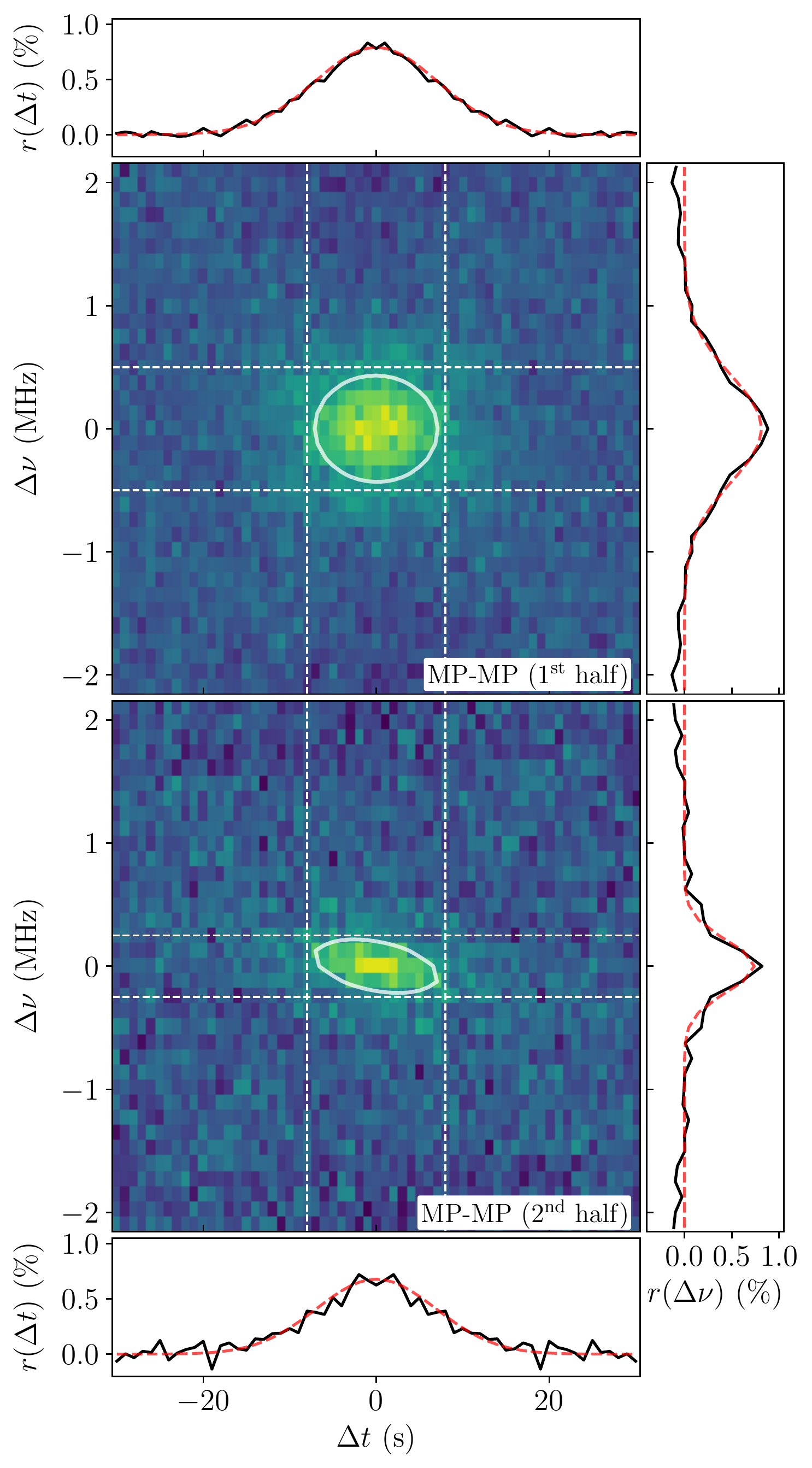}
\caption{
    Cross-correlations of giant pulse spectra between the first ({\em top\/}) and second ({\em bottom\/}) half of MP-MP pairs from EK036~B.
    The frequency resolution here is $125{\rm\;kHz}$ and the layout, time binning and Gaussian fits are as in Figure~\ref{fig:corr_mpmp_mpip_full}.
    One sees that the frequency decorrelation scale decreases drastically, while the amplitude drops a little and the timescale does not change significantly.
    Combined, this suggests the scintillation screen resolves the region in which giant pulses are emitted.
    \label{fig:corr_mpmp_halves}}
\end{figure}

We correlate MP-MP pairs and MP-IP pairs in both time and frequency using power spectra of giant pulses with S/N greater than 8, and without obvious multiple components (for those, see Section~\ref{subsec:correlation_microbursts}).
The power spectra are created using the first $16{\rm\;\mu s}$ of each giant pulse (see Section~\ref{subsec:giant_pulses} where we define the start of a giant pulse), yielding a frequency resolution of $62.5{\rm\;kHz}$.
We show the resulting time-frequency correlations for both the MP-MP and MP-IP pairs from EK036~B in Figure~\ref{fig:corr_mpmp_mpip_full} (a figure set for all observations (4 images) is available in the online journal.)

The binned correlations for all epochs are fitted with bivariate Gaussians, with as parameters the amplitude ($A$), the time ($\sigma_{t}$) and frequency ($\sigma_{\nu}$) widths, the correlation between time and frequency ($\rho_{\text{f,t}}$), and, for the MP-IP correlations, possible time ($\Delta t_{0}$) and frequency ($\Delta\nu_{0}$) offsets (which are relative to the MP, i.e., a positive sign of the time offset indicates that the IP trails the MP),
\begin{gather}
    f(t, \nu) = A\exp \left(-\frac{a^{2} -2\rho_{\text{f,t}}ab + b^{2}}{2(1-\rho_{\text{f,t}}^{2})} \right) \\
    \text{where} \quad a = \frac{\Delta t - \Delta t_{0}}{\sigma_{t}}, \quad b = \frac{\Delta\nu - \Delta\nu_{0}}{\sigma_{\nu}}. \nonumber
\label{eqn:2DGuassian}
\end{gather}

The results of the fits for all four epochs are collected in Table~\ref{table:correlations}, where we also list values from \cite{Cordes2004} and \citetalias{Main2021}.
One sees that the biggest differences between our data and the ones from the literature are seen for the correlation amplitudes, which are very low for all our observations, at $A\simeq0.6\%$ on average, which is about three times lower than found in \citetalias{Main2021}, and a factor $\sim\!50$ lower than what was seen by \cite{Cordes2004}.
In contrast, the scintillation timescale of about $t_{\text{scint}}\approx10{\rm\;s}$ on average is similar to that measured in \citetalias{Main2021}, but substantially shorter than the $\sim\!27{\rm\;s}$ inferred by interpolating in the measurements of \cite{Cordes2004}, while the typical decorrelation bandwidth we find, of $\nu_{\text{decorr}}\simeq0.4{\rm\;MHz}$ on average, is about half what was found in \citetalias{Main2021}, yet similar to what one infers from \cite{Cordes2004}.
We will argue in Section~\ref{sec:emission_regions} that these apparently contradictory trends can be understood from considering differences in effective resolution of the scintillation screens between the different observations.

Like in \citetalias{Main2021}, we find that MP-IP correlations are weaker and broader than the MP-MP correlations.
For EK036~B, we also find a marginal time offset, of $\Delta t_{0}=-0.4\pm 0.4{\rm\;s}$, as well as a slightly more significant frequency offset, of $\Delta\nu_{0}=+0.05\pm 0.02{\rm\;MHz}$.
Both have opposite sign of those found in \citetalias{Main2021}.
The measurements for epochs~A and D are less reliable, but the small frequency offsets are again inconsistent with those of \citetalias{Main2021}.

\subsection{Differences in Correlation between Pulse Halves}\label{subsubsec:correlation_havles}
The high detection rate and large number of strong giant pulses in EK036~B allows us to look for differences in correlation with time of the pulse, and thus use signals that travelled to the scattering screen at increasing distance from the line of sight.
To do this, we divide our giant pulses in half and create spectra from the first $8{\rm\;\mu s}$ and second $8{\rm\;\mu s}$, resulting in $125{\rm\;kHz}$ channels.
We then correlate spectra of the first half (corresponding to the inner regions of the scattering screen) and second half (corresponding to the outer regions of the scattering screen) separately and bin in time.

The resulting averaged correlations are shown in Figure~\ref{fig:corr_mpmp_halves} and the fit parameters listed in Table~\ref{table:correlations}.
From those, most striking is the change in frequency decorrelation width, which nearly halves.
This is not unexpected, as the interference pattern from the outer scattering regions probed at later time will have finer structure.
The decrease in amplitude by about 10\% between first and second half also seems qualitatively consistent with an increase in resolution, as it means two separate pulses are less likely to be in the same resolution element.
Clearly, however, even in the first half, the correlation amplitude of $\sim\!1\%$ is still much less than the $1/3$ expected if giant pulses pass through the same part of the scattering screen.
In contrast to the frequency and amplitude, the time decorrelation remains the same.
That the time scale does not decrease suggests again that it is determined not by the size of the effective resolution element produced by the scintillation screen but rather by the size of the region in which giant pulses occur.

In an attempt to better constrain the MP-IP correlation offset in time, we also correlated the first halves of MP-IP pairs.
We found that while these yielded a slightly higher amplitude than what we inferred from the correlation of spectra from the full, $16\rm{\;mu s}$ windows on the giant pulses, the uncertainties on the offsets were not smaller (see Table~\ref{table:correlations}).

We also attempted to correlate IP-IP pairs for EK036~B, but as there are few giant pulse pairs with sufficiently high S/N, we do not trust the fit parameters and we do not list them in Table~\ref{table:correlations}.

\subsection{Correlation of Nearby Microbursts}\label{subsec:correlation_microbursts}

\begin{figure}
    \centering
    \includegraphics[width=0.47\textwidth,trim=0 0 0 0,clip]{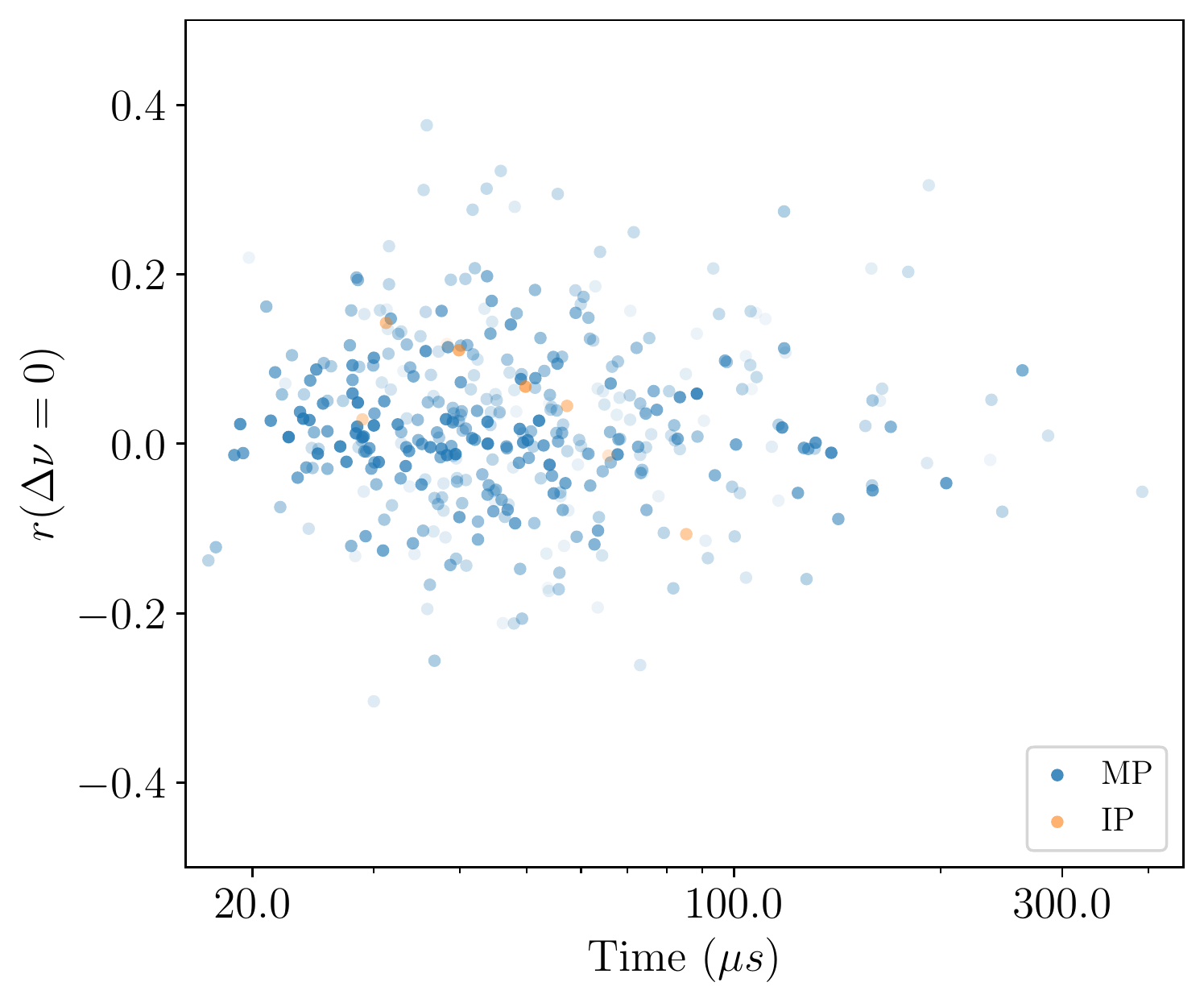}
    \caption{Correlation of microburst pairs within each pulse component, with MP pairs (387 pairs) in blue and IP pairs (8 pairs) in orange.
    We omitted any correlations for which the uncertainty exceeded $0.15$.
    The opacity of the individual points scales with the uncertainty of the correlation.
    The level of correlation is as low as seen for nearby pairs in different rotations, as expected given that the pairs likely not causally related but rather due to chance coincidence.
    \label{fig:correlation_microburts}}
\end{figure}

Sometimes, multiple microbursts occur within one rotation.
So far, we have excluded these from our analysis as we wished to avoid the risk of contamination of the power spectra, but we can check whether they correlate differently.
We focus on EK036~B, where we found 1302 MP and 163 IP cases where we detected multiple giant pulses from the same rotation (with S/N greater than 8 and separated from each other by at least $16{\rm\;\mu s}$): 1185 MP and 151 IP with 2 microbursts, 105 MP and 12 IP with 3 microbursts, 11 MP with 4 microbursts and 1 MP with 5 microbursts.

The expectation for the correlation depends on whether the multiple bursts in a rotation are causally related, i.e., consequences of the same physical events, or whether they are due to multiple observable giant pulses happening to be emitted in a single rotation.
To estimate the fraction of multiples expected by chance, we can use that for EK036~B  we detect a MP or IP giant pulse roughly every 12 and 48 rotations, respectively.
Chance coincidences of $N$ pulses should happen roughly at those rates raised to the power $N$.
Given that we observed for about 177,404 periods, we would thus expect to see about 1299 MP and 78 IP pairs, about 111 MP and 2 IP triples, about 10 MP and 0.03 IP quadruples and about 1 MP and 0.0007 IP quintuples.
For the MP, the observed numbers are roughly consistent with the expected ones, but for the IP they are significantly higher, suggesting some causal connection between multiple IP bursts.

If most pairs are not causally related, we expect the correlation between them to be similar to what is seen between giant pulses that occurred nearby in time.
As can be seen in Figure~\ref{fig:correlation_microburts}, this is indeed the case for the MP: the average weighted correlation amplitude of microbursts within the MP component is $1.8\pm0.4\%$, a bit higher than but still consistent with the MP-MP correlations in Section~\ref{subsec:correlation_pulsepairs}.
The average weighted correlation amplitude of microbursts within the IP component is nominally higher, at $6\pm3\%$, perhaps again suggesting some causal connection between IP microbursts.

For both MP and IP, the correlation amplitude is much less than what was found by \cite{Karuppusamy2010}, consistent with the suggestion that they observed at a time when the scattering time was shorter, the screen's resolution poorer, and the scintillation pattern for different bursts thus more similar.

\subsection{Frequency Correlation Between Polarizations}
\label{subsec:correlation_polarizations}

\begin{figure}
    \centering
    \includegraphics[width=0.47\textwidth,trim=0 0 0 0,clip]{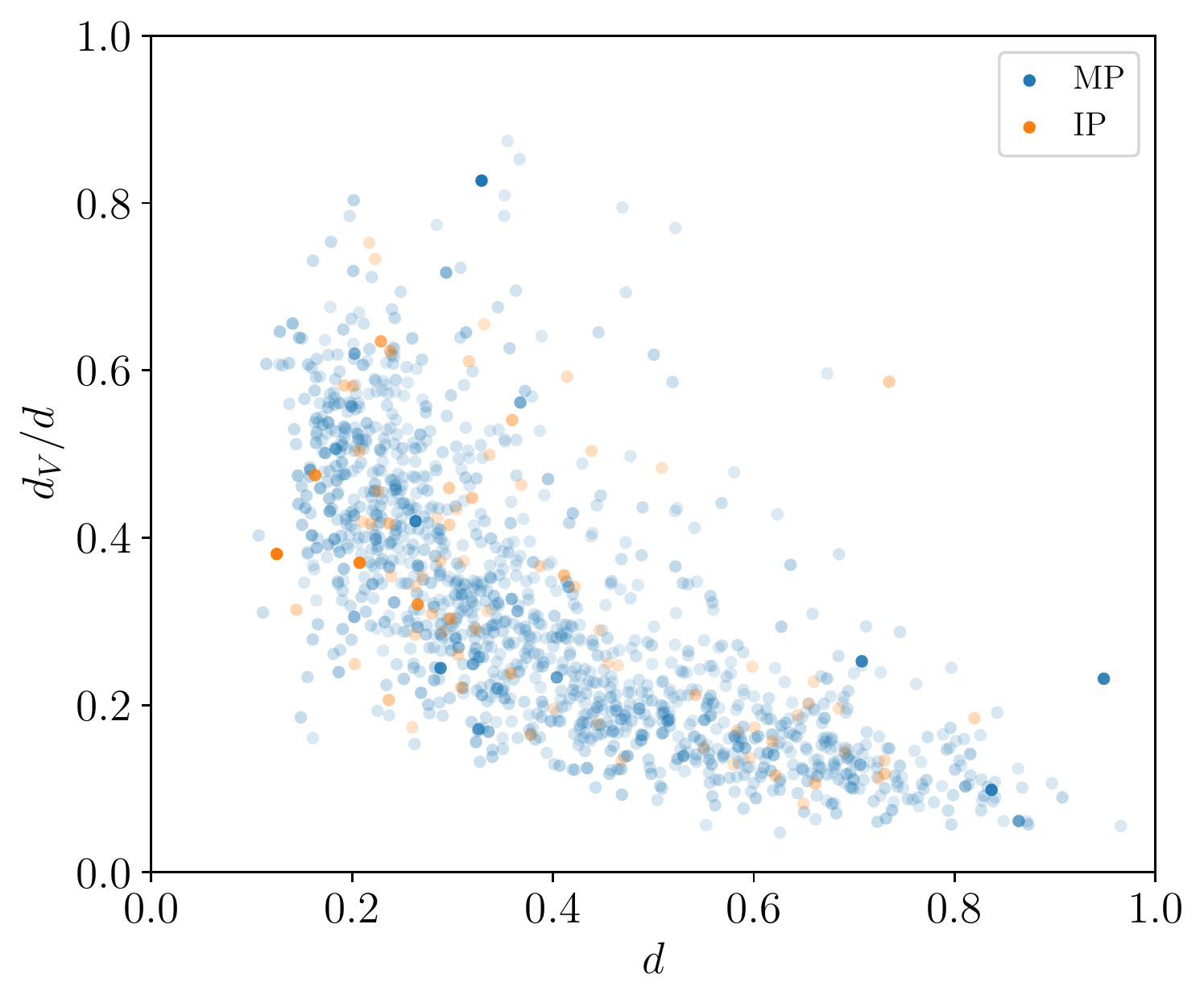}
    \caption{Fractional circular degree of polarization as a function of total degree of polarization for bright MP (blue) and IP (orange) giant pulses.
    The opacity of the individual points scales with the square root of the S/N of the giant pulse.
    One sees that the typical degree of polarization is not very high, consistent with the giant pulses being composed of multiple fully polarized nanoshots.
    The giant pulses are also predominantly linearly polarized (for randomly polarized pulses, one expects $d_{V}/d\simeq1/2$).
    \label{fig:polarization}}
\end{figure}

\begin{figure*}
  \centering
  \includegraphics[width=0.985\textwidth,trim=0 0 0 0,clip]{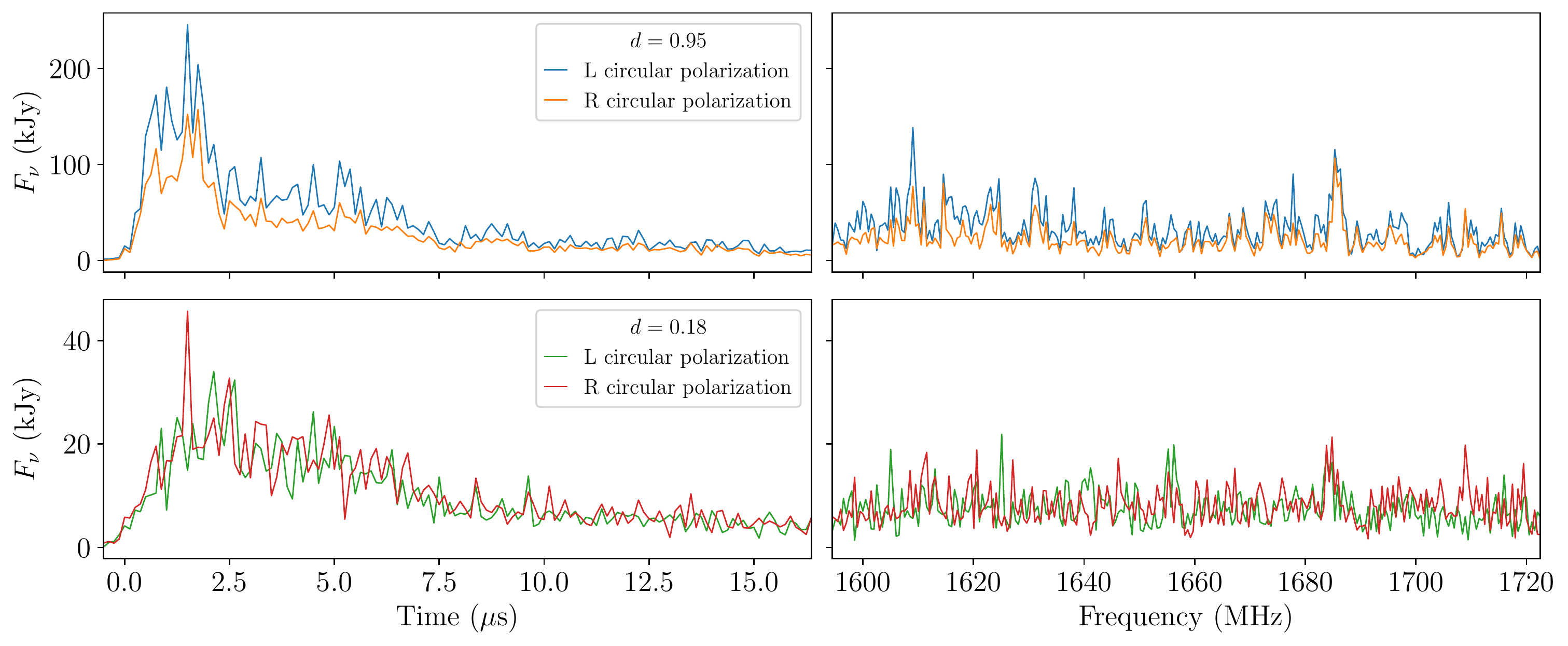}
  \caption{
    Left and right circular polarization pulse profiles and spectra for a giant pulse with high degree of polarization ($d=0.95$; {\em top}) and one with low degree of polarization ($d=0.18$; {\em bottom}).
    {\em Left:\/} Pulses profiles in $125{\rm\;ns}$ bins.
    {\em Right:\/} Pulse spectra for the $16{\rm\;\mu s}$ containing the pulses, in $500{\rm\;kHz}$ channels.
    \label{fig:dop_examples}}
\end{figure*}

At high frequencies, the MP giant pulses are observed to consist generally of multiple nanoshots, which have different center frequencies, bandwidths, and polarization \citep{Hankins2016}.
If the same held at lower frequencies and if all nanoshots passed through the same scattering screen, one would expect the power spectra of giant pulses to have a variance three times larger than the mean squared.
As mentioned in Section~\ref{sec:intro}, this is not the case for the observations of \citetalias{Main2021} and it is not the case for ours either: the variance of the power spectra equals their mean squared (see Appendix~\ref{sec:appendix} and Figure~\ref{fig:mean_std}).
This suggests that if a burst is indeed composed of multiple nanoshots, the shots are imprinted with different IRFs.

Correlating the two polarizations provides a way to verify this conclusion independently of the number of nanoshots, only relying on the fact that scintillation should not depend on the polarization.
This is because the number of nanoshots influences both the degree of polarization of the giant pulse as a whole and the expected strength of the correlation, with the latter depending on whether the nanoshots are imprinted with the same or with different IRFs.

One can see that this would be the case by first considering a single nanoshot: since individual nanoshots are usually very highly polarized, one expects a high degree of polarization.
Furthermore, this single signal is unlikely to be resolved by the screen, so its two polarizations should correlate perfectly.
As one adds more nanoshots, all highly polarized but in roughly random directions, the polarization will start to average out, down to zero for a large number of nanoshots.
The correlation between polarizations will also decrease, but differently depending on whether the nanoshots are all imparted with the same IRF, in which case the limiting correlation coefficient would be $1/3$, or whether they have different IRF, in which case the correlation would tend to zero.

In order to look for this, we calculated Stokes parameters and correlation coefficients between left and right polarization for all pulses with $S/N>100$.
We follow the PSR/IEEE convention \citep{Straten2010} and the Stokes parameters are defined by
\begin{eqnarray}
    I &=& \langle |E_{-}|^{2}\rangle + \langle |E_{+}|^{2}\rangle, \label{eqn:stokesI} \\
    Q &=& 2\Re\{\langle E_{-}^{* }E_{+}\rangle\}, \label{eqn:stokesQ} \\
    U &=& 2\Im\{\langle E_{-}^{* }E_{+}\rangle\}, \label{eqn:stokesU} \\
    V &=& \langle |E_{-}|^{2}\rangle - \langle |E_{+}|^{2}\rangle, \label{eqn:stokesV}
\end{eqnarray}
where $E_{-}$ and $E_{+}$ are the left ($-$) and right ($+$) polarized electric field, $\Re\{\dots\}$ and $\Im\{\dots\}$ indicate the real part of the complex values and the angular brackets $\langle\cdots\rangle$ denote averages over a pulse.
In terms of the Stokes parameters, the total, linear, and circular degree of polarization of a pulse are given by,
\begin{eqnarray}
    d &=& \sqrt{\frac{Q^{2} + U^{2} + V^{2}}{(I - 2)^{2}}}, \label{eqn:dop}\\
    d_{L} &=& \sqrt{\frac{Q^{2} + U^{2}}{(I - 2)^{2}}}, \label{eqn:dopL}\\
    d_{V} &=& \sqrt{\frac{V^{2}}{(I - 2)^{2}}}, \label{eqn:dopV}
\end{eqnarray}
respectively, where we subtract $2$ from the intensities to correct for the contribution from background noise (as appropriate given that we normalized our data by the intensity of the
noise level in each sub-band and polarization; see Section~\ref{subsec:pipeline}).

In Figure~\ref{fig:polarization}, we show the inferred polarization properties.
One sees that the degree of polarization varies over the full range, but with most pulses being only modestly polarized, with a root-mean-squared value of $\langle d^2\rangle^{1/2}\simeq0.45$.
If individual nanoshots are fully but randomly polarized, one would expect that $\langle d^2\rangle\simeq1/n_{s}$, where $n_{s}$ is the number of nanoshots, so we infer that, typically, $n_{s}\simeq5$ nanoshots contribute (or a few more taking into account that the nanoshots will likely have a range in brightness; see  Appendix~\ref{sec:appendix_pol}).
This is of the same order seen by \cite{Hankins2016} at higher frequencies, where nanoshots can be resolved.

One also sees that most pulses have little circular polarization, much less than the $d_{V}/d=1/2$ expected for (an average over) randomly polarized impulses.
This is especially true at high degree of polarization, where a pulse is more likely to be dominated by just a few nanoshots.
It is consistent with what is observed at high frequency, where individual nanoshots have been observed to be strongly linearly polarized \citep{Jessner2010, Hankins2016}.

\begin{figure}
  \centering
  \includegraphics[width=0.47\textwidth,trim=0 0 0 0,clip]{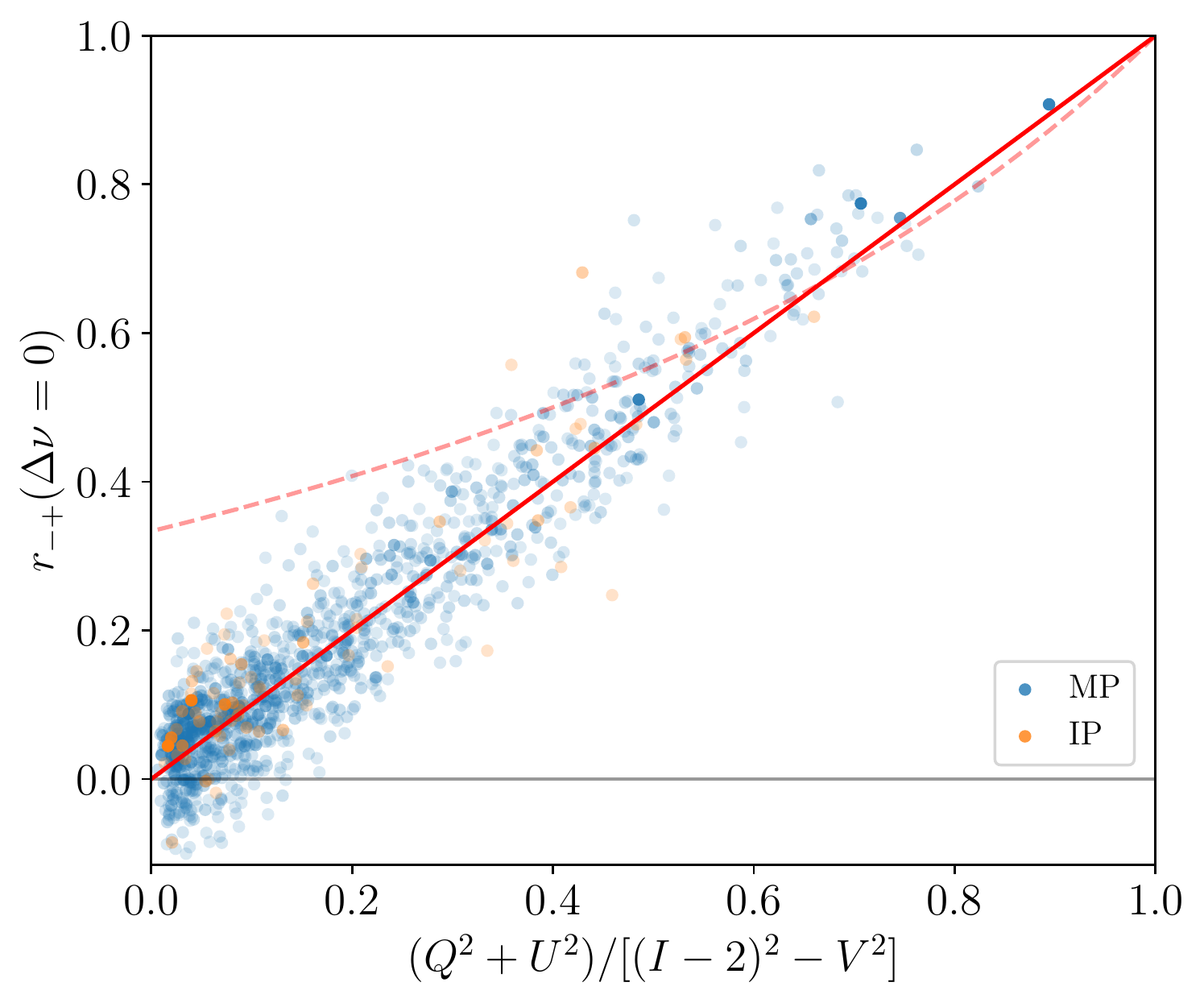}
  \caption{
    Correlation between left and right polarization of MP (blue) and IP (orange) giant pulses as a function of $(Q^2+U^2)/(I^2-V^2)$, with opacity of the individual points scaled with the square root of the S/N of the giant pulse.
    The observations follow the one-to-one correspondence (solid red line) predicted for the case that the nanoshots components of giant pulse have different IRFs.
    If they had passed through the same IRF, the correlations should follow the dashed red line, approaching $1/3$ at low polarization, and have larger scatter (see Appendix \ref{sec:appendix_pol}).
    \label{fig:RLcorrelation_stokes}}
\end{figure}

Turning now to the correlation between left and right circular polarization, we first compare  in Figure~\ref{fig:dop_examples} pulse profiles and power spectra in left and right polarization for one highly polarized and one weakly polarized pulse.
One sees that, as expected, for the highly polarized pulse, left and right are very similar, while for the weakly polarized one the power spectra in particular are quite different.

For the case that each nanoshot is imprinted with a different IRF, it turns out that it is possible to write the correlation coefficient directly in terms of Stokes parameters (see Appendix~\ref{subsec:appendix_pol_diff_irf}), as
\begin{equation}
    r(P_{-}, P_{+}, \Delta\nu=0) = \frac{Q^{2} + U^{2}}{(I - 2)^{2} - V^{2}},
\label{eqn:r_pol}
\end{equation}
where we once again subtract 2 from the intensities to correct for the contribution from background noise.
Hence, the expected dependence is on something close to the degree of linear polarization, not the degree of polarization itself (though for our case, where the giant pulses are mostly linearly polarized, $\langle V\rangle^{2}$ is small and one expects $r(P_{-},P_{+})\simeq d^{2}$).
For the case that all nanoshots are imprinted with the same IRF, no such direct relation is possible, but, as noted above, the expectation is that towards zero polarization, the average correlation will be $1/3$ (see Appendix~\ref{subsec:appendix_pol_same_irf}).

We show the observed correlation coefficients as a function of $(Q^2+U^2)/(I^2-V^2)$ in Figure~\ref{fig:RLcorrelation_stokes}.
We find that the correlation scales nearly perfectly as Equation~\ref{eqn:r_pol}, approaching zero for pulses with small polarization.
This confirms that individual nanoshots are imprinted with different IRFs.

\begin{figure}
  \centering
  \includegraphics[width=0.47\textwidth,trim=0 0 0 0,clip]{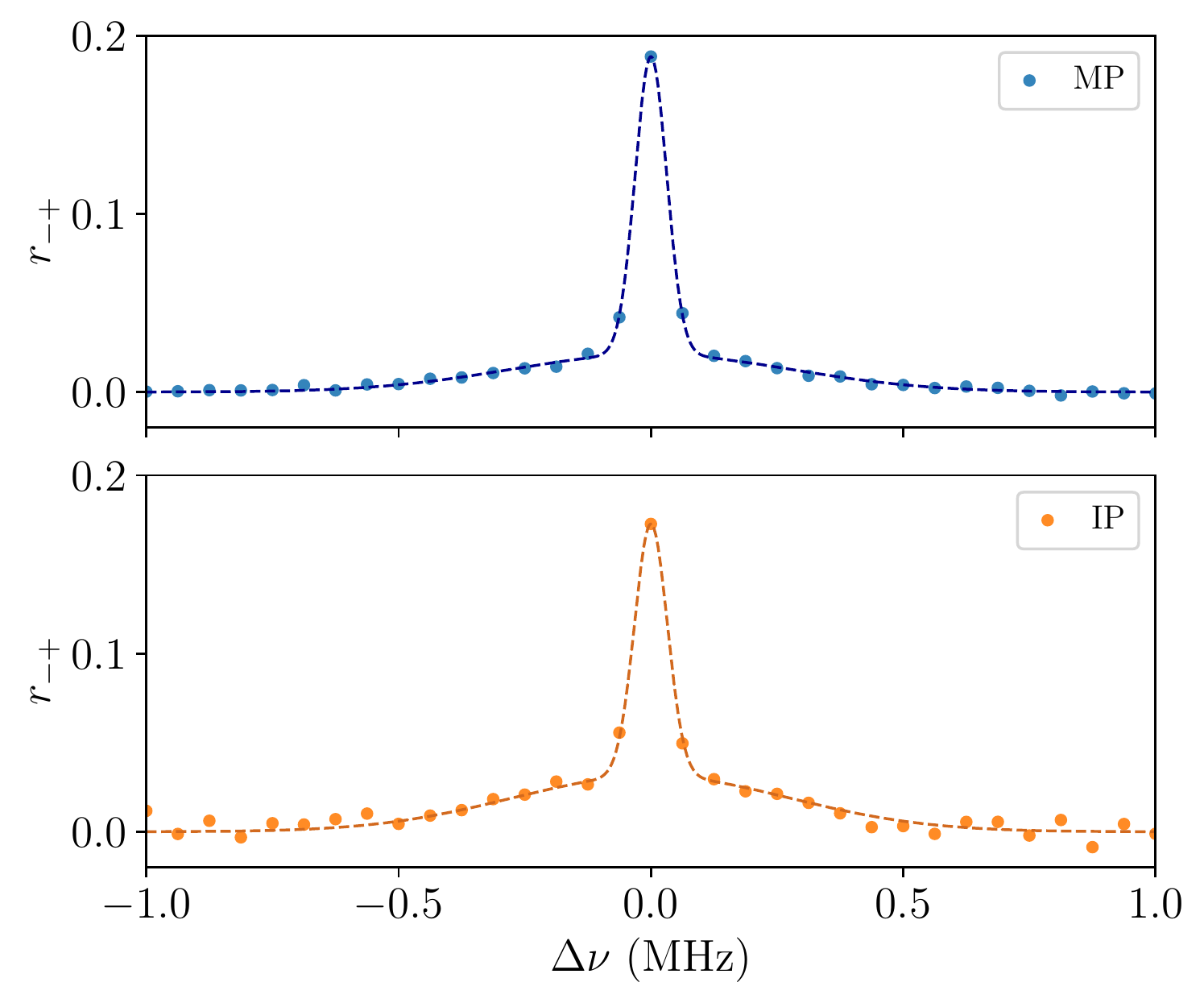}
  \caption{
    Average correlation between left and right polarization of MP (blue) and IP (orange) giant pulses as a function of frequency, with fits using the sum of two Gaussians overdrawn.
    The narrow peak is due to signals from individual nanoshots shared between polarizations, and its width is as expected from the scattering time.
    The wide peak reflects correlation between the nanoshots; the small amplitude and large width show that these are resolved on the sky by the scattering screen.
    \label{fig:RLcorrelation_freq}}
\end{figure}

\begin{deluxetable}{ccccc}[ht]
\tabletypesize{\small}
\setlength\tabcolsep{2.9pt}
\tablecaption{Correlations Between Left and Right Polarization\label{table:pol_corr}}
\tablenum{4}
\tablehead{&
  \colhead{$A_{\text{narrow}}$}&
  \colhead{$\nu_{\text{narrow}}$}&
  \colhead{$A_{\text{wide}}$}&
  \colhead{$\nu_{\text{wide}}$}\\[-1ex]
  \colhead{Comp.}&
  \colhead{(\%)}&
  \colhead{(MHz)}&
  \colhead{(\%)}&
  \colhead{(MHz)}}
\startdata
MP& $16.7\pm0.2$ & $0.0367\pm0.0007$ & $2.11\pm0.13$ & $0.324\pm0.018$ \\
IP& $14.1\pm0.7$ &  $0.038\pm0.002$  &  $3.1\pm0.4$  & $0.32\pm0.03$
\enddata
\end{deluxetable}

As for pulse pairs, we construct weighted averages of the correlations between left and right polarization as a function of frequency, both for MP and IP.
As one can see in Figure~\ref{fig:RLcorrelation_freq}, these show a narrow, strong component and a wide, weak one, which can be reasonably well fit by a combination of two Gaussians (see Table~\ref{table:pol_corr}).
The narrow component is due to the part of the signal from the nanoshots that is shared between polarizations (i.e. the non-circularly polarized parts of individual nanoshots), with amplitude $A_{\text{narrow}}\simeq15\%$ and width $\nu_{\text{narrow}}\simeq0.037{\rm\;MHz}$.
This component's width is consistent with the scattering time ($1/2\pi\nu_{\text{narrow}}\simeq4{\rm\;\mu s}$).
The wide component reflects the correlation between different nanoshots: its small amplitude $A_{\text{wide}}\simeq2\%$ and large width $\nu_{\text{wide}}\simeq0.32{\rm\;MHz}$ show again that the emission region is resolved.
The fact that the amplitude is larger and the width smaller than what is found for pulse pairs, however, suggests that the region in which the nanoshots of a given pulse arise is less resolved and thus smaller than that spanned by pulse pairs.

\section{Emission Regions}\label{sec:emission_regions}

Our analysis suggests that the Crab Pulsar's emissions regions can be resolved by the scintillation screen.
Since the amount and geometry of the scattering material changes over time scales of the order of months \citep{Vandenberg1976}, the extent to which the emission regions are resolved will change as well, and thus different observations can give complementary information.

\begin{figure*}
  \centering
  \includegraphics[width=0.985\textwidth,trim=0 0 0 0,clip]{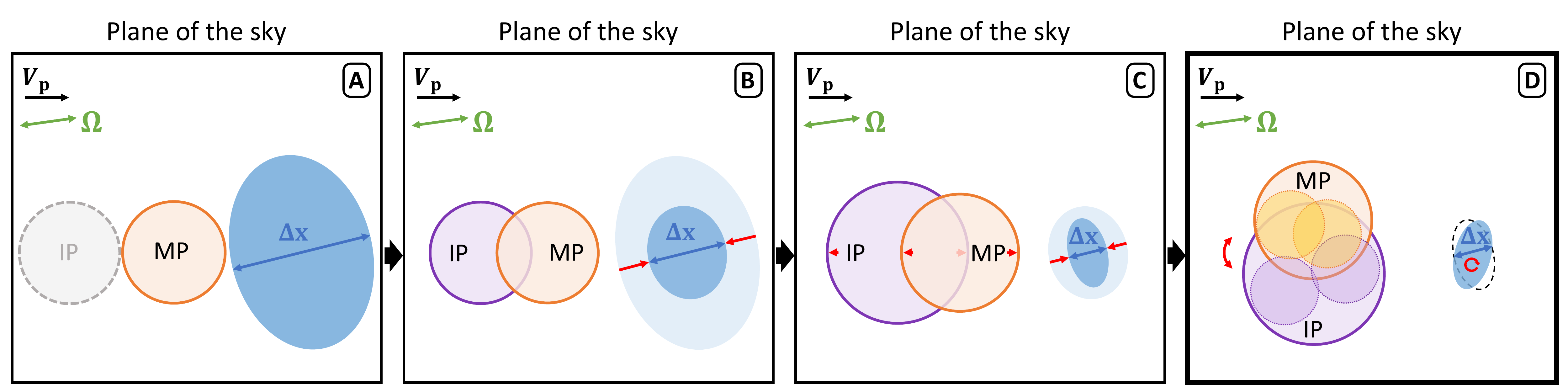}
  \caption{
    Sketch of the giant pulse MP and IP emission regions as projected on the sky (not to scale). The pulsar velocity and spin axis, $\Omega$, are shown on the top left, and the resolution element on the right side of each panel.
    \textbf{A}: The information from \cite{Cordes2004} suggests that at the time of their observations the resolution element covered the full MP emitting region.
    For simplicity, we assume that the emission region is circular.
    Since no information about the IP emitting region size/location is known but there is an IP emission regions somewhere, we show this region in grey.
    \textbf{B}: The clues from \citetalias{Main2021} suggest that the MP and IP emission regions are resolved and that the size of the MP and IP region along the direction of the pulsar motion is larger than the physical resolution at the pulsar, $\Delta x$.
    In addition, the slight positive time offset, $\Delta t_{0}$, indicates that the MP emission region precedes the IP emission region but that the regions likely overlap.
    \textbf{C}: In all our EVN observations, the physical resolution at the pulsar is smaller than in the previous observations.
    We also find that the size of the IP regions is larger than the MP region along the pulsar motion.
    \textbf{D}: In order to have sign changes in $\Delta t_{0}$ and $\Delta\nu_0$, the emission regions cannot be oriented along the direction of the pulsar velocity.
    Rotating them, we arrive at our final picture of the emission regions.
    We use smaller circles within the larger MP and IP emission regions to indicate the regions where nanoshots of a giant pulses may arise.
    \label{fig:cartoon}}
\end{figure*}

\begin{figure}
  \centering
  \includegraphics[width=0.47\textwidth,trim=0 0 0 0,clip]{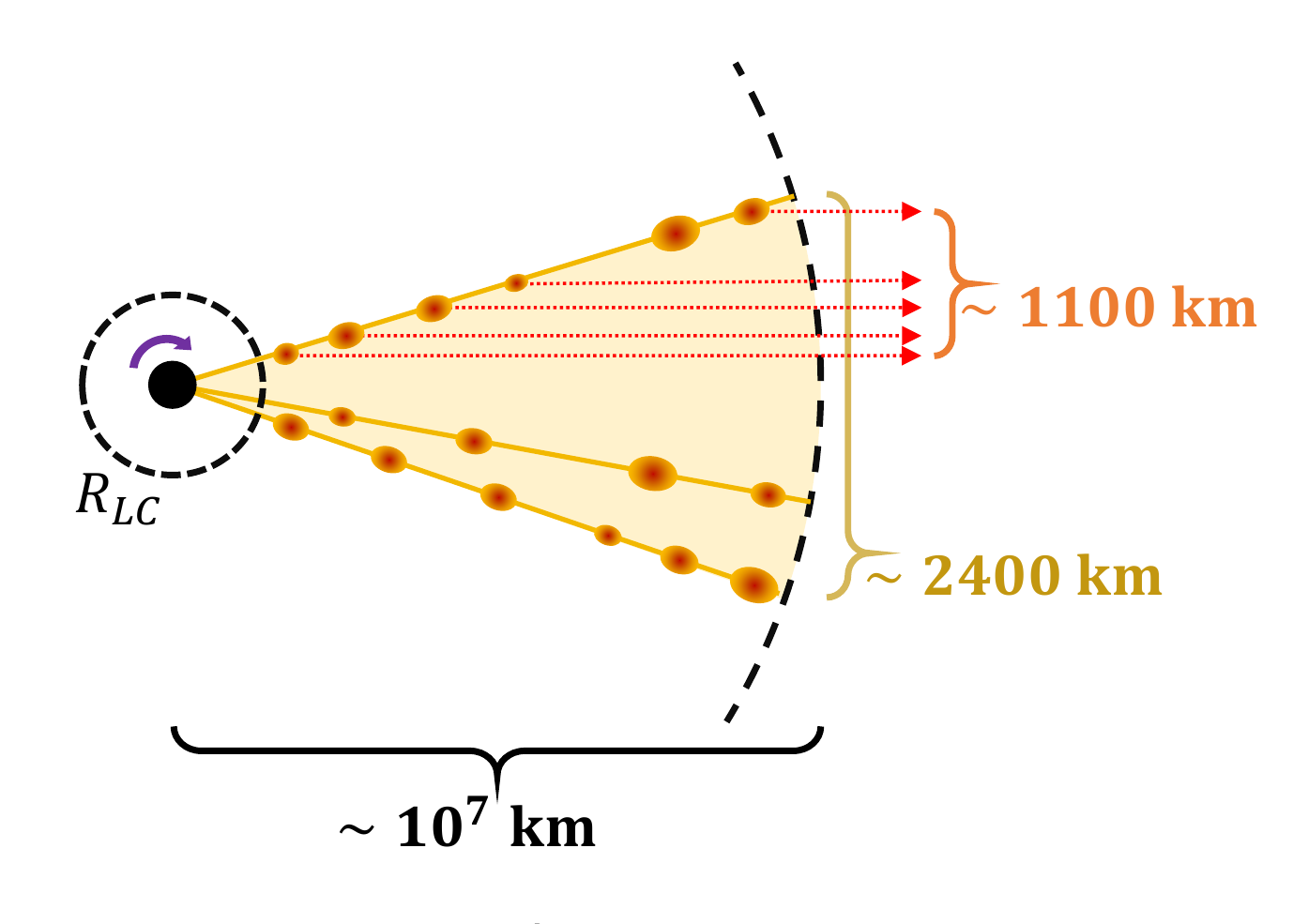}
  \caption{
    Sketch of the giant pulse emission regions as seen from the side (not drawn to scale).
    The black circle on the left indicates the pulsar.
    Blobs of highly relativistic material are ejected from around the light cylinder, at radius, $R_{LC}$, and emit nanoshots at some specific locations along their trajectory, as indicated with the orange circles.
    Projected on the sky, these occur within a smaller part, of $\sim\!1100{\rm\;km}$, of the full $\sim\!2400{\rm\;km}$ emission region.
    Along the line of sight, giant pulse emission occurs over $\sim\!10^{7}{\rm\;km}$.
    \label{fig:nanoshot_cartoon}}
\end{figure}

\begin{figure}
  \centering
  \includegraphics[width=0.47\textwidth,trim=0 0 0 0,clip]{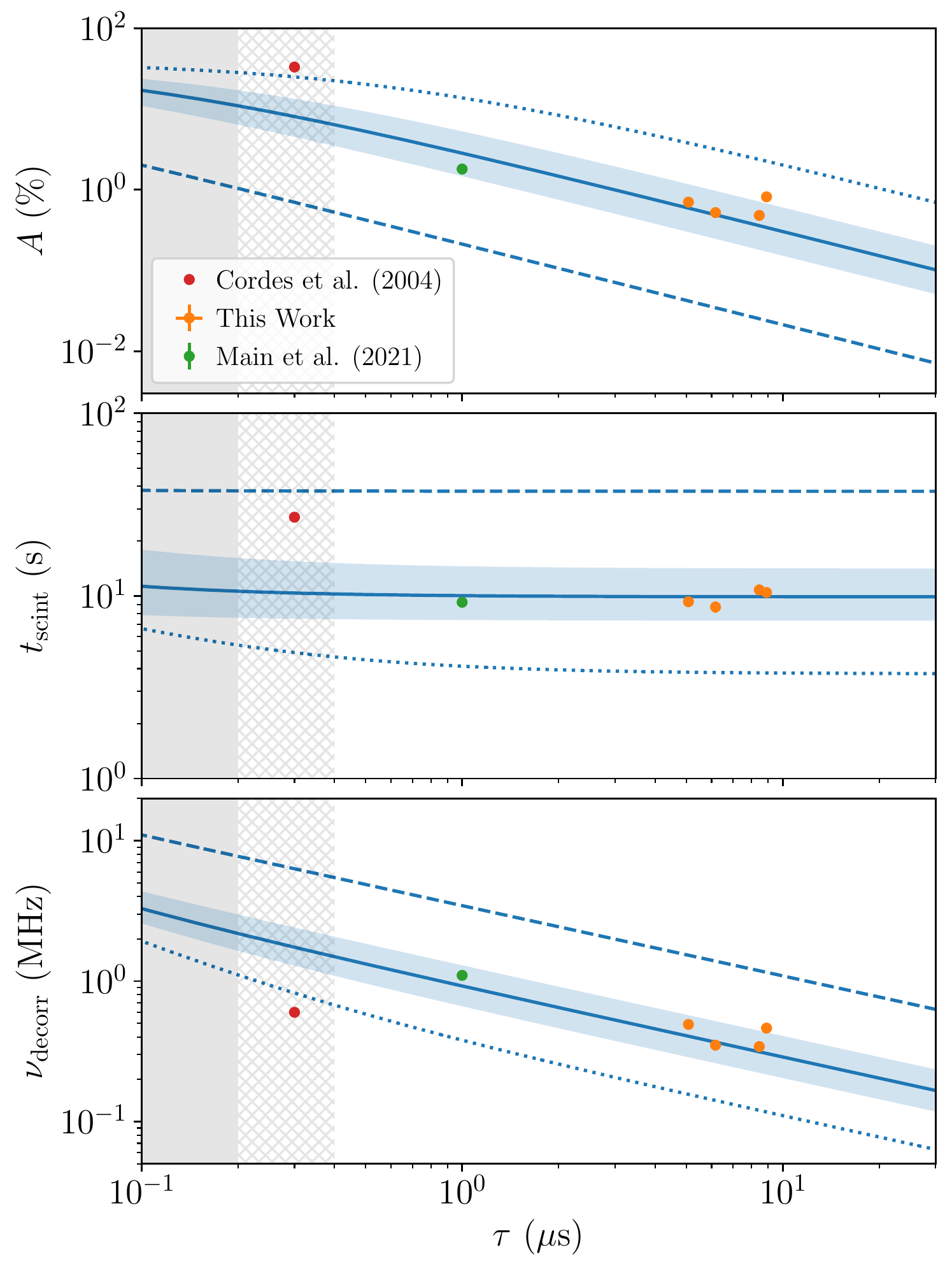}
  \caption{
    The correlation amplitude $A$, scintillation time $t_{\text{scint}}$ and decorrelation bandwidth $\nu_{\text{decorr}}$ as a function of scattering time $\tau$.
    The measurements for the MP are shown by the points (see Tables~\ref{table:profiles} and \ref{table:correlations}).
    Overdrawn with a solid line is what is expected (Eqs~\ref{eqn:A_n_irf}--\ref{eqn:t_scint} and Appendix~\ref{sec:appendix_irf}) for an emission region size $\sigma_{\text{MP}}=835{\rm\;km}$, using a pulsar distance $d_{p}=2{\rm\;kpc}$ and distance between the screen and the pulsar of $d_{p}-d_{s}=1{\rm\;pc}$; the shading shows the effect of varying these within $1.4\leq d_{p}\leq2.7{\rm\;kpc}$ and $0.5\leq d_{p} - d_{s}\leq2{\rm\;pc}$.
    For comparison, the dotted and dashed blue lines show the expectations for  emission sizes of $\sigma_{\text{MP}}=10^{2.5}$ and $10^{3.5}{\rm\;km}$.
    At long $\tau$, the size of the resolution element $\sigma_{x}\propto1/\sqrt\tau$ is smaller than the emission region size and hence the scintillation time is constant, while the amplitude and decorrelation bandwidth decrease as $1/\tau$ and $1/\sqrt\tau$, respectively.
    Conversely, at short $\tau$, the resolution is poorer and the scintillation time is proportional to $1/\sqrt\tau$, while the amplitude approaches 1/3 and the decorrelation bandwidth approaches $1/2\pi\tau$.
    The model's assumption of negligible contribution from interstellar scattering breaks down at short $\tau$: the solid gray region indicates where interstellar scattering likely dominates and the hatched grey one where it contributes significantly.
    A significant contribution from interstellar scattering means the model cannot reproduce the measurements of \citet{Cordes2004}.
    \label{fig:model}}
\end{figure}

In this section, we aim to sketch out what the emission regions may look like as projected on the sky, trying to follow what a consistent picture would look like starting from the measurements of \citet{Cordes2004}, and then including those of \citetalias{Main2021} and ourselves.
Below, we focus on the two emission regions from which the giant pulses that comprise the MP and IP originate.
In addition, we can constrain the properties of the parts of the regions from which the nanoshots originate that make up a given giant pulse, since we found in Section~\ref{subsec:correlation_polarizations} that these parts are also resolved in our observations.
We show our sketch of the constraints on the emission regions in Figure~\ref{fig:cartoon}, beginning with the results of \cite{Cordes2004} in panel~A, and ending with what we infer from all observations in panel~D; a side view of the latter is shown in Figure~\ref{fig:nanoshot_cartoon}.
In Figure~\ref{fig:model} we show how the MP-MP correlation parameters allow us to place constraints on the emission region size from which MPs originate.

For our numerical estimates, we calculate resolutions from the scattering time $\tau$ using Equation~\ref{eqn:resolution}, assuming $d_{\text{p}}-d_{\text{s}}\simeq1{\rm\;pc}$ (which is much smaller than $d_{\text{p}}\simeq2{\rm\;kpc}$, and thus the term $d_{\text{p}}/d_{\text{s}}\simeq1$).
We will ignore dependencies of the prefactor in Equation~\ref{eqn:resolution} on the degree of anisotropy of the scattering screen.
We will also use the following relations, derived in Appendix~\ref{sec:appendix_irf}, that link the scattering time $\tau$, resolution $\sigma_{x}$, the size of the emission region $\sigma_{s}$, the typical number $n_{s}$ of nanoshots per giant pulse, and the relative velocity between the pulsar and screen $v_{\text{rel}}$ with our observables, the amplitude $A$, decorrelation bandwidth $\nu_{\text{decorr}}$ and scintillation timescale $t_{\text{scint}}$:
\begin{eqnarray}
  \sigma_{c} &=& \sqrt{\sigma_{x}^{2} + (2\sigma_{s})^{2}}\label{eqn:corr_scale},\\
  n_{g} &=& \left(\frac{\sigma_{c}}{\sigma_{x}}\right)^{2}=1+\frac{(2\sigma_{s})^{2}}{\sigma_{x}^{2}}\label{eqn:n_g},\\
  A &=& \frac{1}{n_{g} + 2(1-1/n_{s})}\label{eqn:A_n_irf},\\
  \nu_{\text{decorr}} &=& \frac{\sqrt{n_{g}}}{2\pi\tau}\label{eqn:nu_decorr},\\
  t_{\text{scint}} &=& \frac{\sigma_{c,\parallel}}{v_{\text{rel}}\sqrt{2}}\label{eqn:t_scint}.
\end{eqnarray}
Here, all sizes are defined as standard deviations of normal distributions.
The intermediate quantities $\sigma_{c}$ and $n_{g}$ are the effective spatial correlation scale and effective number of resolution elements, respectively; for unresolved emission, $\sigma_{c}\simeq\sigma_{x}$ and $n_{g}\simeq1$, while for highly resolved emission, $\sigma_{c}\simeq2\sigma_{s}$ and $n_{g}\simeq(2\sigma_{s}/\sigma_{x})^2$.
We write $\sigma_{c,\parallel}$ as a reminder that the scintillation time depends on the sizes parallel to the direction of motion.

For the pulsar velocity, we use the value inferred by \cite{Kaplan2008} from the proper motion ($12.5{\rm\;mas\;yr^{-1}}$ at an angle of $290\arcdeg$ east of north): $v_{\text{p}}=120(d_{\text{p}}/{2{\rm\;kpc}}){\rm\;km\;s^{-1}}$.
For the screen velocity, we will assume that it has negligible average velocity with respect to the pulsar, but that it can vary, with velocities ranging up to $\sim\!70{\rm\;km\;s^{-1}}$ like for the optical filaments \citep{Trimble1968}.

\subsection{Constraints from Unresolved Observations}

From the average correlation amplitude of $\sim\!1/3$ of MP spectra at small time delays, \cite{Cordes2004} suggests that the MP emission region is one where giant pulses are comprised of multiple nanoshots which pass through the same scintillation element, i.e., the emission region is not resolved.

In the limit of unresolved emission, the scintillation timescale is determined by the speed with which the pulsar crosses a scintillation resolution element, $t_{\text{scint}}\simeq (\sigma_{x,\parallel}/\sqrt{2})/v_{\text{p}}$ (see Equation~\ref{eqn:t_scint}), where $\sigma_{x, \parallel}$ is the resolution along the direction of motion.
\cite{Cordes2004} measure $t_{\text{scint}}=25\pm5$ and $35\pm5{\rm\;s}$ at $1.48$ and $2.33{\rm\;GHz}$, respectively, which yields $\sigma_{x,\parallel}=4240$ and $5940{\rm\;km}$, respectively.
Since $\sigma_x$ scales linearly with wavelength (see Equation~\ref{eqn:resolution}), the two results are consistent with each other.

Another estimate of the resolution can be made from the scattering time $\tau$ (Section~\ref{sec:interstellar_interferometry}).
\cite{Cordes2004} do not measure $\tau$ directly, but for an unresolved source it can be inferred from the decorrelation bandwidth.
At $1.48{\rm\;GHz}$, they can only set a limit, $\nu_{\text{decorr}}\lesssim0.8{\rm\;MHz}$, which implies  $\tau_{1480}\simeq1/2\pi\nu_{\text{decorr}}\gtrsim0.2{\rm\;\mu s}$, but at $2.33{\rm\;GHz}$, they measure $\nu_{\text{decorr}}\simeq2.3{\rm\;MHz}$, which implies $\tau_{2330}\simeq0.07{\rm\;\mu s}$.\footnote{In \citetalias{Main2021}, the scattering time was inferred from the value of $\tau_{610}\simeq0.1{\rm\;ms}$ observed at $610{\rm\;MHz}$ by \cite{McKee2018} at the time of the \cite{Cordes2004} observation.
  Using the usual $\nu^{-4}$ scaling, that value implies $\tau_{2330}=0.5{\rm\;\mu s}$.
  We noticed, however, that \citeauthor{McKee2018} find few lower scattering times from their fits to the average profiles, while from our own studies of the profiles of individual giant pulses in the same data (Serafin-Nadeau et al., in preparation), is clear that lower values, down to $0.01{\rm\;ms}$, are common.
  This suggests that values of $\tau_{610}\lesssim0.1{\rm\;ms}$ should be treated as upper limits; we confirmed with McKee (2022, pers.\ comm.) that this is possible.
  And indeed, at $111{\rm\;MHz}$, \cite{Losovsky2019} find $\tau_{111}=15{\rm\;ms}$ at the time of the \citeauthor{Cordes2004} observation, which implies $\tau_{2330}\simeq0.08{\rm\;\mu s}$, consistent with what we infer from the decorrelation bandwidth.}
If this scattering were dominated by the nebula, one would infer, from Equation~\ref{eqn:resolution}, $\sigma_x\lesssim730{\rm\;km}$ and $\sigma_x\simeq790{\rm\;km}$ at 1.48 and $2.33{\rm\;GHz}$, respectively (again consistent with each other given the dependence on wavelength).
This is much smaller than inferred above from the scintillation time.
This might be taken to suggest a very anisotropic screen, but more likely is the assumption that the nebula dominates scattering does not hold.
As estimated in Section~\ref{sec:interstellar_interferometry}, the interstellar screen should have $\tau\simeq0.2{\rm\;\mu s}$ at $1.66{\rm\;GHz}$.
Hence, the interstellar screen likely contributed to the scattering (though it cannot have dominated it completely, since that would imply a scintillation time of several minutes).

We conclude from the \cite{Cordes2004} observations that the emission region is small enough to not contribute to the scintillation time.
From Equation~\ref{eqn:corr_scale}, one sees that for given size $\sigma_s$ of the region in which giant pulses are emitted, the relevant ratio that determines whether it contributes is  $2\sigma_s/\sigma_x$ (see Appendix~\ref{sec:appendix_irf} and \cite{Gwinn1998} for derivations).
Taken this ratio to be smaller than unity and using the smaller of the two constraints on $\sigma_{x,\parallel}$ inferred from the scintillation timescale, we conclude that the emission region has a size $\sigma_{s,\parallel}\lesssim\!4240/2 = 2120{\rm\;km}$.

\subsection{The Emission Regions Resolved}\label{subsec:emission_region_resolved}

The observations presented in \citetalias{Main2021} and here were both taken at times that the scattering time was substantially longer than those of \cite{Cordes2004}, and therefore dominated by the nebular screen.
The scattering times of $\tau_{1660}\simeq1$ and $5{\rm\;\mu s}$, respectively, imply  resolutions of $\sigma_x=290$ and $130{\rm\;km}$.
In both observations, the maximum correlation amplitudes are greatly reduced compared to the $1/3$ found by \cite{Cordes2004}, to $\sim\!1.8$ and $0.6\%$, respectively, and the decorrelation bandwidths are substantially larger, at $1.1$ and $0.4{\rm\;MHz}$ respectively, than expected from the scattering time ($1/2\pi\tau=0.16$ and $0.03{\rm\;MHz}$, respectively).

Both these indicate the emission regions are resolved.
If two giant pulses, even close in time, arise from a region much larger than the scintillation elements, they are imprinted with typically different IRFs and hence the correlation amplitudes will be low.
Given that they will correlate best at the start of the scattering tail, with the paths still close to the line of sight and the effective resolution poor, and poorly correlated at later times when the resolution is higher, only the coarser frequency structure in the power spectra will correlate and thus the decorrelation bandwidth will be relatively large (as derived for resolved sources by \citealt{Gwinn1998}).
Following the same logic, the even lower amplitude of the MP-IP correlation suggests that the IP emission region is even more resolved and thus somewhat larger.

We also found that the regions from which nanoshots of given giant pulses arise are resolved.
This is surprising, since the nanoshots are causally related and occur within a few microsecond.
We return to this in Section~\ref{subsec:superluminal_velocity}, but here recall that if they were not resolved, i.e., if all nanoshots were imprinted with the same IRF, the variance of the power spectra should be three times larger than the mean squared. In contrast, in both our observations and those of \citetalias{Main2021}, the variance is equal to the mean squared as expected if nanoshots are imprinted with different IRFs (see Appendix~\ref{sec:appendix}).
Furthemore, the correlations between the left and right circular polarization approach zero for pulses with low polarization (see Section~\ref{subsec:correlation_polarizations}), which can only be understood if the nanoshots are imprinted with different responses, i.e., if their physical separations as projected on the sky are resolved (see Appendix~\ref{sec:appendix_pol}).

We can use Equation~\ref{eqn:A_n_irf} to estimate the number of scintillation resolution elements $n_g$ from the observed amplitudes and the number of nanoshots per giant pulse, $n_s\simeq5$ (see Section~\ref{subsec:correlation_polarizations}).
We find that the emission region covers approximately $55$ and $165$ resolution elements in the data from \citetalias{Main2021} and presented here, respectively.

As a consistency check, we can use the decorrelation bandwidth, which for low correlation amplitude depends on the amplitude and scattering time as $\nu_{\text{decorr}}\simeq(1/\sqrt{A})/2\pi\tau$ (see Equation~\ref{eqn:nu_decorr}). Inserting numbers, we find $\nu_{\text{decorr}}=1.2$ and $0.4{\rm\;MHz}$, respectively, reasonably consistent with what we observe ($1.1$ and $0.4{\rm\;MHz}$, respectively; see Table~\ref{table:correlations}).

Following the same logic for the correlations between the two polarizations of individual pulses, for which we found an amplitude of $\sim\!2\%$ for the wide component for epoch~B, we infer that the regions in which nanoshots from individual pulses arise cover about 50 resolution elements.
The implied decorrelation bandwidth is $\nu_{\text{decorr}}=0.23{\rm\;MHz}$, reasonably consistent with the observed width ($\sim\!0.32{\rm\;MHz}$; see Table~\ref{table:pol_corr}), although not as close as we found for the pulse pairs.

\subsection{Emission Region Sizes}\label{subsec:emission_region_sizes}

If the emission regions are larger than the resolution elements, the time decorrelation scale should reflect the time required for an emission region to move by its own size, independent of resolution.
Indeed, we find that the time decorrelation scale is similar between the observations of \citetalias{Main2021} and ours, at $t_{\text{scint, MP-MP}}\simeq10{\rm\;s}$.
Thus, along the direction of motion, the size of the MP emitting region should be $\sigma_{\text{MP}, \parallel}\simeq v_{p}t_{\text{scint, MP-MP}}/\sqrt{2} \approx850{\rm\;km}$ (see Equation~\ref{eqn:t_scint}).
Note that here we ignore any contribution of the screen motion, which, at up to $70{\rm\;km\;s^{-1}}$, introduces an uncertainty of about 30\%.
Indeed, variations in screen velocity likely are responsible for the fact that the individual scintillation times are not quite the same within the measurement errors (see Table~\ref{table:correlations}, in particular epoch~C).

Above, we found that the resolution elements had size $\sigma_x\simeq290$ and $130{\rm\;km}$ in the observations of \citetalias{Main2021} and here, respectively, and that the MP emitting region spans about $n_g\simeq55$ to $165$ resolution elements, respectively.
Since in the limit of well-resolved emission, $n_g\simeq(2\sigma_{\text{MP}}/\sigma_x)^2$ (Equation~\ref{eqn:n_g}),
one infers $\sigma_{\text{MP}}\simeq\frac{1}{2}\sigma_x\sqrt{n_g}=1075$ and $835{\rm\;km}$, respectively.
These numbers are pleasingly close to what we inferred from the scintillation time above, with the small differences perhaps due to changes in screen distance within the nebula, or deviations from isotropy.

The low amplitude and increased time and frequency widths of the MP-IP correlations in both our datasets and that of \citetalias{Main2021} suggests that the MP emission size is slightly smaller than the IP emission size.
We can estimate the IP emission size along the direction of the pulsar velocity from the scintillation times for MP-MP and MP-IP, by $\sigma_{\text{IP}, \parallel} = v_{p}(t_{\text{scint, MP-IP}}^2 - \frac{1}{2}t_{\text{scint, MP-MP}}^2)^{1/2}\simeq 1040{\rm\;km}$.
Unfortunately there are insufficient pairs of IP pulses to construct an IP-IP correlation meaningful enough for a complementary measurement, though for completeness we note that the very noisy result is certainly consistent with what is implied by the above size: $n_g\simeq(2\sigma_{\text{IP},\parallel}/\sigma_x)^2\simeq255$ and thus $A_{\text{IP-IP}}\simeq1/n_g\simeq0.4\%$ and $\nu_{\text{decorr,IP-IP}}\simeq\sqrt{n_g}/2\pi\tau\simeq0.5{\rm\;MHz}$

As noted, the correlations between polarizations imply that the screen also resolves the region in which the nanoshots that comprise individual giant pulses originate.
Using $n_g\simeq50$, we infer $\sigma_{\text{nano}}\simeq460{\rm\;km}$.

Overall, we infer that the MP and IP emission regions, if described by normal distributions, have sizes $\sigma_{\text{MP}}\simeq835{\rm\;km}$ and $\sigma_{\text{IP}}\simeq1040{\rm\;km}$, or in terms of full width at half maximum (FWHM), about $2000$ and $2400{\rm\;km}$, respectively.
The constituent nanoshots arise in somewhat smaller regions, $\sigma_{\text{nano}}\simeq460{\rm\;km}$ (or FWHM of $\sim\!1100{\rm\;km}$).
In Figures \ref{fig:cartoon}D and \ref{fig:nanoshot_cartoon} we show how we envisage the nanoshot emission regions to fit within the larger ones.

\subsection{Orientation and Separation between Emission Regions}\label{subsec:emission_region_orientation_separation}

Time and frequency offsets in the MP-IP correlations can only occur if the respective emissions regions are offset as projected on the sky.
Time offsets are easiest to interpret, as they reflect mostly how one region will move through the same resolution element at a slightly different time than the other.
In both our observations and those of \citetalias{Main2021}, marginally significant time offsets up to $\sim1{\rm\;s}$ are detected.
Treating those as upper limits suggests that the MP and IP emission regions are coincident along the pulsar motion to within a few $100{\rm\;km}$.
If we instead take them as measurements, we need to consider that the sign is different.
This is possible only if the two regions are not aligned along the pulsar motion and the scattering screen is anisotropic on both occasions, but with a different orientation (see Figure~\ref{fig:cartoon}D).
In that case, for a given angle $\psi$ between the short axis of the resolution element (long axis of the scattering screen) and the proper motion, the projected separation would be $\sim v_{p}\Delta t_{0}/\cos\psi$, i.e., $\sim\!170{\rm\;km}$ for a typical angle $\psi=45\arcdeg$.

Independent evidence for anisotropy comes from the measurement of significant time-frequency covariance $\rho_{f, t}$ in the correlations.
Furthermore, frequency offsets can only happen for an anisotropic screen.
For those, one requires knowledge of the screen to infer a spatial offset, but a lower limit is given by $(w/2)\Delta\nu_0/\nu_{\text{decorr}}$ (where $w$ is the full width at half maximum of the resolution element), which implies offsets of $\gtrsim\!80$ and $\gtrsim\!20{\rm\;km}$ for \citetalias{Main2021} and epoch~B, respectively.

Overall, we conclude that the MP and IP emission regions overlap substantially, but are likely offset by a few $100{\rm\;km}$, i.e., about 10\% of their diameter, in a direction different from the direction of the proper motion.

\subsection{Evidence for Superluminal Velocity}\label{subsec:superluminal_velocity}

From our observations, it seems clear the individual nanoshots that make up a giant pulse are resolved by the scattering screen, arising in a region on the sky with width $\sigma_{\text{nano}}\simeq460{\rm\;km}$.
Yet, the nanoshots in a given giant pulse are clearly causally related, and arrive within $\sigma_{t}\simeq0.6{\rm\;\mu s}$ (Table~\ref{table:profiles}), which at face value suggests a physical separation of $\lesssim\!0.18{\rm\;km}$.

As mentioned in Section~\ref{sec:intro}, a similar problem was encountered by \cite{Bij2021}, who found that in order to reproduce the drifting bands seen during the scattering tail of one particular giant pulse, the screen had to resolve the constituent nanoshots, which therefore had to be separated by $\sim\!60{\rm\;km}$.
\citeauthor{Bij2021} suggest that the discrepancy could be resolved if the nanoshots are emitted by blobs of material moving highly relativistically, with $\gamma\simeq10^4$.
Highly relativistic motion would also naturally explain why the nanoshots are resolved in our observation; they require $\gamma\gtrsim3\times10^3$.

One implication of relativistic motion is that the nanoshots are emitted over a region that is extended along the line of sight by another factor $\gamma$, i.e., several ${\rm Gm}$,
or of order $10^{3}$ light cylinder radii.
Given that, our measured sizes would be upper limits to the sizes of the regions where the plasma causing the emission originates, and the differences in apparent size between the interpulse and main pulse that we find may reflect differences in $\gamma$ rather than true size.

A better measure of the true size may be the observed pulse phase widths of $\sim\!7.6\arcdeg$ and $\sim\!9.4\arcdeg$ for MP and IP, respectively.
These are much larger than the beaming angle implied by the relativistic motion, of $\sim\!1/\gamma\lesssim1\arcmin$, suggesting that the blobs are emitted in a small range of directions, and thus also from a range of positions.
Assuming the source is near the light cylinder radius, the implied widths perpendicular to the spin axis (and also roughly perpendicular to the direction of the pulsar velocity) are $\sim\!210$ and $260{\rm\;km}$, respectively.

\section{Conclusions and Ramifications}\label{sec:future_work}
We find that in our observations, when the scattering was relatively strong and dominated by the nebular screen, the physical resolution at the pulsar was $\sim\!130{\rm\;km}$.
From correlations between spectra of our large numbers of giant pulses, it is clear that this resolves the giant pulse emission regions.

We infer apparent diameters of $\sim\!2000$ and $2400{\rm\;km}$ for the emission regions of the MP and IP components.
This strongly favors emission arising beyond the light cylinder radius.
Slight time and frequency offsets in the MP-IP correlations suggest that MP and IP emission regions overlap significantly but not completely, with changes in sign suggesting that they are not aligned along the direction of the pulsar motion.

The largest surprise is that we also resolve the parts of the emission region from which the $\sim\!5$ nanoshots that comprise a given giant pulse arise: this is clear both from the statistics of the giant pulse power spectra as well as from the dependence of the correlation between polarizations on polarization properties.
From the frequency dependence of the polarization correlations, we infer that the nanoshots arise in a region with diameter of $\sim\!1100{\rm\;km}$, smaller than but of the same order as the size of the region in which giant pulses occur.
Since nanoshots are causally related, the simplest solution seems to be that the plasma emitting them move at highly relativistic speeds with $\gamma\simeq10^{4}$, generalizing to all pulses what was found for a single pulse from drifting drifting bands in its scattering tail \cite{Bij2021}.
It thus provides additional support for models that require highly relativistic motion (e.g., \citealt{Lyutikov2021}).

\subsection{Implications for FRB-substructure}
Several lines of evidence suggest that at least some repeating fast radio bursts (FRBs) are generated by young magnetars \citep[see e.g. ][for a recent review]{petroff_2022_aarv}.
Observational support for such a scenario is given by, e.g., the FRB-like signals from the Galactic magnetar SGR~1935+2154 \citep{chime_2020_natur_galacticfrb, bochenek_2020_natur} and the extreme magneto-ionic environment of FRB~20121102A \citep{michilli_2018_natur, spitler_2016_natur}.

Additionally, substructure on microsecond and nanosecond scales has been reported for FRB~20180916B \citep{Nimmo_2021_NatAs} and FRB~20200120E \citep{Nimmo_2022_NatAs}, respectively.
Such short timescale variation translates to light travel times of order $1{\rm\;km}$ and below, i.e., similar in extent to what one would expect given the time separation between nanoshots making up an individual giant pulse of the Crab Pulsar (Table~\ref{table:profiles}, Section~\ref{subsec:superluminal_velocity}).
The evidence for superluminal motion discussed above may also play a role in FRB emission, suggesting that relativistic effects cannot be neglected when modeling FRBs.
In fact, relativistic beaming would help explain the high brightness temperatures of FRBs, and may even be directly responsible for the emission, if the plasma blobs can act as a magnetic mirror \citep{Yalinewich2022}.

\subsection{Future work}
To improve our understanding of the orientation and sizes of the giant pulse emission regions, further observations may help.
In particular, probing the emission regions at different scattering timescales will further constrain the size of the emission regions and place bounds on the transverse velocities of the optical filaments.
Similarly, observations at higher frequencies will result in a lower resolution at the pulsar which can also help to quantify the size of the emitting regions.
With additional MP-IP correlations, we can more confidently determine whether there is a spatial offset between the emission regions and also map changes in the shapes of the resolution elements.

Our beamformed data shows the greatly improved sensitivity gained using multiple telescopes.
With even more telescopes, we can perhaps detect more IPs and obtain more meaningful IP-IP correlations, which will constrain the size of the IP region directly.
As already mentioned in \citetalias{Main2021}, some of the uncertainty in our determination for the resolution element size comes from not having well constrained distances to the Crab Pulsar and the optical filaments; a good parallax distance would resolve this.

\subsection*{Acknowledgements}

We thank Aard Keimpena for very helpful advice on fringe stopping and other VLBI aspects of our data reduction, the Toronto scintillometry group for useful discussions, and the anonymous referee for their careful reading.
Computations were performed on the Niagara supercomputer at the SciNet HPC Consortium \citep{Loken2010, Ponce2019}.
SciNet is funded by the Canada Foundation for Innovation, the Government of Ontario, the Ontario Research Fund -- Research Excellence, and the University of Toronto.
MHvK is supported by the Natural Sciences and Engineering Research Council of Canada (NSERC) via discovery and accelerator grants, and by a Killam Fellowship.
U.-L.P receives support from Ontario Research Fund-Research Excellence Program (ORF-RE), NSERC [funding reference number RGPIN-2019-067, CRD 523638-18, 555585-20], Canadian Institute for Advanced Research (CIFAR), the National Science Foundation of China (Grant No. 11929301), Alexander von Humboldt Foundation, and the National Science and Technology Council (NSTC) of Taiwan (111-2123-M-001, -008, and 111-2811-M-001, -040).

\facility{
  The European VLBI Network (EVN) is a joint facility of independent European, African, Asian, and North American radio astronomy institutes.
  Scientific results from data presented in this publication are derived from the following EVN project codes: EK036~A-D.}

\software{
  astropy \citep{AstropyCollaboration2013, AstropyCollaboration2018, AstropyCollaboration2022},
  Baseband \citep{VanKerkwijk2020},
  CALC10 \citep{Ryan1980},
  numpy \citep{Harris2020},
  matplotlib \citep{Hunter2007},
  pulsarbat \citep{Mahajan2022},
  scipy \citep{Gommers2022},
  SFXC \citep{Keimpema2015},
  tempo2 \citep{Hobbs2012}.}

\appendix
\section{Optimal Estimates of Intrinsic Correlations from Power Spectra}\label{sec:appendix}

For two observed power spectra $P_{1}(\nu)$ and $P_{2}(\nu)$ sampled at $k$ frequencies $\nu_{i}$, the sample correlation coefficient is given by,
\begin{equation}
  r(P_{1}, P_{2}) = \frac{\frac{1}{k - 1}\sum_{i=1}^{k} (P_{1,i} - m_{1})(P_{2,i} - m_{2})}{s_{1} s_{2}},
\label{eqn:r}
\end{equation}
where $m$ and $s$ are the sample mean and standard deviation\footnote{We use Greek letters to indicate population statistics and Latin numerals to indicate sample statistics.} of the power spectra, given by $m=\frac{1}{k}\sum_{i} P_{i}$ and  $s^{2}=\frac{1}{k-1}\sum_{i} (P_{i}-m)^{2}$, respectively.

We wish to determine the intrinsic degree of correlation, $\rho$, between pairs of power spectra from multiple pairs of measurements, correcting the individual estimates for biases due to noise, and combining them using optimal weights, i.e., for a set of pairs $ij$ to determine an optimal
\begin{equation}
  \bar{r}_{\text{opt}} = \frac{\sum_{ij} w_{ij}c_{ij}r_{ij}}{\sum_{ij} w_{ij}},
\label{eqn:r_opt}
\end{equation}
where in general one would expect that the weights $w_{ij}$ and correction factors $c_{ij}$ would depend primarily on the signal-to-noise ratios of the input power spectra.
\citetalias{Main2021} already addressed the correction factors, but as we need those for the optimal weights as well, we briefly repeat the logic here.

The power spectra we are considering are Fourier transforms of the measured electric field of a giant pulse, given by
\begin{equation}
    E(\nu) = S(\nu) + N(\nu),
\label{eqn:electric_field}
\end{equation}
where $S(\nu)$ and $N(\nu)$ are the pulse signal convolved with the IRF and the measurement noise, respectively. Since we are working with complex data, the measured power is,
\begin{equation}
  P(\nu) = E(\nu)E^{\ast}(\nu)
  = |S(\nu)|^{2} + |N(\nu)|^{2} + 2\Re\{S(\nu)N^{\ast}(\nu)\},
\label{eqn:power_spectra}
\end{equation}
where $|\dots|$ indicates absolute values and $\Re\{\dots\}$ indicates the real part of the complex values.

We wish to determine the correction to the correlation coefficient to make it an estimate of the intrinsic correlation $\rho(S_{1}, S_{2})$.
As noted in \citetalias{Main2021}, the presence of (uncorrelated) noise does not bias the sample covariance, which is the numerator in Equation~\ref{eqn:r}, but it does bias the standard deviations in the denominator.
Hence, $\rho(S_{1}, S_{2})$ will be related to the expectation value $\rho(P_{1}, P_{2})$ for the correlation coefficient by,
\begin{equation}
  \rho(S_{1}, S_{2}) = \frac{{\text{Cov}}(S_{1}, S_{2})}{\sigma_{S,1}\sigma_{S,2}}
  = \frac{{\text{Cov}}(P_{1}, P_{2})}{\sigma_{S,1}\sigma_{S,2}}
  = \rho(P_{1}, P_{2}) \frac{\sigma_{P,1}\sigma_{P,2}}{\sigma_{S,1}\sigma_{S,2}}.
\label{eqn:rho1}
\end{equation}
Thus, to calculate the bias we need to relate the variances $\sigma_{P}^{2}$ and $\sigma_{S}^{2}$.
The expectation values $\mu_{P}$ and $\sigma_{P}^{2}$ of the sample mean and variance $m_{P}$ and $s_{P}^{2}$ are given by,
\begin{eqnarray}
  \mu_{P} &=& \mu_{S} + \mu_{N}, \label{eqn:power_spectra_mean}\\
  \sigma_{P}^{2} &=& \sigma_{S}^{2} + \sigma_{N}^{2} + 2\mu_{S}\mu_{N},
\label{eqn:power_spectra_variance}
\end{eqnarray}
where we assumed that $S(\nu)$ and $N(\nu)$ are independent and used $\mu_{S}$, $\mu_{N}$, $\sigma_{S}^{2}$, and $\sigma^{2}_{N}$ to represent their means and variances.
Note that while the cross-term between signal and noise that originates from the squaring of the voltages does not contribute to the mean, it does contribute to the variance.

\begin{figure}
\centering
\includegraphics[width=0.985\textwidth,trim=0 0 0 0,clip]{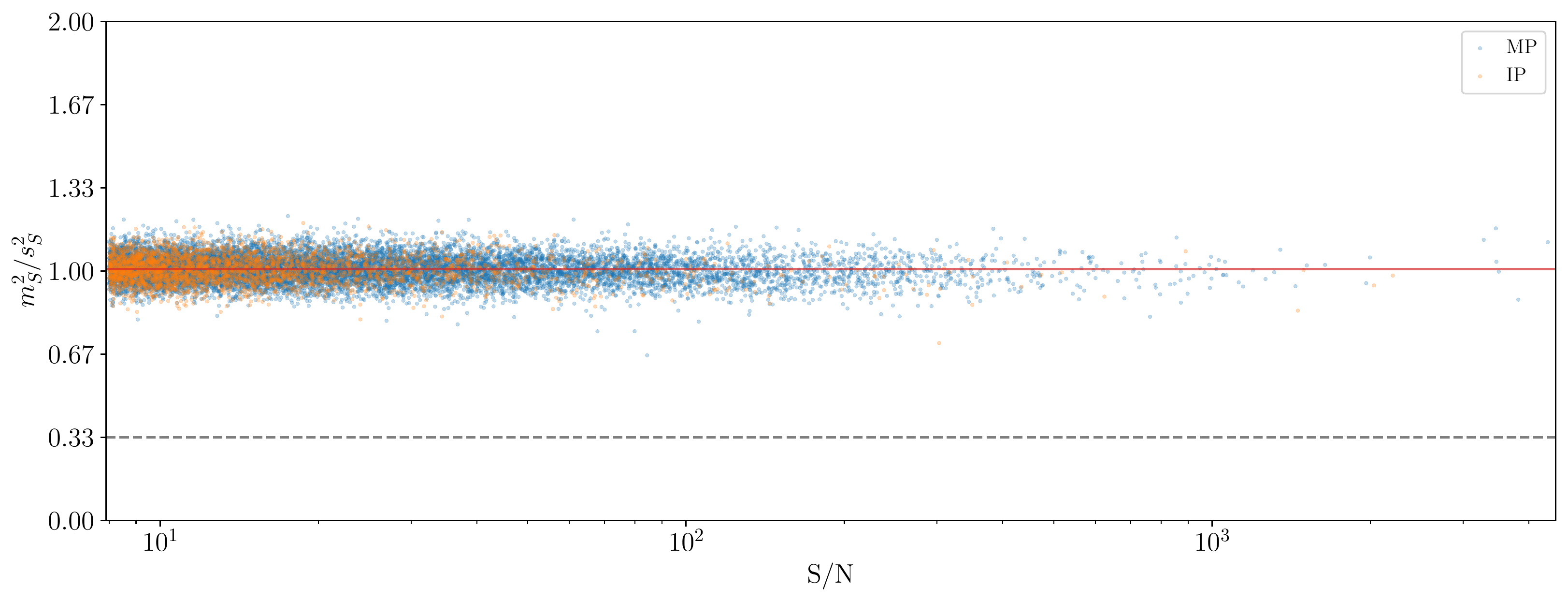}
\caption{
    Left polarization power spectra mean and standard deviation ratios of giant pulses detected in EK036~B. MPs are shown in blue and IPs are shown in orange. The ratio averages to 1, indicated by the red line, and is independent of S/N. The dashed gray line shows the expected $m_{S}^{2}/s_{S}^{2} = 1/3$ if nanoshots in $S(\nu)$ have the same IRF.
    \label{fig:mean_std}}
\end{figure}

With the above, one sees that one could make an estimate of $\sigma_{S}^{2}$ by $s_{P}^{2}-s_{N}^{2}-2(m_{P}-m_{N})m_{N}$, but as noted in \citetalias{Main2021}, this will lead to difficulties since $s_{P}$ is a noisy estimate of $\sigma_{P}$ (the estimate of $m_{P}$ is better, and those of $\mu_{N}$ and $\sigma_{N}$ much better, since these are based on more data).
A better estimate is possible using information on the distribution of $S(\nu)$ and $N(\nu)$.
In particular, if both are normally distributed, with zero mean but different variances, their powers will be distributed like scaled $\chi^{2}$ distributions with 2 degrees of freedom (real and imaginary parts).
Hence, one will have $\mu_{S}^{2}=\sigma_{S}^{2}$ and $\mu_{N}^{2}=\sigma_{N}^{2}$, which implies $\sigma_{P}^{2}=(\sigma_{S}+\sigma_{N})^{2}=(\mu_{S}+\mu_{N})^{2}=\mu_{P}^{2}$.
If $S(\nu)$ is not $\chi^{2}$ distributed, as in the case where each giant pulse is comprised of many nanoshots with the same IRF, then $\sigma_{S}^{2}=3\mu_{S}^{2}$ (see Section~\ref{sec:appendix_pol}).
We show that $S(\nu)$ is indeed $\chi^{2}$ distributed in Figure~\ref{fig:mean_std}.
This allows one to estimate $c_{12}$ more simply,
\begin{equation}
  \rho(S_{1}, S_{2}) = \rho(P_{1}, P_{2}) c_{12}
  = \rho(P_{1}, P_{2}) \left(\frac{\mu_{P,1}}{\mu_{P,1} - \mu_{N,1}}\right)\left(\frac{\mu_{P,2}}{\mu_{P,2} - \mu_{N,2}}\right).
\label{eqn:rho2}
\end{equation}

To determine the weights $w_{ij}$ for an optimal estimate of $\bar{r}_{\text{opt}}$, we need to know the uncertainties in the sample correlation coefficients.
The general case depends on both $\rho$ and the nature of the distributions, but for the case of no intrinsic correlation (i.e., $\rho=0$), the variance of $\rho$ is simply ${\text{Var}}(\rho)=1/(k-1)$, i.e., independent of the signal-to-noise ratio.
In our case, we expect this to hold roughly as well, since we have low $\rho$.
Hence, for our estimate of the intrinsic correlation, the expected variance is,
\begin{equation}
  {\text{Var}}(\rho(S_{1}, S_{2})) \simeq \frac{1}{k-1}c_{12}^{2}
  = \frac{1}{k - 1} \left(\frac{\mu_{P,1}}{\mu_{P,1} - \mu_{N,1}}\right)^{2}\left(\frac{\mu_{P,2}}{\mu_{P,2} - \mu_{N,2}}\right)^{2}.
\label{eqn:rho_variance}
\end{equation}
Given this variance, if we choose weights equal to its inverse,
\begin{equation}
    w_{12} = (k-1)\times\frac{(\mu_{P,1} - \mu_{N,1})^{2}(\mu_{P,2} - \mu_{N,2})^{2}}{\mu_{P,1}^{2}\mu_{P,2}^{2}},
\label{eqn:rho_weights}
\end{equation}
we will have an optimal estimate $\bar{r}_{\text{opt}}$ (even for higher $\rho$, this likely is reasonably optimal).
Note that this implies that for high signal-to-noise pulses, with $\mu_{P}\gg\mu_{N}$, the weight approaches a constant: all such pulses bring equal information, independent of exactly how high their signal-to-noise ratio is.
But for low signal-to-noise pulses, with $\mu_{P}\lesssim2\mu_{N}$ (i.e., signal-to-noise ratio per channel dropping below 1), the weight decreases as they bring little information.

\section{Correlation between Circular Polarizations and Stokes Parameters}\label{sec:appendix_pol}

If individual nanoshots are fully polarized, but with a random direction, then the ensemble should be less polarized. With a random walk like picture, one expects that while the intensity will be just be the sum of the intensities of the individual pulses, $I = \sum_{i} I_{i}$, the polarized intensity will be $I_{p} \simeq (\sum_{i} I_{i}^{2})^{1/2}$ and thus the degree of polarization $d = I_{p}/I \simeq (\sum_{i} (I_{i}/I)^{2})^{1/2}$.
For $n_{s}$ equal-intensity nanoshots, one would thus expect $d\simeq1/\sqrt{n_{s}}$.

We describe polarizations using the Poincar\'e sphere, with latitude $2\chi$ ($\pm\frac{\pi}{2}$ being fully left ($-$) and right ($+$) polarized) and longitude $2\psi$ ($2\psi=0$, $\pi/2$, $\pi$, $3\pi/2$ being horizontal, diagonal, vertical, and anti-diagonal, respectively), and write the Jones vector for a single pulse in left-right basis as,
\begin{equation}
    \left(\begin{matrix}
        A_{-} \exp(i\phi_{-})\\
        A_{+} \exp(i\phi_{+})
        \end{matrix}\right) =
        \left(\begin{matrix}
        A\cos(\chi + \frac{\pi}{4}) \exp(i(\phi - \psi))\\
        A\sin(\chi + \frac{\pi}{4}) \exp(i(\phi + \psi))
    \end{matrix}\right),
\label{eqn:Jones_matrix}
\end{equation}
where $A^{2}=A_{-}^{2}+A_{+}^{2}$ is the total amplitude and $\phi$ is a random angle of the wave.
For random polarizations, $2\psi$ will be uniform in $[0,2\pi)$, while $\sin2\chi$ will be uniform in $[-1, 1]$.
Since, $\sin^{2}(\chi+\frac{\pi}{4}) = \frac{1}{2}+\frac{1}{2}\sin2\chi$ and $\cos^{2}(\chi+\frac{\pi}{4})=\frac12-\frac{1}{2}\sin2\chi$, we see that for random polarization, $(A_{+}/A)^{2}$ is uniform in [0,1] and $(A_{-}/A)^{2}=1-(A_{+}/A)^{2}$.

For multiple nanoshots, the voltage spectra are given by,
\begin{equation}
  E_{\pm}(\nu) = \sum_{i} A_{\pm,i}\exp[i(\phi_{i}(\nu)\pm\psi_{i})],
  \label{eqn:electric_field_pol}
\end{equation}
where the sum is over the number of shots $n_{s}$.
The corresponding Stokes parameters (see Equations~\ref{eqn:stokesI}--\ref{eqn:stokesV}) following the PSR/IEEE convention \citep{Straten2010} are
\begin{eqnarray}
  I &=& \sum_{i} I_{i} = \sum_{i} I_{-,i} +  \sum_{i} I_{+,i}, \label{eqn:StokeI_pol} \\
  Q &=& \sum_{i} Q_{i} = \sum_{i} I_{i}\cos(2\chi_{i})\cos(2\psi_{i}), \label{eqn:StokeQ_pol} \\
  U &=& \sum_{i} U_{i} = \sum_{i} I_{i}\cos(2\chi_{i})\sin(2\psi_{i}), \label{eqn:StokeU_pol} \\
  V &=& \sum_{i} V_{i} = \sum_{i} I_{i}\sin(2\chi_{i}), \label{eqn:StokeV_pol}
\end{eqnarray}
where we defined $I_{\pm,i}\equiv A_{\pm,i}^{2}$, and assumed everywhere that differences in angles $\phi_{i}(\nu)$ causes the cross terms between nanoshots to cancel.
For the polarized flux, one finds,
\begin{equation}
  I_{p}^{2} = Q^{2} + U^{2} + V^{2}
  = \sum_{i} I_{i}^{2} + 2\sum_{i<j} I_{i}I_{j}\left(\cos(2\chi_{i})\cos(2\chi_{j})\cos(2\Delta\psi_{ij}) + \sin(2\chi_{i})\sin(2\chi_{j})\right)
  \simeq \sum_{i} I_{i}^{2},
\label{eqn:polarized_flux}
\end{equation}
where the approximate equality uses the assumption that the polarizations of the individual shots are random, so that the cross-terms cancel.
Hence, we reproduce the expectation above, that the expected degree of polarization is $d^{2} \simeq \sum_{i} (I_{i}^{2}/I^{2})\simeq 1/n_{s}$.

In order to see the effect of scattering of the pulses in a screen, we should include an IRF $g_i(\nu)$ in Equation~\ref{eqn:electric_field_pol} for the voltages.
We will assume that all $g_i(\nu)$ have the same total power, i.e., the same  $\mu_g=\langle|g(\nu)|^2\rangle$, and that the real and imaginary parts are normally distributed, so that $|g(\nu)|^2$ is distributed as a $\chi^2$ distribution with 2 degrees of freedom and hence $\sigma_g^{2}=\langle(|g(\nu)|^2-\mu_g)^2\rangle=\mu_g^{2}$.
For the Stokes parameters, which are frequency averages of linear additions of the nanoshots, this will introduce a factor $\mu_{g}$ in all equations, which we will absorb in the definition of the intensities, i.e., $I_{\pm,i}=\mu_{g}A_{\pm,i}^{2}$.
Below we discuss two cases, one where all nanoshots have the same IRF (Appendix~\ref{subsec:appendix_pol_same_irf}) and one were they all have different IRF (Appendix~\ref{subsec:appendix_pol_diff_irf}).

\subsection{Same Impulse Response Function}\label{subsec:appendix_pol_same_irf}

If all nanoshots are affected by the same IRF, then the left and right pulse power spectra are
\begin{equation}
  P_{\pm}(\nu) = |g(\nu)|^{2}\left(\sum_{i} A_{\pm,i}^{2} + 2\sum_{i<j} A_{\pm,i}A_{\pm,j}\cos(\Delta\phi_{\pm,ij}(\nu))\right), \label{eqn:power_spectra_same_irf}
\end{equation}
where we have abbreviated $\Delta\phi_{\pm,ij}(\nu) = (\phi_{i}(\nu) - \phi_{j}(\nu))\pm(\psi_{i} - \psi_{j})$.
The means and variances of these power spectra are
\begin{eqnarray}
  \mu_{\pm} &=& \mu_{g}\sum_{i} A_{\pm,i}^{2} = I_{\pm}, \label{eqn:power_spectra_mean_same_irf} \\
  \sigma_{\pm}^{2} &=& \sigma_{g}^{2}\left(\sum_{i} A_{\pm,i}^{2}\right)^{2} + 2(\sigma_{g}^{2} + \mu_{g}^{2})\sum_{i<j} A_{\pm,i}^{2}A_{\pm,j}^{2} = 3I_{\pm}^{2} - 2\sum_{i} I_{\pm,i}^{2} \simeq \left(3 - \frac{2}{n_{s}}\right)I_{\pm}^{2}, \label{eqn:power_spectra_variance_same_irf}
\end{eqnarray}
where in the approximate equality we used that $I_{\pm,i}\simeq I_{\pm}/n_{s}$.
For a large $n_{s}$, one thus has $\sigma_{\pm}^{2}=3\mu_{\pm}^{2}$.

The covariance between the two polarizations is given by
\begin{eqnarray}
  \text{Cov}(P_{-}, P_{+}) &=& \sigma_{g}^{2}\sum_{i} A_{-,i}^{2}\sum_{j} A_{+,j}^{2}
    + 2\mu_{g}^{2}\sum_{i<j} A_{-,i}A_{+,i}A_{-,j}A_{+,j}\cos(2\Delta\psi_{ij}) \nonumber\\
                           &=& I_{-}I_{+} + \frac{1}{2}\sum_{i<j} \sqrt{(I_{i}^{2} - V_{i}^{2})(I_{j}^{2} - V_{j}^{2})}\cos(2\Delta\psi_{ij}) \nonumber\\
                           &=& \frac{1}{4}\left(I^{2} - V^{2}\right) + \frac{1}{4}\left(Q^{2} + U^{2}\right) - \frac{1}{4}\sum_{i} \left(Q_{i}^{2} + U_{i}^{2}\right) \nonumber\\
                           &\simeq& \frac{1}{4}\left(I^{2}-V^{2}\right),
\label{eqn:covariance_same_irf}
\end{eqnarray}
where we simplified using $\langle\cos(\Delta\phi_{-})\cos(\Delta\phi_{+})\rangle = \frac{1}{2}\cos(2\Delta\psi)$, $I_{\pm} = \frac{1}{2}(I \pm V)$, and $\mu_{g}^{2}A_{-,i}^{2}A_{+,i}^{2} = I_{-,i}I_{+,i} = \frac{1}{4}(I_{i}^{2} - V_{i}^{2})=\frac{1}{4}(Q_{i}^2 + U_{i}^2)$.
For the approximate equality we used that for random polarizations, $Q^{2}+U^{2}\simeq\sum_{i}(Q_{i}^{2}+U_{i}^{2})$.
Combined with the product of the variances, $\sigma_{-}\sigma_{+}\simeq(3-2/n_{s})I_{-}I_{+}=\frac{1}{4}(3-2/n_{s})(I^{2}-V^{2})$, the expected correlation coefficient is,
\begin{equation}
  \rho(P_{-},P_{+}) \simeq \frac{1}{3 - 2/n_{s}}.
  \label{eqn:rho_same_irf}
\end{equation}
For large $n_{s}$, $\rho\simeq1/3$, as noted by \cite{Cordes2004}.

\begin{figure}
  \centering
  \includegraphics[width=0.985\textwidth,trim=0 0 0 0,clip]{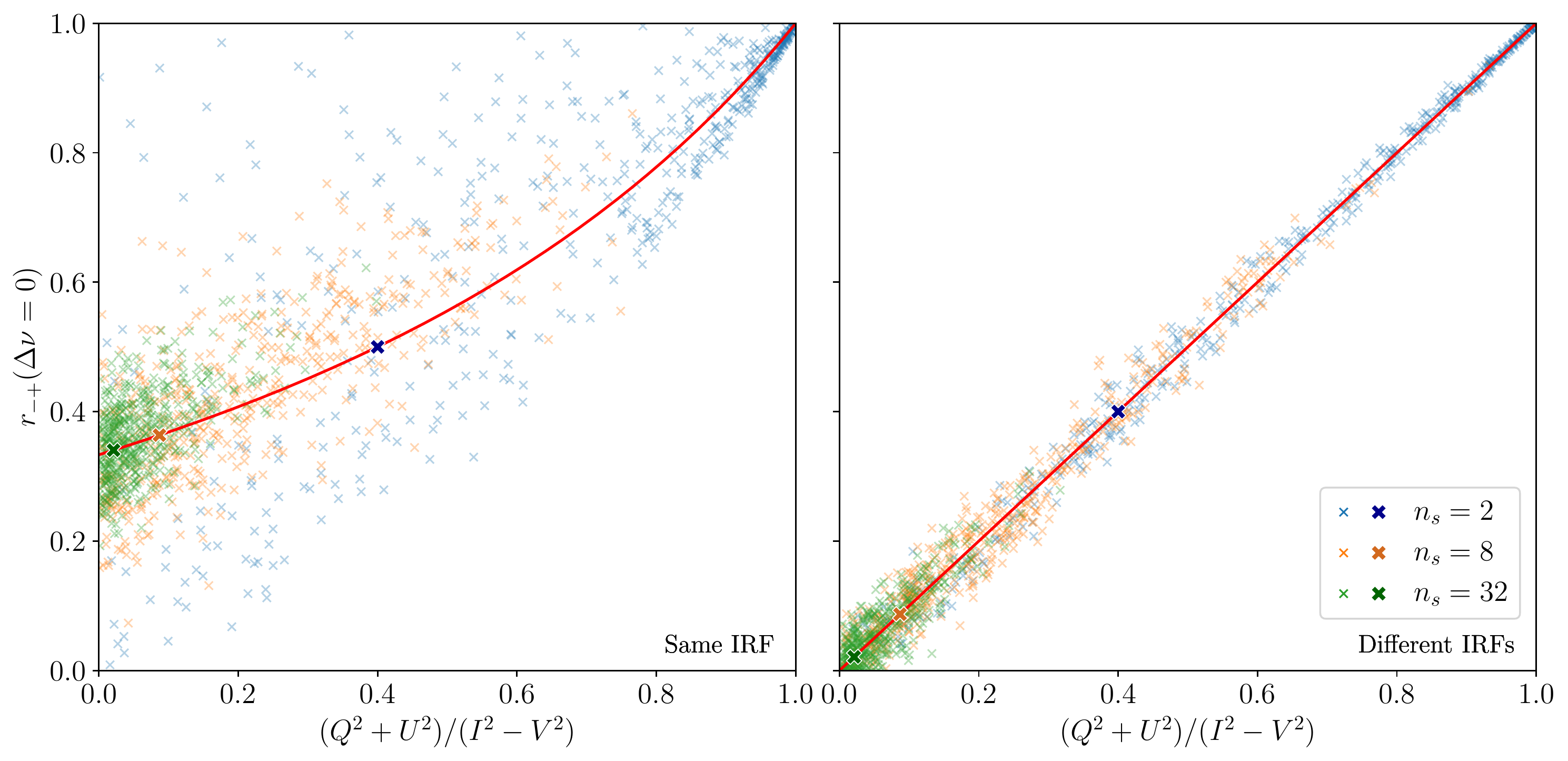}
  \caption{
    Correlation between left and right polarization powers for simulated pulses as a function of polarization properties.
    In the simulations, a giant pulse consists of $n_{s}$ nanoshots with different polarization, imprinted with response functions which either are all the same ({\em left\/}) or are all different ({\em right\/}).
    The polarization of each nanoshot is drawn randomly from the full Poincar\'e sphere, and its amplitude drawn from a power law with index $\alpha=-2.8$, mimicking what we see in our data and others \citep{Bera2019}; the frequency bandwidth, number of channels and scattering timescale are chosen to be as in our EK036~B dataset.
    We show 500 randomly selected giant pulses for $n_{s}=2$, $8$, and $32$ (as indicated).
    The red lines mark expectations for large samples of pulses, with the bold markers indicating where $n_{s}=2$, $8$ and $32$.
    One sees that the simulations confirm the expectations.
    The scatter is much larger for the case that the response functions are the same, because the expectation is only for the average over many pulses, in which polarization properties average out.
    In contrast, the expectation for different response functions is for individual giant pulses, only relying on averaging out random wave angles.
    \label{fig:sim_IRF}}
\end{figure}

Of course, we do not know $n_{s}$ a priori, but we can use the degree of polarization as an estimate.
In Figure~\ref{fig:sim_IRF}, we compare simulations with the prediction for a combination akin to the degree of linear polarization, $\tilde{d}=(Q^2+U^2)/(I^2-V^2)$, which we find below is expected to be a good approximation for the correlation coefficient for the case that nanoshots are imprinted with different IRF.
For our case, given that $Q^{2} \simeq U^{2} \simeq V^{2} \simeq  I^{2}/3n_{s}$,
one expects $\tilde{d}\simeq2/(3n_{s}-1)$ and thus $\rho\simeq(2+\tilde{d})/3(2-\tilde{d})$.
One sees that the simulation confirms the expectation, albeit with fairly large scatter for individual simulated pulses.
This is because the correlation coefficient for a given pulse depends on the polarizations of its constituent nanoshots; the approximations only hold for the average over a large number of pulses.

\subsection{Different Impulse Response Functions}\label{subsec:appendix_pol_diff_irf}

Turning now to the case that each nanoshot is imprinted with a different IRF, $g_{i}(\nu)$, the left and right pulse power spectra are
\begin{equation}
  P_{\pm}(\nu) = \sum_{i} A_{\pm,i}^{2}|g_{i}(\nu)|^{2} + 2\sum_{i<j} A_{\pm,i}A_{\pm,j}|g_{i}(\nu)||g_{j}(\nu)|\cos(\Delta\phi_{\pm,ij}(\nu) + \Delta\phi_{g,ij}(\nu)), \label{eqn:power_spectra_diff_irf}
\end{equation}
with means and variances
\begin{eqnarray}
  \mu_{\pm} &=& \mu_{g}\sum_{i} A_{\pm,i}^{2} = I_{\pm}, \label{eqn:power_spectra_mean_diff_irf} \\
  \sigma_{\pm}^{2} &=& \sigma_{g}^{2}\sum_{i} A_{\pm,i}^{4} + 2\mu_{g}^{2}\sum_{i<j} A_{\pm,i}^{2}A_{\pm,j}^{2} = \sum_{i} I_{\pm,i}^{2} + 2\sum_{i<j} I_{\pm,i}I_{\pm,j} = I_{\pm}^{2}. \label{eqn:power_spectra_variance_diff_irf}
\end{eqnarray}
Here, the variances are smaller because the cross-terms between the different nanoshots now cancel.
Similar cancellation happens for the covariance between left and right polarizations, leading to,
\begin{eqnarray}
  \text{Cov}(P_{-}, P_{+}) &=& \sigma_{g}^{2}\sum_{i} A_{-,i}^{2}A_{+,i}^{2} + 2\mu_{g}^{2}\sum_{i<j} A_{-,i}A_{+,i}A_{-,j}A_{+,j}\cos(2\Delta\psi_{ij}) \nonumber\\
                           &=& \frac{1}{4}\sum_{i} (I_{i}^{2} - V_{i}^{2}) + \frac{1}{2}\sum_{i<j} \sqrt{(I_{i}^{2} - V_{i}^{2})(I_{j}^{2} - V_{j}^{2})}\cos(2\Delta\psi_{ij}) \nonumber\\
                           &=& \frac{1}{4}(Q^{2} + U^{2}).
\label{eqn:covariance_diff_irf}
\end{eqnarray}
where used the same simplifications as in Equation~\ref{eqn:covariance_same_irf} as well as that
$\langle\cos(\Delta\phi_{-} + \Delta\phi_{g})\cos(\Delta\phi_{+} + \Delta\phi_{g})\rangle = \frac{1}{2}\cos(2\Delta\psi)$.
The product of the left and right standard deviations $\sigma_{-}\sigma_{+} = \frac14(I^{2}-V^{2})$,
and thus the expected correlation coefficient is,
\begin{equation}
  \rho(P_{-},P_{+}) = \frac{Q^{2} + U^{2}}{I^{2} - V^{2}}\simeq \frac{2}{3n_{s}-1},
\label{eqn:rho_diff_irf}
\end{equation}
where for the approximate equality we used that $Q^{2} \simeq U^{2} \simeq V^{2} \simeq  I^{2}/3n_{s}$.
Note that unlike for the same-IRF case, for a given pulse with measured Stokes parameters, one does not have to rely on the last approximation -- the result is known independent of any approximation for the individual properties of the nanoshots.
Hence, comparing with simulations in Figure~\ref{fig:sim_IRF}, one sees that even for individual pulses, the predictions are good; there is only some scatter due to the random wave angles, which causes imperfect cancellation of cross-terms.

\section{Effects of Resolving the Emission Region}\label{sec:appendix_irf}

The correlation between giant pulses is low, much lower even than the $\rho=1/3$ expected for giant pulses consisting of large numbers of nanoshots.
Furthermore, the ratio of mean to variance of pulse spectra as well as the correlation between polarizations suggests that the different nanoshots generally are imprinted with different IRFs.
Here, we derive expected correlation coefficients assuming each giant pulse consists of $n_{s}$ nanoshots, which arise randomly in some larger emission region, which is resolved by the screen.
We will assume that the nanoshots are true impulses, without spectral structure, which randomly sample the whole emission region without any spatial correlation within a pulse.

For a first, simple derivation, we assume the emission region is resolved by the screen into $n_{g}$ different patches, each of which have completely independent IRFs.
We write the electric field as in Equation~\ref{eqn:electric_field_pol}, as a sum of impulses with amplitudes $A_{i}$ seen through possibly different IRFs $g_{i}(\nu)$.
Inspecting the results above for the power spectra for the case of nanoshots that all share the same IRF or all have different ones (Equations~\ref{eqn:power_spectra_same_irf}--\ref{eqn:power_spectra_variance_same_irf} and \ref{eqn:power_spectra_diff_irf}--\ref{eqn:power_spectra_variance_diff_irf}, respectively), one sees that the mean, $\mu_{P}$, is always the same, but the variance, $\sigma_{P}^{2}$, will depend on the number of pairs $n_{\text{corr}}$ that share their IRF (and thus correlate fully with each other), as,
\begin{equation}
    \sigma_{P}^{2} = \mu_{P}^{2}\left(1 + \frac{4n_{\text{corr}}}{n_{s}^{2}}\right) = \mu_{P}^{2}\left(1 + \frac{2\left(1 - 1/n_{s}\right)}{n_{g}}\right).
\label{eqn:n_irf_var}
\end{equation}
where we assumed that all nanoshots have the same amplitude $A_{i}=\sqrt{\mu_{P}/n_{s}}$ (for different amplitudes, one could rewrite using some effective number of pulses and pairs), and where in the second equality we used that for any given pair of nanoshots, the probability that they share the same IRF is simply the inverse of the number of patches, so that $n_{\text{corr}} = n_{s}(n_{s}-1)/2n_{g}$.

Similarly, for the covariance between two giant pulse spectra, one finds
\begin{equation}
  \text{Cov}(P_{1}, P_{2}) = \mu_{1}\mu_{2}\frac{n_{\text{corr}}}{n_{s,1}n_{s,2}} = \mu_{1}\mu_{2}\frac{1}{n_{g}},
\label{eqn:n_irf_cov}
\end{equation}
where in the second equality we again used that the probability for a given pair to share the same IRF is $1/n_{g}$ and thus that $n_{\text{corr}} = n_{s,1}n_{s,2}/n_{g}$.

Combining the above, one finds that the correlation coefficient for two separate pulses is expected to be,
\begin{equation}
  \rho(P_{1}, P_{2}) = \frac{1}{n_{g} + 2\left(1 - 1/n_{s}\right)}.
\label{eqn:n_irf_corrcoeff}
\end{equation}
Thus, for two single-shot pulses, one has the expected $\rho(P_{1}, P_{2})=1/n_{g}$ and for pulses with large $n_{s}$, $\rho(P_{1}, P_{2})\simeq1/(n_{g}+2)$.

The above derivation is a simplification in that it assumes that the screen resolves the region in a fixed number of patches, while in reality the resolution of the screen varies throughout the scattering tail, from very poor early on, when the paths along which radiation travels are all close to the line of sight, to much better later on.
Indeed, in Section~\ref{subsec:correlation_pulsepairs}, we found that spectra taken in the first half of pulses correlated more strongly with each other than those taken in the second half.

In order to take this into account, we first start by rewriting the above in terms of a fraction of the IRF that is correlated.
Next, to understand not just the amplitude of the correlations at zero lag, but also the correlation at other $\Delta\nu$ and thus the decorrelation width $\nu_{\text{decorr}}$, we consider the behaviour of the IRF in the time domain, $g(t)$.

To take into account that part of the IRF of nanoshot $i$ is shared with that of other nanoshots, we write the IRF in terms of a shared (correlated) part $g_{\text{c}}(\nu)$ that contributes a fraction $f_{\text{c}}$ of its power, and a different (uncorrelated) part $g_{\text{u}, i}(\nu)$ that contributes the rest, i.e.,
\begin{equation}
  g_{i}(\nu) = g_{\text{c}}(\nu)\sqrt{f_{\text{c}}(\nu)} + g_{\text{u}, i}(\nu)\sqrt{1-f_{\text{c}}(\nu)}.
\end{equation}
Note that we implicitly assume all IRFs are normalized the same way, i.e., all have the same $\int_{\nu}|g(\nu)|^{2}d\nu$.
With this,
\begin{eqnarray}
  \sigma_{P}^{2}
  &=& \sigma_{g}^{2}\left[\bar{f}_{\text{c}}^{2}\left(\sum_{i}A_{i}^{2}\right)^2
      + (1-\bar{f}_{\text{c}}^2)\sum_{i}A_{i}^{4}\right]
      + 2\left((\sigma_{g}^{2}+\mu_{g}^{2})\bar{f}_{\text{c}}^{2}
      + \mu_{g}^{2}(1-\bar{f}_{\text{c}}^{2})\right)\sum_{i<j} A_{i}^{2}A_{j}^{2}\nonumber\\
  &=& \mu_{P}^{2}\left[1
      + 2\bar{f}_{\text{c}}^2\left(1-\frac{\sum_{i} \mu_{i}^{2}}{\mu_{P}^{2}}\right)\right],
      \label{eqn:irf_frac_var}
\end{eqnarray}
where we use $\bar{f}_{\text{c}}$ to indicate an appropriate average over frequency of $f_{\text{c}}(\nu)$ (note that for relatively narrow-band observations like ours, the dependence on frequency can be safely ignored).
Thus, identifying $n_{g}=1/\bar{f}_{\text{c}}^2$ and $n_{s}=\mu_{P}^2/\sum_i \mu_i^2$ one recovers Equation~\ref{eqn:n_irf_var}.
Similarly, for the correlation coefficient (assuming both pulses have the same effective $n_s$),
\begin{equation}
  \rho\left(P_1, P_2\right)
  = \frac{\sum_{i,j}A_{1,i}^{2}A_{2,j}^{2}\text{Cov}\left(|g_{1,i}(\nu)|^2, |g_{2,j}(\nu)|^2\right)}
  {\sigma_1\sigma_2}
  = \frac{\bar{f}_{\text{c}}^{2}}{1 + 2\bar{f}_{\text{c}}^2\left(1-1/n_{s}\right)}.
  \label{eqn:irf_frac_rho}
\end{equation}

Turning now to the time domain, without much loss of generality we can take $g(t)$ to be amplitude-modulated noise, i.e.,
\begin{equation}
  g(t) = a(t) {\cal CN}(0, 1),
\end{equation}
where ${\cal CN}(0, 1)$ is the complex normal distribution with mean zero and variance one, and $a(t)$ is normalized to give unit total power, i.e., $\int p(t){\text{d}}t=1$ with $p(t)=a^{2}(t)$.
For two impulses arising from different locations, the fraction $f_{\text{c}}(t)$ of the response that is shared (and thus correlates) will decrease with increasing time, i.e., the part phased coherently will have reduced power $p_{\text{c}}(t)=p(t)f_{\text{c}}(t)$.

We can relate this to the frequency domain by noting that generalizing the cross-correlation coefficient in Equation~\ref{eqn:irf_frac_rho} to a function of $\Delta\nu$, the covariance term becomes a cross-covariance, which requires the auto-correlation of $|g_{\text{c}}(\nu)|^2$.
Using the cross-correlation theorem, we can write the latter as,
\begin{equation}
  |g_{\text{c}}(\nu)|^2 \star |g_{\text{c}}(\nu)|^2
  = {\cal F}\left\{\left|{\cal F}^{-1}\left\{|g_{\text{c}}(\nu)|^2\right\}\right|^2\right\}
  = {\cal F}\left\{\left|g_{\text{c}}(t) \star g_{\text{c}}(t)\right|^2\right\},
\end{equation}
where ${\cal F}$ indicates a Fourier transform and ${\cal F}^{-1}$ its inverse.
Using that the expectation for the auto-correlation power between the part of impulses that is shared in the time domain is given by the autocorrelation of their power envelopes, the expectation value for the auto-correlation is,
\begin{equation}
  \left\langle|g_{\text{c}}(\nu)|^2 \star |g_{\text{c}}(\nu)|^2\right\rangle
  = {\cal F}\left\{p_{\text{c}}(t) \star p_{\text{c}}(t)\right\}
  = \left|p_{\text{c}}(\Delta\nu)\right|^2.
\end{equation}

The precise shape of the autocorrelation of $|g_{\text{c}}(\nu)|^2$ will depend on $p(t)$ and $f_{\text{c}}(t)$, but one can gain insight using a simple assumption, that both $p(t)$ and $f_{\text{c}}(t)$ are exponentials.
For $p(t)$, this is reasonable, as the pulse profiles can be modelled fairly well as the convolution of a Gaussian arising from nanoshots with an exponential scattering tail (see Section~\ref{sec:scattering_timescale}).
It is also the expected profile for an isotropic scattering screen in which the scattering points are distributed normally.
Since the profile has to be normalized, one has $p(t) = \exp(-t/\tau)/\tau$ (for $t\ge0$) where $\tau$ is the scattering timescale.

For the shared fraction $f_{\text{c}}(t)$ averaged over many pulses, an exponential distribution may also be plausible: if the locations of the individual pulses are distributed roughly isotropically with a normal distribution in each direction, then squared distances between pulse pairs will follow an exponential distribution.
Hence, while for close pairs $f_{\text{c}}(t)$ will extend further in time than for distant pairs, on average one may expect $f_{\text{c}}(t)\simeq\exp(-t/\tau_{\text{c}})$, with $\tau_{\text{c}}$ corresponding to the time delay for which the IRF of two typical pulses will start to differ significantly.
Thus, the correlated power envelope will be,
\begin{equation}
  p_{\text{c}}(t) = \frac{1}{\tau}\exp(-t/\tau)\exp(-t/\tau_{\text{c}})
  = \frac{1}{\tau}\exp(-t/\tau^\prime),
\end{equation}
with $\tau^\prime = \tau\tau_{\text{c}}/(\tau+\tau_{\text{c}})$.
The autocorrelation in time is $(p_{\text{c}}\star p_{\text{c}})(\Delta t) = (\tau^\prime/2\tau^2)\exp(-|\Delta t|/\tau^\prime)$, and hence in the frequency domain, one expects a Lorentzian,
\begin{equation}
  |p_{\text{c}}(\Delta\nu)|^2 = \left(\frac{\tau^\prime}{\tau}\right)^2
  \frac{1}{1+(2\pi\Delta\nu\tau^\prime)^2}.
  \label{eqn:irf_frac_lorentzian}
\end{equation}
Comparing with Equation~\ref{eqn:irf_frac_rho}, we can identify $\bar{f}_{\text{c}}=\tau^\prime/\tau$.
The half-width at half maximum is given by $\nu_{\text{decorr}}=1/2\pi\tau^\prime$, which is larger than the $1/2\pi\tau$ one would infer from the measured scattering time $\tau$ by a factor $1/\bar{f}_{\text{c}}$.
For well-resolved emission, i.e., small $\bar{f}_{\text{c}}$, the dependence on number of pulses drops out, and one expects a scaling with amplitude $A$ of the decorrelation width as $\nu_{\text{decorr}}\simeq(1/\sqrt{A})/2\pi\tau$.

As noted above, an exponential for $f_{\text{c}}(t)$ is expected if the nanoshots are distributed roughly isotropically on the sky, with a normal distribution.
Assuming they have a variance $\sigma_s^2$ in each direction, differences in position between pairs will have a variance $2\sigma_s^2$ in each direction.
Hence, the distance squared $r^2$ between pairs will follow an exponential distribution $\exp(-r^2/4\sigma_s^2)$ and the typical distance is $2\sigma_s$.
Thus, like \cite{Gwinn1998}, we find that $2\sigma_s/\sigma_x$ is the relevant measure of the degree to which the emission region is resolved.
We confirmed using simulations that the average amplitude for a pair of pulses equals $(\tau^\prime/\tau)^2 = (1+(2\sigma_s/\sigma_x)^2)^{-1}$.

The amplitude is inversely related to the effective number of resolution elements (see Equation~\ref{eqn:n_irf_corrcoeff}), which suggests defining an effective spatial correlation scale $\sigma_{\text{c}} = (\sigma_x^2+(2\sigma_s)^2)^{1/2}$, such that one has $n_g=(\sigma_{\text{c}}/\sigma_x)^2$ (and thus $\tau^\prime/\tau=\bar{f}_{\text{c}}=\sigma_x/\sigma_{\text{c}}=n_g^{-1/2}$ and $\nu_{\rm decorr}=\sqrt{n_g}/2\pi\tau$).
The correlation scale $\sigma_{\text{c}}$ can be related to the spatial separation $\ell_{\text{scint}}$ between giant pulses at which the correlation will decrease by $1/e$, which is useful for comparing with the scintillation time $t_{\text{scint}}=\ell_{\text{scint}}/v_{\text{rel}}$.
Since amplitudes vary with separation $\ell$ as $\exp(-\ell^2/2\sigma_{\text{c}}^2)$, the correlation, which is a fourth-order product, will fall off as $\exp(-2\ell^2/\sigma_{\text{c}}^2)$.
Hence, one infers $\ell_{\text{scint}}=\sigma_{\text{c}}/\sqrt2$ (which we also confirmed with simulations).

\bibliographystyle{aasjournal}
\bibliography{crab_bib}

\begin{thebibliography}{}
\expandafter\ifx\csname natexlab\endcsname\relax\def\natexlab#1{#1}\fi
\providecommand{\url}[1]{\href{#1}{#1}}
\providecommand{\dodoi}[1]{doi:~\href{http://doi.org/#1}{\nolinkurl{#1}}}
\providecommand{\doeprint}[1]{\href{http://ascl.net/#1}{\nolinkurl{http://ascl.net/#1}}}
\providecommand{\doarXiv}[1]{\href{https://arxiv.org/abs/#1}{\nolinkurl{https://arxiv.org/abs/#1}}}

\bibitem[{Akima(1970)}]{Akima1970}
Akima, H. 1970, J. ACM, 17, 589, \dodoi{10.1145/321607.321609}

\bibitem[{{Astropy Collaboration} {et~al.}(2013){Astropy Collaboration},
  {Robitaille}, {Tollerud}, {Greenfield}, {Droettboom}, {Bray}, {Aldcroft},
  {Davis}, {Ginsburg}, {Price-Whelan}, {Kerzendorf}, {Conley}, {Crighton},
  {Barbary}, {Muna}, {Ferguson}, {Grollier}, {Parikh}, {Nair}, {Unther},
  {Deil}, {Woillez}, {Conseil}, {Kramer}, {Turner}, {Singer}, {Fox}, {Weaver},
  {Zabalza}, {Edwards}, {Azalee Bostroem}, {Burke}, {Casey}, {Crawford},
  {Dencheva}, {Ely}, {Jenness}, {Labrie}, {Lim}, {Pierfederici}, {Pontzen},
  {Ptak}, {Refsdal}, {Servillat}, \& {Streicher}}]{AstropyCollaboration2013}
{Astropy Collaboration}, {Robitaille}, T.~P., {Tollerud}, E.~J., {et~al.} 2013,
  \aap, 558, A33, \dodoi{10.1051/0004-6361/201322068}

\bibitem[{{Astropy Collaboration} {et~al.}(2018){Astropy Collaboration},
  {Price-Whelan}, {Sip{\H{o}}cz}, {G{\"u}nther}, {Lim}, {Crawford}, {Conseil},
  {Shupe}, {Craig}, {Dencheva}, {Ginsburg}, {VanderPlas}, {Bradley},
  {P{\'e}rez-Su{\'a}rez}, {de Val-Borro}, {Aldcroft}, {Cruz}, {Robitaille},
  {Tollerud}, {Ardelean}, {Babej}, {Bach}, {Bachetti}, {Bakanov}, {Bamford},
  {Barentsen}, {Barmby}, {Baumbach}, {Berry}, {Biscani}, {Boquien}, {Bostroem},
  {Bouma}, {Brammer}, {Bray}, {Breytenbach}, {Buddelmeijer}, {Burke},
  {Calderone}, {Cano Rodr{\'\i}guez}, {Cara}, {Cardoso}, {Cheedella}, {Copin},
  {Corrales}, {Crichton}, {D'Avella}, {Deil}, {Depagne}, {Dietrich}, {Donath},
  {Droettboom}, {Earl}, {Erben}, {Fabbro}, {Ferreira}, {Finethy}, {Fox},
  {Garrison}, {Gibbons}, {Goldstein}, {Gommers}, {Greco}, {Greenfield},
  {Groener}, {Grollier}, {Hagen}, {Hirst}, {Homeier}, {Horton}, {Hosseinzadeh},
  {Hu}, {Hunkeler}, {Ivezi{\'c}}, {Jain}, {Jenness}, {Kanarek}, {Kendrew},
  {Kern}, {Kerzendorf}, {Khvalko}, {King}, {Kirkby}, {Kulkarni}, {Kumar},
  {Lee}, {Lenz}, {Littlefair}, {Ma}, {Macleod}, {Mastropietro}, {McCully},
  {Montagnac}, {Morris}, {Mueller}, {Mumford}, {Muna}, {Murphy}, {Nelson},
  {Nguyen}, {Ninan}, {N{\"o}the}, {Ogaz}, {Oh}, {Parejko}, {Parley}, {Pascual},
  {Patil}, {Patil}, {Plunkett}, {Prochaska}, {Rastogi}, {Reddy Janga},
  {Sabater}, {Sakurikar}, {Seifert}, {Sherbert}, {Sherwood-Taylor}, {Shih},
  {Sick}, {Silbiger}, {Singanamalla}, {Singer}, {Sladen}, {Sooley},
  {Sornarajah}, {Streicher}, {Teuben}, {Thomas}, {Tremblay}, {Turner},
  {Terr{\'o}n}, {van Kerkwijk}, {de la Vega}, {Watkins}, {Weaver}, {Whitmore},
  {Woillez}, {Zabalza}, \& {Astropy Contributors}}]{AstropyCollaboration2018}
{Astropy Collaboration}, {Price-Whelan}, A.~M., {Sip{\H{o}}cz}, B.~M., {et~al.}
  2018, \aj, 156, 123, \dodoi{10.3847/1538-3881/aabc4f}

\bibitem[{{Astropy Collaboration} {et~al.}(2022){Astropy Collaboration},
  {Price-Whelan}, {Lim}, {Earl}, {Starkman}, {Bradley}, {Shupe}, {Patil},
  {Corrales}, {Brasseur}, {N{\"o}the}, {Donath}, {Tollerud}, {Morris},
  {Ginsburg}, {Vaher}, {Weaver}, {Tocknell}, {Jamieson}, {van Kerkwijk},
  {Robitaille}, {Merry}, {Bachetti}, {G{\"u}nther}, {Aldcroft},
  {Alvarado-Montes}, {Archibald}, {B{\'o}di}, {Bapat}, {Barentsen},
  {Baz{\'a}n}, {Biswas}, {Boquien}, {Burke}, {Cara}, {Cara}, {Conroy},
  {Conseil}, {Craig}, {Cross}, {Cruz}, {D'Eugenio}, {Dencheva}, {Devillepoix},
  {Dietrich}, {Eigenbrot}, {Erben}, {Ferreira}, {Foreman-Mackey}, {Fox},
  {Freij}, {Garg}, {Geda}, {Glattly}, {Gondhalekar}, {Gordon}, {Grant},
  {Greenfield}, {Groener}, {Guest}, {Gurovich}, {Handberg}, {Hart},
  {Hatfield-Dodds}, {Homeier}, {Hosseinzadeh}, {Jenness}, {Jones}, {Joseph},
  {Kalmbach}, {Karamehmetoglu}, {Ka{\l}uszy{\'n}ski}, {Kelley}, {Kern},
  {Kerzendorf}, {Koch}, {Kulumani}, {Lee}, {Ly}, {Ma}, {MacBride}, {Maljaars},
  {Muna}, {Murphy}, {Norman}, {O'Steen}, {Oman}, {Pacifici}, {Pascual},
  {Pascual-Granado}, {Patil}, {Perren}, {Pickering}, {Rastogi}, {Roulston},
  {Ryan}, {Rykoff}, {Sabater}, {Sakurikar}, {Salgado}, {Sanghi}, {Saunders},
  {Savchenko}, {Schwardt}, {Seifert-Eckert}, {Shih}, {Jain}, {Shukla}, {Sick},
  {Simpson}, {Singanamalla}, {Singer}, {Singhal}, {Sinha}, {Sip{\H{o}}cz},
  {Spitler}, {Stansby}, {Streicher}, {{\v{S}}umak}, {Swinbank}, {Taranu},
  {Tewary}, {Tremblay}, {Val-Borro}, {Van Kooten}, {Vasovi{\'c}}, {Verma}, {de
  Miranda Cardoso}, {Williams}, {Wilson}, {Winkel}, {Wood-Vasey}, {Xue},
  {Yoachim}, {Zhang}, {Zonca}, \& {Astropy Project
  Contributors}}]{AstropyCollaboration2022}
{Astropy Collaboration}, {Price-Whelan}, A.~M., {Lim}, P.~L., {et~al.} 2022,
  \apj, 935, 167, \dodoi{10.3847/1538-4357/ac7c74}

\bibitem[{Backer {et~al.}(2000)Backer, Wong, \& Valanju}]{Backer2000}
Backer, D.~C., Wong, T., \& Valanju, J. 2000, \apj, 543, 740,
  \dodoi{10.1086/317150}

\bibitem[{Bassa {et~al.}(2016)Bassa, Janssen, Karuppusamy, Kramer, Lee, Liu,
  McKee, Perrodin, Purver, Sanidas, Smits, \& Stappers}]{Bassa2016}
Bassa, C.~G., Janssen, G.~H., Karuppusamy, R., {et~al.} 2016, \mnras, 456,
  2196, \dodoi{10.1093/mnras/stv2755}

\bibitem[{Bera \& Chengalur(2019)}]{Bera2019}
Bera, A., \& Chengalur, J.~N. 2019, \mnras, 490, L12,
  \dodoi{10.1093/mnrasl/slz140}

\bibitem[{Bietenholz {et~al.}(1997)Bietenholz, Kassim, Frail, Perley, Erickson,
  \& Hajian}]{Bietenholz1997}
Bietenholz, M.~F., Kassim, N., Frail, D.~A., {et~al.} 1997, \apj, 490, 291,
  \dodoi{10.1086/304853}

\bibitem[{{Bij} {et~al.}(2021){Bij}, {Lin}, {Li}, {van Kerkwijk}, {Pen}, {Lu},
  {Main}, {Peterson}, {Quine}, \& {Vanderlinde}}]{Bij2021}
{Bij}, A., {Lin}, H.-H., {Li}, D., {et~al.} 2021, \apj, 920, 38,
  \dodoi{10.3847/1538-4357/ac1589}

\bibitem[{{Bochenek} {et~al.}(2020){Bochenek}, {Ravi}, {Belov}, {Hallinan},
  {Kocz}, {Kulkarni}, \& {McKenna}}]{bochenek_2020_natur}
{Bochenek}, C.~D., {Ravi}, V., {Belov}, K.~V., {et~al.} 2020, \nat, 587, 59,
  \dodoi{10.1038/s41586-020-2872-x}

\bibitem[{Chevalier(1977)}]{Chevalier1977}
Chevalier, R.~A. 1977, \araa, 15, 175,
  \dodoi{10.1146/annurev.aa.15.090177.001135}

\bibitem[{{Chevalier} \& {Gull}(1975)}]{ChevalierGull1975}
{Chevalier}, R.~A., \& {Gull}, T.~R. 1975, \apj, 200, 399,
  \dodoi{10.1086/153802}

\bibitem[{{CHIME/FRB Collaboration} {et~al.}(2020){CHIME/FRB Collaboration},
  {Andersen}, {Bandura}, {Bhardwaj}, {Bij}, {Boyce}, {Boyle}, {Brar},
  {Cassanelli}, {Chawla}, {Chen}, {Cliche}, {Cook}, {Cubranic}, {Curtin},
  {Denman}, {Dobbs}, {Dong}, {Fandino}, {Fonseca}, {Gaensler}, {Giri}, {Good},
  {Halpern}, {Hill}, {Hinshaw}, {H{\"o}fer}, {Josephy}, {Kania}, {Kaspi},
  {Landecker}, {Leung}, {Li}, {Lin}, {Masui}, {McKinven}, {Mena-Parra},
  {Merryfield}, {Meyers}, {Michilli}, {Milutinovic}, {Mirhosseini},
  {M{\"u}nchmeyer}, {Naidu}, {Newburgh}, {Ng}, {Patel}, {Pen},
  {Pinsonneault-Marotte}, {Pleunis}, {Quine}, {Rafiei-Ravandi}, {Rahman},
  {Ransom}, {Renard}, {Sanghavi}, {Scholz}, {Shaw}, {Shin}, {Siegel}, {Singh},
  {Smegal}, {Smith}, {Stairs}, {Tan}, {Tendulkar}, {Tretyakov}, {Vanderlinde},
  {Wang}, {Wulf}, \& {Zwaniga}}]{chime_2020_natur_galacticfrb}
{CHIME/FRB Collaboration}, {Andersen}, B.~C., {Bandura}, K.~M., {et~al.} 2020,
  \nat, 587, 54, \dodoi{10.1038/s41586-020-2863-y}

\bibitem[{Cordes {et~al.}(2004)Cordes, Bhat, Hankins, McLaughlin, \&
  Kern}]{Cordes2004}
Cordes, J., Bhat, N., Hankins, T., McLaughlin, M., \& Kern, J. 2004, Astrophys.
  J., 612, 375, \dodoi{10.1086/422495}

\bibitem[{Cordes \& Rickett(1998)}]{cordes1998}
Cordes, J.~M., \& Rickett, B.~J. 1998, \apj, 507, 846, \dodoi{10.1086/306358}

\bibitem[{Cordes {et~al.}(1983)Cordes, Weisberg, \& Boriakoff}]{Cordes1983}
Cordes, J.~M., Weisberg, J.~M., \& Boriakoff, V. 1983, \apj, 268, 370,
  \dodoi{10.1086/160961}

\bibitem[{Crossley {et~al.}(2007)Crossley, Eilek, \& Hankins}]{Crossley2007}
Crossley, J.~H., Eilek, J.~A., \& Hankins, T.~H. 2007, in Astronomical Society
  of the Pacific Conference Series, Vol. 365, SINS - Small Ionized and Neutral
  Structures in the Diffuse Interstellar Medium, ed. M.~{Haverkorn} \& W.~M.
  {Goss}, 271.
\newblock \doarXiv{astro-ph/0612109}

\bibitem[{Driessen {et~al.}(2019)Driessen, Janssen, Bassa, Stappers, \&
  Stinebring}]{Driessen2019}
Driessen, L.~N., Janssen, G.~H., Bassa, C.~G., Stappers, B.~W., \& Stinebring,
  D.~R. 2019, \mnras, 483, 1224, \dodoi{10.1093/mnras/sty3192}

\bibitem[{Eilek \& Hankins(2016)}]{Eilek2016}
Eilek, J.~A., \& Hankins, T.~H. 2016, Journal of Plasma Physics, 82, 635820302,
  \dodoi{10.1017/S002237781600043X}

\bibitem[{Enoto {et~al.}(2021)Enoto, Terasawa, Kisaka, Hu, Guillot,
  Lewandowska, Malacaria, Ray, Ho, Harding, Okajima, Arzoumanian, Gendreau,
  Wadiasingh, Markwardt, Soong, Kenyon, Bogdanov, Majid, G{\"u}ver, Jaisawal,
  Foster, Murata, Takeuchi, Takefuji, Sekido, Yonekura, Misawa, Tsuchiya, Aoki,
  Tokumaru, Honma, Kameya, Oyama, Asano, Shibata, \& Tanaka}]{Enoto2021}
Enoto, T., Terasawa, T., Kisaka, S., {et~al.} 2021, Science, 372, 187,
  \dodoi{10.1126/science.abd4659}

\bibitem[{Gommers {et~al.}(2022)Gommers, Virtanen, Burovski, Weckesser,
  Oliphant, Cournapeau, Haberland, Reddy, Alexbrc, Peterson, Nelson, Wilson,
  Endolith, Mayorov, Polat, Van Der~Walt, Laxalde, Brett, Larson, Millman,
  Lars, Peterbell10, Roy, Van~Mulbregt, Carey, Eric-Jones, Sakai, Moore, Kai,
  \& Kern}]{Gommers2022}
Gommers, R., Virtanen, P., Burovski, E., {et~al.} 2022, scipy/scipy: SciPy
  1.8.0, v1.8.0, Zenodo,  Zenodo, \dodoi{10.5281/zenodo.595738}

\bibitem[{Gupta {et~al.}(1999)Gupta, Bhat, \& Rao}]{Gupta1999}
Gupta, Y., Bhat, N. D.~R., \& Rao, A.~P. 1999, The Astrophysical Journal, 520,
  173, \dodoi{10.1086/307442}

\bibitem[{Gwinn {et~al.}(1998)Gwinn, Britton, Reynolds, Jauncey, King,
  McCulloch, Lovell, \& Preston}]{Gwinn1998}
Gwinn, C.~R., Britton, M.~C., Reynolds, J.~E., {et~al.} 1998, \apj, 505, 928,
  \dodoi{10.1086/306178}

\bibitem[{Hankins \& Eilek(2007)}]{Hankins2007}
Hankins, T.~H., \& Eilek, J.~A. 2007, \apj, 670, 693, \dodoi{10.1086/522362}

\bibitem[{Hankins {et~al.}(2016)Hankins, Eilek, \& Jones}]{Hankins2016}
Hankins, T.~H., Eilek, J.~A., \& Jones, G. 2016, \apj, 833, 47,
  \dodoi{10.3847/1538-4357/833/1/47}

\bibitem[{{Hankins} {et~al.}(2003){Hankins}, {Kern}, {Weatherall}, \&
  {Eilek}}]{Hankins2003}
{Hankins}, T.~H., {Kern}, J.~S., {Weatherall}, J.~C., \& {Eilek}, J.~A. 2003,
  \nat, 422, 141, \dodoi{10.1038/nature01477}

\bibitem[{Harding {et~al.}(2008)Harding, Stern, Dyks, \&
  Frackowiak}]{Harding2008}
Harding, A.~K., Stern, J.~V., Dyks, J., \& Frackowiak, M. 2008, \apj, 680,
  1378, \dodoi{10.1086/588037}

\bibitem[{{Harris} {et~al.}(2020){Harris}, {Millman}, {van der Walt},
  {Gommers}, {Virtanen}, {Cournapeau}, {Wieser}, {Taylor}, {Berg}, {Smith},
  {Kern}, {Picus}, {Hoyer}, {van Kerkwijk}, {Brett}, {Haldane}, {del R{\'\i}o},
  {Wiebe}, {Peterson}, {G{\'e}rard-Marchant}, {Sheppard}, {Reddy}, {Weckesser},
  {Abbasi}, {Gohlke}, \& {Oliphant}}]{Harris2020}
{Harris}, C.~R., {Millman}, K.~J., {van der Walt}, S.~J., {et~al.} 2020, \nat,
  585, 357, \dodoi{10.1038/s41586-020-2649-2}

\bibitem[{Hobbs \& Edwards(2012)}]{Hobbs2012}
Hobbs, G., \& Edwards, R. 2012, Tempo2: Pulsar Timing Package, Astrophysics
  Source Code Library, record ascl:1210.015.
\newblock \doeprint{1210.015}

\bibitem[{Hunter(2007)}]{Hunter2007}
Hunter, J.~D. 2007, Computing in Science \& Engineering, 9, 90,
  \dodoi{10.1109/MCSE.2007.55}

\bibitem[{{Jessner} {et~al.}(2010){Jessner}, {Popov}, {Kondratiev}, {Kovalev},
  {Graham}, {Zensus}, {Soglasnov}, {Bilous}, \& {Moshkina}}]{Jessner2010}
{Jessner}, A., {Popov}, M.~V., {Kondratiev}, V.~I., {et~al.} 2010, \aap, 524,
  A60, \dodoi{10.1051/0004-6361/201014806}

\bibitem[{Jun(1998)}]{Jun1998}
Jun, B.-I. 1998, \apj, 499, 282, \dodoi{10.1086/305627}

\bibitem[{Kaplan {et~al.}(2008)Kaplan, Chatterjee, Gaensler, \&
  Anderson}]{Kaplan2008}
Kaplan, D.~L., Chatterjee, S., Gaensler, B.~M., \& Anderson, J. 2008, \apj,
  677, 1201, \dodoi{10.1086/529026}

\bibitem[{Karuppusamy {et~al.}(2010)Karuppusamy, Stappers, \& van
  Straten}]{Karuppusamy2010}
Karuppusamy, R., Stappers, B.~W., \& van Straten, W. 2010, \aap, 515, A36,
  \dodoi{10.1051/0004-6361/200913729}

\bibitem[{Keimpema {et~al.}(2015)Keimpema, Kettenis, Pogrebenko, Campbell,
  Cim{\'o}, Duev, Eldering, Kruithof, van Langevelde, Marchal,
  Molera~Calv{\'e}s, Ozdemir, Paragi, Pidopryhora, Szomoru, \&
  Yang}]{Keimpema2015}
Keimpema, A., Kettenis, M.~M., Pogrebenko, S.~V., {et~al.} 2015, Experimental
  Astronomy, 39, 259, \dodoi{10.1007/s10686-015-9446-1}

\bibitem[{Lawrence {et~al.}(1995)Lawrence, MacAlpine, Uomoto, Woodgate, Brown,
  Oliversen, Lowenthal, \& Liu}]{Lawrence1995}
Lawrence, S.~S., MacAlpine, G.~M., Uomoto, A., {et~al.} 1995, \aj, 109, 2635,
  \dodoi{10.1086/117477}

\bibitem[{Loken {et~al.}(2010)Loken, Gruner, Groer, Peltier, Bunn, Craig,
  Henriques, Dempsey, Yu, Chen, Dursi, Chong, Northrup, Pinto, Knecht, \&
  Zon}]{Loken2010}
Loken, C., Gruner, D., Groer, L., {et~al.} 2010, Journal of Physics: Conference
  Series, 256, 012026, \dodoi{10.1088/1742-6596/256/1/012026}

\bibitem[{{Losovsky} {et~al.}(2019){Losovsky}, {Dumsky}, \&
  {Belyatsky}}]{Losovsky2019}
{Losovsky}, B.~Y., {Dumsky}, D.~V., \& {Belyatsky}, Y.~A. 2019, Astronomy
  Reports, 63, 830, \dodoi{10.1134/S1063772919090051}

\bibitem[{Lyne {et~al.}(1993)Lyne, Pritchard, \& Graham~Smith}]{Lyne1993}
Lyne, A.~G., Pritchard, R.~S., \& Graham~Smith, F. 1993, \mnras, 265, 1003,
  \dodoi{10.1093/mnras/265.4.1003}

\bibitem[{Lyne {et~al.}(2001)Lyne, Pritchard, \& Graham-Smith}]{Lyne2001}
Lyne, A.~G., Pritchard, R.~S., \& Graham-Smith, F. 2001, \mnras, 321, 67,
  \dodoi{10.1046/j.1365-8711.2001.03998.x}

\bibitem[{{Lyubarskii}(1996)}]{Lyubarskii1996}
{Lyubarskii}, Y.~E. 1996, \aap, 311, 172

\bibitem[{Lyutikov(2021)}]{Lyutikov2021}
Lyutikov, M. 2021, The Astrophysical Journal, 922, 166,
  \dodoi{10.3847/1538-4357/ac1b32}

\bibitem[{Mahajan \& Lin(2022)}]{Mahajan2022}
Mahajan, N., \& Lin, R. 2022, theXYZT/pulsarbat: pulsarbat 0.0.8, v0.0.7,
  Zenodo, \dodoi{10.5281/zenodo.6934355}

\bibitem[{{Main} {et~al.}(2021){Main}, {Lin}, {van Kerkwijk}, {Pen},
  {Rudnitskii}, {Popov}, {Soglasnov}, \& {Lyutikov}}]{Main2021}
{Main}, R., {Lin}, R., {van Kerkwijk}, M.~H., {et~al.} 2021, \apj, 915, 65,
  \dodoi{10.3847/1538-4357/ac01c6}

\bibitem[{Martin {et~al.}(2021)Martin, Milisavljevic, \& Drissen}]{Martin2021}
Martin, T., Milisavljevic, D., \& Drissen, L. 2021, \mnras, 502, 1864,
  \dodoi{10.1093/mnras/staa4046}

\bibitem[{{McKee} {et~al.}(2018){McKee}, {Lyne}, {Stappers}, {Bassa}, \&
  {Jordan}}]{McKee2018}
{McKee}, J.~W., {Lyne}, A.~G., {Stappers}, B.~W., {Bassa}, C.~G., \& {Jordan},
  C.~A. 2018, \mnras, 479, 4216, \dodoi{10.1093/mnras/sty1727}

\bibitem[{{Michilli} {et~al.}(2018){Michilli}, {Seymour}, {Hessels}, {Spitler},
  {Gajjar}, {Archibald}, {Bower}, {Chatterjee}, {Cordes}, {Gourdji}, {Heald},
  {Kaspi}, {Law}, {Sobey}, {Adams}, {Bassa}, {Bogdanov}, {Brinkman},
  {Demorest}, {Fernandez}, {Hellbourg}, {Lazio}, {Lynch}, {Maddox}, {Marcote},
  {McLaughlin}, {Paragi}, {Ransom}, {Scholz}, {Siemion}, {Tendulkar}, {van
  Rooy}, {Wharton}, \& {Whitlow}}]{michilli_2018_natur}
{Michilli}, D., {Seymour}, A., {Hessels}, J.~W.~T., {et~al.} 2018, \nat, 553,
  182, \dodoi{10.1038/nature25149}

\bibitem[{Moffett \& Hankins(1996)}]{Moffett1996}
Moffett, D.~A., \& Hankins, T.~H. 1996, \apj, 468, 779, \dodoi{10.1086/177734}

\bibitem[{Muslimov \& Harding(2004)}]{Muslimov2004}
Muslimov, A.~G., \& Harding, A.~K. 2004, The Astrophysical Journal, 617, 471,
  \dodoi{10.1086/425227}

\bibitem[{{Nimmo} {et~al.}(2021){Nimmo}, {Hessels}, {Keimpema}, {Archibald},
  {Cordes}, {Karuppusamy}, {Kirsten}, {Li}, {Marcote}, \&
  {Paragi}}]{Nimmo_2021_NatAs}
{Nimmo}, K., {Hessels}, J.~W.~T., {Keimpema}, A., {et~al.} 2021, Nature
  Astronomy, 5, 594, \dodoi{10.1038/s41550-021-01321-3}

\bibitem[{{Nimmo} {et~al.}(2022){Nimmo}, {Hessels}, {Kirsten}, {Keimpema},
  {Cordes}, {Snelders}, {Hewitt}, {Karuppusamy}, {Archibald}, {Bezrukovs},
  {Bhardwaj}, {Blaauw}, {Buttaccio}, {Cassanelli}, {Conway}, {Corongiu},
  {Feiler}, {Fonseca}, {Forss{\'e}n}, {Gawro{\'n}ski}, {Giroletti}, {Kharinov},
  {Leung}, {Lindqvist}, {Maccaferri}, {Marcote}, {Masui}, {Mckinven},
  {Melnikov}, {Michilli}, {Mikhailov}, {Ng}, {Orbidans}, {Ould-Boukattine},
  {Paragi}, {Pearlman}, {Petroff}, {Rahman}, {Scholz}, {Shin}, {Smith},
  {Stairs}, {Surcis}, {Tendulkar}, {Vlemmings}, {Wang}, {Yang}, \&
  {Yuan}}]{Nimmo_2022_NatAs}
{Nimmo}, K., {Hessels}, J.~W.~T., {Kirsten}, F., {et~al.} 2022, Nature
  Astronomy, 6, 393, \dodoi{10.1038/s41550-021-01569-9}

\bibitem[{Osterbrock(1957)}]{Osterbrock1957}
Osterbrock, D.~E. 1957, \pasp, 69, 227, \dodoi{10.1086/127053}

\bibitem[{Pen {et~al.}(2014)Pen, Macquart, Deller, \& Brisken}]{Pen2014}
Pen, U.~L., Macquart, J.~P., Deller, A.~T., \& Brisken, W. 2014, \mnras, 440,
  L36, \dodoi{10.1093/mnrasl/slu010}

\bibitem[{{Petroff} {et~al.}(2022){Petroff}, {Hessels}, \&
  {Lorimer}}]{petroff_2022_aarv}
{Petroff}, E., {Hessels}, J.~W.~T., \& {Lorimer}, D.~R. 2022, \aapr, 30, 2,
  \dodoi{10.1007/s00159-022-00139-w}

\bibitem[{Philippov {et~al.}(2019)Philippov, Uzdensky, Spitkovsky, \&
  Cerutti}]{Philippov2019}
Philippov, A., Uzdensky, D.~A., Spitkovsky, A., \& Cerutti, B. 2019, \apjl,
  876, L6, \dodoi{10.3847/2041-8213/ab1590}

\bibitem[{Ponce {et~al.}(2019)Ponce, van Zon, Northrup, Gruner, Chen, Ertinaz,
  Fedoseev, Groer, Mao, Mundim, Nolta, Pinto, Saldarriaga, Slavnic, Spence, Yu,
  \& Peltier}]{Ponce2019}
Ponce, M., van Zon, R., Northrup, S., {et~al.} 2019, CoRR, abs/1907.13600.
\newblock \doarXiv{1907.13600}

\bibitem[{Rankin \& Counselman(1973)}]{Rankin1973}
Rankin, J.~M., \& Counselman, C.~C., I. I.~I. 1973, \apj, 181, 875,
  \dodoi{10.1086/152099}

\bibitem[{Rickett(1990)}]{Rickett1990}
Rickett, B.~J. 1990, \araa, 28, 561,
  \dodoi{10.1146/annurev.aa.28.090190.003021}

\bibitem[{Romani \& Yadigaroglu(1995)}]{Romani1995}
Romani, R.~W., \& Yadigaroglu, I.~A. 1995, \apj, 438, 314,
  \dodoi{10.1086/175076}

\bibitem[{{Rudnitskii} {et~al.}(2016){Rudnitskii}, {Karuppusamy}, {Popov}, \&
  {Soglasnov}}]{Rudnitskii2016}
{Rudnitskii}, A.~G., {Karuppusamy}, R., {Popov}, M.~V., \& {Soglasnov}, V.~A.
  2016, Astronomy Reports, 60, 211, \dodoi{10.1134/S1063772916020116}

\bibitem[{Ryan \& Vandenberg(1980)}]{Ryan1980}
Ryan, J.~W., \& Vandenberg, N.~R. 1980, in Bulletin of the American
  Astronomical Society, Vol.~12, 457.
\newblock \url{https://ui.adsabs.harvard.edu/abs/1980BAAS...12..457R}

\bibitem[{Sallmen {et~al.}(1999)Sallmen, Backer, Hankins, Moffett, \&
  Lundgren}]{Sallmen1999}
Sallmen, S., Backer, D.~C., Hankins, T.~H., Moffett, D., \& Lundgren, S. 1999,
  \apj, 517, 460, \dodoi{10.1086/307183}

\bibitem[{Shearer {et~al.}(2003)Shearer, Stappers, O'Connor, Golden, Strom,
  Redfern, \& Ryan}]{Shearer2003}
Shearer, A., Stappers, B., O'Connor, P., {et~al.} 2003, Science, 301, 493,
  \dodoi{10.1126/science.1084919}

\bibitem[{Shearer {et~al.}(2012)Shearer, Collins, Naletto, Barbieri, Zampieri,
  Germana, Gradari, Verroi, \& Stappers}]{Shearer2012}
Shearer, A., Collins, S., Naletto, G., {et~al.} 2012, in Astronomical Society
  of the Pacific Conference Series, Vol. 466, Electromagnetic Radiation from
  Pulsars and Magnetars, ed. W.~{Lewandowski}, O.~{Maron}, \& J.~{Kijak}, 11.
\newblock \url{https://ui.adsabs.harvard.edu/abs/2012ASPC..466...11S}

\bibitem[{Smirnova {et~al.}(1996)Smirnova, Shishov, \& Malofeev}]{Smirnova1996}
Smirnova, T.~V., Shishov, V.~I., \& Malofeev, V.~M. 1996, in Astronomical
  Society of the Pacific Conference Series, Vol. 105, IAU Colloq. 160: Pulsars:
  Problems and Progress, ed. S.~{Johnston}, M.~A. {Walker}, \& M.~{Bailes},
  475.
\newblock \url{https://ui.adsabs.harvard.edu/abs/1996ASPC..105..475S}

\bibitem[{Smits {et~al.}(2017)Smits, Bassa, Janssen, Karuppusamy, Kramer, Lee,
  Liu, McKee, Perrodin, Purver, Sanidas, Stappers, \& Zhu}]{Smits2017}
Smits, R., Bassa, C.~G., Janssen, G.~H., {et~al.} 2017, Astronomy and
  Computing, 19, 66, \dodoi{10.1016/j.ascom.2017.02.002}

\bibitem[{Sobey {et~al.}(2019)Sobey, Bilous, Grie{\ss}meier, Hessels,
  Karastergiou, Keane, Kondratiev, Kramer, Michilli, Noutsos, Pilia, Polzin,
  Stappers, Tan, van Leeuwen, Verbiest, Weltevrede, Heald, Alves, Carretti,
  En{\ss}lin, Haverkorn, Iacobelli, Reich, \& Van~Eck}]{Sobey2019}
Sobey, C., Bilous, A.~V., Grie{\ss}meier, J.~M., {et~al.} 2019, \mnras, 484,
  3646, \dodoi{10.1093/mnras/stz214}

\bibitem[{{Spitler} {et~al.}(2016){Spitler}, {Scholz}, {Hessels}, {Bogdanov},
  {Brazier}, {Camilo}, {Chatterjee}, {Cordes}, {Crawford}, {Deneva}, {Ferdman},
  {Freire}, {Kaspi}, {Lazarus}, {Lynch}, {Madsen}, {McLaughlin}, {Patel},
  {Ransom}, {Seymour}, {Stairs}, {Stappers}, {van Leeuwen}, \&
  {Zhu}}]{spitler_2016_natur}
{Spitler}, L.~G., {Scholz}, P., {Hessels}, J.~W.~T., {et~al.} 2016, \nat, 531,
  202, \dodoi{10.1038/nature17168}

\bibitem[{Staelin \& Reifenstein(1968)}]{Staelin1968}
Staelin, D.~H., \& Reifenstein, Edward~C., I. I.~I. 1968, Science, 162, 1481,
  \dodoi{10.1126/science.162.3861.1481}

\bibitem[{Thompson {et~al.}(2017)Thompson, Moran, \& Swenson}]{Thompson2017}
Thompson, A.~R., Moran, J.~M., \& Swenson, George~W., J. 2017, Interferometry
  and Synthesis in Radio Astronomy, 3rd Edition (Springer),
  \dodoi{10.1007/978-3-319-44431-4}

\bibitem[{Trimble(1968)}]{Trimble1968}
Trimble, V. 1968, \aj, 73, 535, \dodoi{10.1086/110658}

\bibitem[{Trimble(1973)}]{Trimble1973}
---. 1973, \pasp, 85, 579, \dodoi{10.1086/129507}

\bibitem[{Van~Kerkwijk {et~al.}(2020)Van~Kerkwijk, Zhu, Guo, Lin, Mahajan,
  McKenna, Main, Simard, Stein, \& Van~Lieshout}]{VanKerkwijk2020}
Van~Kerkwijk, M.~H., Zhu, C.~C., Guo, S.~G., {et~al.} 2020, mhvk/baseband,
  v4.0.3,  Zenodo, \dodoi{10.5281/zenodo.4292543}

\bibitem[{van Straten {et~al.}(2010)van Straten, Manchester, Johnston, \&
  Reynolds}]{Straten2010}
van Straten, W., Manchester, R.~N., Johnston, S., \& Reynolds, J.~E. 2010,
  Publications of the Astronomical Society of Australia, 27, 104,
  \dodoi{10.1071/as09084}

\bibitem[{Vandenberg(1976)}]{Vandenberg1976}
Vandenberg, N.~R. 1976, \apj, 209, 578, \dodoi{10.1086/154753}

\bibitem[{Wolszczan \& Cordes(1987)}]{Wolszczan1987}
Wolszczan, A., \& Cordes, J. 1987, Astrophys. J., 320, L35,
  \dodoi{10.1086/184972}

\bibitem[{{Yalinewich} \& {Pen}(2022)}]{Yalinewich2022}
{Yalinewich}, A., \& {Pen}, U.-L. 2022, \mnras, 515, 5682,
  \dodoi{10.1093/mnras/stac2087}

\end{thebibliography}

\end{document}